\documentclass[a4paper,11pt]{article}
\pdfoutput=1 

\usepackage{jcappub} 
\usepackage[T1]{fontenc} 
\usepackage{tikz-feynhand}
\usepackage{graphicx}
\usepackage{xcolor}
\usepackage{caption,subcaption}
\usepackage{mathrsfs,mathtools}
\usepackage{physics,amssymb}
\usepackage{bm}
\usepackage{braket}
\usepackage{listings}
\usepackage{cases}
\usepackage{comment}
\usepackage{soul}
\usepackage{cancel}
\usepackage{cases}
\usepackage[utf8]{inputenc}
\usepackage{url}
\usepackage[normalem]{ulem}
\usepackage{xspace}
\usepackage{aas_macros}
\usepackage{acronym}
\usepackage{siunitx}
\usepackage {diagbox}

\hypersetup{colorlinks=true
,urlcolor=DARKBLUE
,anchorcolor=DARKBLUE
,citecolor=DARKBLUE
,filecolor=DARKBLUE
,linkcolor=DARKBLUE
,menucolor=DARKBLUE
,linktocpage=true
,pdfproducer=medialab
,pdfa=true
}

\usepackage{xcolor}
\definecolor{bleudefrance}{rgb}{0.19, 0.55, 0.91}
\definecolor{desyblue}{HTML}{009EE2}
\definecolor{desyorange}{HTML}{FD8800}
\definecolor{dark_red}{rgb}{0.7, 0., 0.}
\definecolor{light_pink}{rgb}{1,0.4,0.4}
\definecolor{lblue}{rgb}{0.384602,0.117763,0.973947}
\usepackage{color}
\input{colordvi.tex}
\allowdisplaybreaks[1]
\usepackage{url}
\usepackage{hyperref}
\hypersetup{
colorlinks=false,
hidelinks}
\newcommand*{\D}{{\rm d}}
\newcommand*{\mpl}{M_{\rm Pl}}

\definecolor{MONZA}{HTML}{CF000F}
\definecolor{DARKBLUE}{HTML}{00008b}
\definecolor{DARKMAGENTA}{HTML}{8b008b}
\definecolor{DARKCYAN}{HTML}{00cfc0}

\begin{document}
\title{Primordial full bispectra from the general bounce cosmology}


\author{Shingo Akama}


\affiliation{Faculty of Physics, Astronomy and Applied Computer Science, Jagiellonian University, 30-348 Krakow, Poland}

\emailAdd{shingo.akama@uj.edu.pl}

\date{\today}

\abstract{Primordial non-Gaussianities are key quantities to test early universe scenarios. In this paper, we compute full bispectra of scalar and tensor perturbations generated during a contracting phase in a general bounce model. The general bounce model consists of two branches: one realizes scale-invariant scalar and tensor power spectra from perturbations whose amplitudes become constant on superhorizon scales, as in de Sitter inflation, while the other realizes scale-invariant power spectra from perturbations whose amplitudes grow on superhorizon scales, as in matter bounce cosmology. We study the auto- and cross-bispectra originating from the scale-invariant scalar and tensor perturbations in these two branches. We investigate the amplitudes and shapes of non-Gaussianities and find that the differences between the two branches manifest for equilateral and squeezed momentum triangle configurations. In particular, one of the branches in which the superhorizon perturbations are conserved reproduces the so-called Maldacena's consistency relation. By examining perturbativity conditions and considering current observational constraints on primordial non-Gaussianities, we also find a viable parameter space in which both theoretical and observational constraints are satisfied simultaneously. }

\maketitle
\flushbottom

\section{Introduction and summary}

Explaining the observed cosmic microwave background (CMB) anisotropies is one of the key components for successful early universe scenarios. 
So far, inflation~\cite{Guth:1980zm,Sato:1981qmu,Linde:1983gd} has been the most successful scenario, and future CMB experiments are expected to clarify which model in that leading paradigm described the early epoch of our universe. On the other hand, bounces~\cite{Battefeld:2014uga,Brandenberger:2016vhg} and galilean genesis~\cite{Creminelli:2010ba} have also been studied as alternatives that avoid the initial singularity~\cite{Borde:1996pt} predicted in an inflationary spacetime.\footnote{See also Refs.~\cite{Yoshida:2018ndv,Nomura:2021lzz,Lesnefsky:2022fen,Geshnizjani:2023hyd,Easson:2024uxe,Easson:2024fzn,Garcia-Saenz:2024ogr} for recent discussions on the geodesic incompletness and the initial singularity in the inflationary spacetime.} Similarly to inflation, its nonsingular alternatives that can explain the observed CMB fluctuations such as a small tensor-to-scalar ratio and a small scalar non-Gaussianity have been found, e.g., in a general single-field scalar-tensor framework in~\cite{Akama:2019qeh}. Observationally distinguishing inflation from its alternatives by future experiments is thus important to clarify which early universe scenario or model is correct and ultimately test whether inflation indeed occurred or not. This motivates us to investigate observational predictions of the alternatives to inflation as well as those of inflation to explore the predictions that can make each model observationally conclusive and also to clarify a model space in which their observational signatures are degenerate.

When amplitudes of cosmological perturbations become constant on large scales as in the conventional slow-roll inflation or grow on large scales as in the matter bounce scenario in which a contracting phase is described by a matter-dominated phase~\cite{Wands:1998yp,Finelli:2001sr,Brandenberger:2012zb}, their power spectra are scale invariant. In general, not only a scale factor but also coefficients of quadratic actions of perturbations affect the time evolution of linear perturbations, and hence a different time evolution of a background spacetime can yield the same time evolution of linear perturbations depending on the time dependence of the coefficients of the quadratic actions, which can be understood by invoking a conformal or disformal transformation. For example, (nearly) scale-invariant fluctuations on a non-de Sitter background do not necessarily grow (see e.g., Refs~\cite{Khoury:2008wj,Khoury:2009my,Khoury:2010gw,Baumann:2011dt,Nishi:2015pta,Nishi:2016ljg,Nandi:2019xag,Akama:2019qeh,Nandi:2020sif,Nandi:2020szp,Nandi:2022twa,Ganz:2022zgs}). Since it is nontrivial whether the same time evolution of linear perturbations in different models yields the same non-Gaussian signatures, it is important to study the similarities and differences at the level of non-Gaussianities. As quantities to characterize the non-Gaussianities, we study three-point correlation functions that have been extensively studied in the context of inflation (see, e.g., Refs.~\cite{Bartolo:2004if,Chen:2010xka} for reviews), while have been less studied in the context of alternatives to inflation. In particular, the study of primordial cross-bispectra of scalar and tensor perturbations is lacking in the latter context. In light of the expected future detection of a CMB B-mode polarization~\cite{SimonsObservatory:2018koc, Abazajian:2019eic, Moncelsi:2020ppj, SPIDER:2021ncy, LiteBIRD:2022cnt}, it is important to study the cross-bispectra as well.

Among the nonsingular alternatives, a general class of galilean genesis suffers from inconsistency with the current CMB experiments due to a large scalar non-Gaussianity~\cite{Akama:2022usl}. Bounce models have suffered from a similar issue known as a no-go theorem stating that all single-field matter-dominated contracting models in k-essence theories predict either a large tensor-to-scalar ratio or a large scalar non-Gaussianity~\cite{Li:2016xjb,Quintin:2015rta}, but this discrepancy with the CMB experiments can be resolved in beyond k-essence theories~\cite{Akama:2019qeh}. 
In this paper, we study bouncing cosmologies in a more general framework. To extensively study the observational predictions, it is useful to use a general framework instead of working out an individual model. So far, the authors of Ref.~\cite{Akama:2022usl} have constructed a general framework of bouncing cosmologies in the Horndeski theory which is the most general single-field scalar-tensor theory with second-order field equations~\cite{Horndeski:1974wa,Deffayet:2011gz,Kobayashi:2011nu} (see Ref.~\cite{Kobayashi:2019hrl} for a review). That general framework thus makes it possible to study a vast class of bouncing cosmologies that reside in single-field scalar-tensor theories, including the matter bounce scenario, in a unified way.

In this paper, similarly to Ref.~\cite{Akama:2019qeh}, we study primordial two-point and three-point correlation functions during a contracting phase and ignore any impacts of subsequent bouncing and expanding phases that follow the contracting phase. Whether this is justified or not is model-dependent. For example, in a single matter field case, the later phases can change not only the two-point function but also the three-point function, e.g., for a relatively long bouncing phase during which amplification of curvature perturbations cannot be ignored~\cite{Quintin:2015rta}.\footnote{This amplification was motivated by the fact that the tensor-to-scalar ratio $r$ during a matter-dominated contracting phase is $\mathcal{O}(10)$ in a k-essence theory~\cite{Quintin:2015rta,Li:2016xjb} if curvature perturbations are almost Gaussian. Such a contracting model is excluded by the current constraint ($r\leq\mathcal{O}(10^{-2})$~\cite{Planck:2018jri}) unless $r$ is suppressed during the subsequent phases. In this case, one of the ways to suppress $r$ was to amplify the amplitude of curvature perturbations during the bouncing phase, which enhanced a scalar non-Gaussianity. However, that amplification is not always necessary for models whose contracting phase is described by a more general scalar-tensor theory such as a cubic Galileon theory in which the tensor-to-scalar ratio during the contracting phase can be much smaller than unity, while keeping the curvature perturbations almost Gaussian~\cite{Akama:2019qeh}.} See also Ref.~\cite{Chowdhury:2015cma} in which overall amplitudes of both tensor perturbations and their auto-bispectrum can change during the subsequent phases. (Note, however, that scale dependence of a non-linearity parameter of a tensor non-Gaussianity in their model is determined during a contracting phase~\cite{Chowdhury:2015cma,Akama:2019qeh}.) On the other hand, a sufficiently short bouncing phase can make it possible to match cosmological perturbations at the end of a contracting phase to those at the beginning of an expanding phase after bounce (see, e.g., Ref.~\cite{Gao:2009wn}), that is, the effects of the short bouncing phase can be naively ignored depending on its concrete realization.
Furthermore, there exists an example of a bounce model, Ref.~\cite{Nandi:2019xag}, in which two-point and three-point statistics of tensor perturbations are mainly determined during a contracting phase. In the present paper, we assume that the subsequent phases on the primordial correlation functions can be ignored (due to an instantaneous transition from the contracting phase to the subsequent expanding phase) and compare the theoretical predictions from the contracting phase with observations.

Let us present the main results of this paper. \\
{\leftline{$\bullet${$\it Non$}-{$\it Gaussianity\ consistency\ relation\ and\ its\ violation$}}}\
The general bounce cosmology has two branches: one is conformally equivalent to (single-clock) inflation, and the other is conformally equivalent to matter bounce. The first branch in which the superhorizon modes are conserved reproduces the so-called Maldacena's consistency relation for the full bispectra under the exact scale-invariant limit, which means that primordial non-Gaussianities from this branch are suppressed for several squeezed momentum triangle configurations. On the other hand, the second branch in which the superhorizon modes grow can predict nonvanishing signals and sharp (divergent) peaks for those momentum triangle configurations due to the violation of the non-Gaussianity consistency relation. In particular, these findings for scalar-tensor cross-bispectra are obtained for the first time in the context of bouncing cosmologies in the present paper.\\
{\leftline{$\bullet${$\it Distinctive \ features\ in\ the\ shapes\ of\ non$}-{$\it Gaussianities$}}}\\
The consistency relation for the first branch affects the shapes of non-Gaussianities in such a way that some of the peak behaviors that could appear at the squeezed limits vanish. Indeed, potentially divergent peaks in a scalar auto-bispectrum and scalar-tensor-tensor cross-bispectra disappear. Also, some of the bispectra peak for an equilateral momentum triangle configuration. On the other hand, due to the violation of the non-Gaussianity consistency relation, the second branch predicts squeezed non-Gaussianities over a wider parameter range than the first branch. However, the second branch does not predict any peaks at the equilateral limit. Therefore, the equilateral and squeezed non-Gaussianities are distinctive features of the first and second branches, respectively.\\
{\leftline{$\bullet${$\it Viable\ parameter\ space\ avoiding\ theoretical\ and\ observational\ constraints$}}}\\
As theoretical constraints, we study perturbativity conditions such that cubic actions (Lagrangians) are smaller than quadratic actions (Lagrangians) at horizon-cross and superhorizon scales. The ratios of the cubic Lagrangians to the quadratic ones characterize non-linear corrections to linear perturbations. As observational constraints, we take into account the current constraints on primordial non-Gaussianities obtained through the CMB experiments. We find a viable parameter space in which the perturbativity conditions and the observational constraints on the primordial non-Gaussianities are satisfied simultaneously.

The rest of the present paper is organized as follows. In the following section, we review the general framework of bouncing cosmologies. In Sec.~\ref{Sec: powerspectra}, we show explicit forms of primordial scalar and tensor power spectra. In the same section, we show the relation of the general model to inflation and matter bounce by using a conformal transformation. In Sec.~\ref{Sec: bispectra}, we compute the full bispectra from the general bounce cosmology by focusing on the exact scale-invariant fluctuations. Then, we study the similarity between inflation and one of the branches in which the superhorizon modes are conserved from the viewpoint of non-Gaussianity consistency relations. In the same section, we study the shapes of bispectra. (The analysis of the shape for the branch which is conformally equivalent to inflation is mostly overlapped with that for inflation, and hence we summarize the shapes for that branch in Appendix~\ref{Appendix: shape}.)
In Sec.~\ref{Sec: const}, we study the perturbativity conditions and the observational constraints on the primordial non-Gaussianities and clarify a viable parameter space in which both theoretical and observational constraints are satisfied simultaneously. Our conclusion is drawn in Sec.~\ref{Sec: summary}. In Appendix~\ref{Sec: app-perturbed-actions}, we summarize both quadratic and cubic actions derived from the Horndeski action. We also show total-time-derivative terms involving a time derivative. These terms for the scalar-tensor cubic interactions are derived from the full Horndeski action for the first time. In Appendix~\ref{Sec: app-boundary}, we compute the contributions from field redefinitions and total-time-derivative terms to the scalar auto-bispectrum and the scalar-tensor cross-bispectra. In Appendix~\ref{Sec: app-polarization-tensor}, we summarize explicit expressions of polarization tensors for tensor perturbations and these products. In Appendix~\ref{Sec: app-consistency}, we summarize the detailed drivation of the non-Gaussianity consistency relation. In Appendix~\ref{app: example}, we present an example of a contracting model in which the perturbations are conserved on superhorizon scales.

\section{Framework of general bounce cosmology}\label{Sec: setup}

We study the contracting phase with a spatially flat Friedmann-Lema\^{i}tre-Robertson-Walker (FLRW) metric:
\begin{align}
\D s^2&=-\D t^2+a^2(t)\delta_{ij}\D x^i \D x^j\notag\\
&=a^2(\eta)(-\D\eta^2+\delta_{ij}\D x^i \D x^j), \label{eq: bgmetric}
\end{align}
with the conformal time $\eta$ defined by $\D\eta:=\D t/a$. As notations of differentiations with respect to the cosmic time $t$ and the conformal time $\eta$, we use a dot and a prime, respectively. Throughout the contracting phase studied in this paper, the scale factor evolves with time according to the following power-law dependence:
\begin{align}
a=\biggl(\frac{-t}{-t_b}\biggr)^n=\biggl(\frac{-\eta}{-\eta_b}\biggr)^{n/(1-n)}, \label{eq: scale-factor}
\end{align}
where the time at the end of the contracting phase is represented by the cosmic time as $t_b$ or the conformal time as $\eta_b$, both of which are negative. The variable $n$ in the above takes a value in the range $0<n<1$.
Then, the Hubble parameter $H:=\dot a/a$ takes $H=n/t$ during the contracting phase. For later convenience, we here show a relation between $H$ and $\eta$ at the end of the contracting phase:
\begin{align}
aH|_{\eta=\eta_b}=H|_{\eta=\eta_b}=\frac{n}{1-n}\frac{1}{\eta_b}. \label{eq: H-etab}
\end{align}
In a single-field framework, a contracting phase can be studied with a canonical scalar field or a k-essence field, but matter-dominated contracting models realizing scale-invariant power spectra in k-essence theories cannot explain the observed CMB fluctuations~\cite{Li:2016xjb}, which is not the case in beyond k-essence theories in general. Also, as we will explicitly show, a certain class of bounces having non-minimal coupling between a scalar field and gravity can lead to  distinctive features in primordial non-Gaussianities such as the non-Gaussianity consistency relation. See Ref.~\cite{Chowdhury:2015cma} for the consistency relation for a tensor auto-bispectrum. To study primordial correlation functions in such a scalar-tensor framework involving non-minimal coupling, we use the full Horndeski action~\cite{Horndeski:1974wa,Deffayet:2011gz,Kobayashi:2011nu}
\begin{align}
S&=\int\D^4 x\sqrt{-g}\biggl[G_2(\phi,X)-G_3(\phi,X)\Box\phi+G_4(\phi,X)R+G_{4X}[(\Box\phi)^2-(\nabla_{\mu}\nabla_{\nu}\phi)^2]\notag\\
&\quad +G_5(\phi,X)G_{\mu\nu}\nabla^{\mu}\nabla^{\nu}\phi-\frac{1}{6}G_{5X}[(\Box\phi)^3-3\Box\phi(\nabla_{\mu}\nabla_{\nu}\phi)^2+2(\nabla_{\mu}\nabla_{\nu}\phi)^3]\biggr] \, , \label{eq: Horndeski action}
\end{align}
where $X:=-g^{\mu\nu}\nabla_{\mu}\phi\nabla_{\nu}\phi/2$, and $G_i(\phi,X)$ are arbitrary functions of $\phi$ and $X$. Since the Horndeski theory is the most general single-field scalar-tensor theory with second-order field equations, one can study a wide variery of bouncing cosmologies with the above action. (See, e.g., Ref.~\cite{Easson:2011zy} and Ref.~\cite{Cai:2012va} for examples of bounces with a stable null energy condition violation in cubic Galileon theories. Note, however, that gradient instabilties occur at some moment.)
Here, let us mention the use of the Horndeski action for nonsingular cosmologies. So far, it has been shown that nonsingular cosmologies characterized by $a>{\rm const.}>0$, independently of the concrete models, exhibit gradient instabilities in a scalar sector (i.e., an exponential growth of scalar perturbations around the FLRW background) if the whole time evolution of the universe, i.e., the time evolution from the past infinity to the future infinity, was described by the Horndeski theory~\cite{Libanov:2016kfc,Kobayashi:2016xpl}. (See also Refs.~\cite{Kolevatov:2016ppi,Creminelli:2016zwa,Akama:2017jsa} for an extention of this statement to multi-field scalar-tensor theories.) The essential inequality for the proof of this no-go theorem for the nonsingular cosmologies is~\cite{Libanov:2016kfc,Kobayashi:2016xpl}\footnote{The convergence condition in Eq.~(\ref{eq: no-go-ineq}) has been found to be interpreted as geodesic incompleteness for gravitons with working in the so-called Einstein frame~\cite{Cai:2016thi,Creminelli:2016zwa} and without working in that frame~\cite{Akama:2017jsa}.}
\begin{align}
\int_{-\infty}^t\D t' a(t')\mathcal{F}_T(t')<\infty\ \  {\rm or}\ \int_t^{\infty}\D t' a(t')\mathcal{F}_T(t')<\infty, \label{eq: no-go-ineq}
\end{align}
where $\mathcal{F}_T$ will be defined later. For the nonsingular cosmological solutions, the way to avoid the gradient instabilities within the Horndeski theory is to take $\mathcal{F}_T\to0$ in the past or future infinity. (See e.g., Refs~\cite{Kobayashi:2016xpl,Ageeva:2024knc} for such examples.) In the models in which $\mathcal{F}_T$ is not asymptotic to $0$, some beyond-Horndeski higher-derivative terms may be invoked~\cite{Kobayashi:2015gga,Cai:2016thi, Creminelli:2016zwa, Misonoh:2016btv, Cai:2017tku, Yoshida:2017swb, Cai:2017dyi, Kolevatov:2017voe, Cai:2017dxl, Cai:2017pga, Kolevatov:2018mhu, Santoni:2018rrx, Heisenberg:2018wye, Mironov:2018oec, Ye:2019frg, Fukushima:2019yww, Ye:2019sth, Mironov:2019qjt, Mironov:2019haz,Cai:2019hge, Mironov:2019mye, Volkova:2019jlj, Ilyas:2020qja, Mironov:2020mfo, Mironov:2020pqh, Cai:2020qpu, Zhu:2021whu, Zhu:2021ggm, Mironov:2022ffa, Mironov:2024umy, Mironov:2024pjt, An:2025xeb}. (See also Refs.~\cite{Quintin:2019orx, Kim:2020iwq, HosseiniMansoori:2022xnq, Ganz:2022zgs} in which stable nonsingular solutions without introducing higher-derivative terms have been found outside the Horndeski theory.) However, the no-go theorem does not predict when the gradient instabilities occur during the whole time evolution of non-singular universes. In the present paper, we study the two cases: one where $\mathcal{F}_T$ is asymptotic to $0$ in the past infinity, and the other where $\mathcal{F}_T$ is not asymptotic to $0$ in the past infinity. For the latter case, we assume that at least the contracting phase of general bounce cosmology is governed by the Horndeski action, and at some point during the whole time evolution, a term beyond Horndeski acts to avoid the no-go theorem.

In the Horndeski theory, the gravitational field equations, corresponding to the Friedmann equation ($\mathcal{E}=0$) and the evolution equation ($\mathcal{P}=0$), are of the form~\cite{Kobayashi:2011nu}
\begin{align}
\mathcal{E}=\sum_{i=2}^5\mathcal{E}_i=0,\ \mathcal{P}=\sum_{i=2}^5\mathcal{P}_i=0,
\end{align}
where $\mathcal{E}_i$ and $\mathcal{P}_i$ stem from $G_i$ and take the following forms,
\begin{align}
\mathcal{E}_2&=2XG_{2X}-G_2,\\
\mathcal{E}_3&=6X\dot\phi{H}G_{3X}-2XG_{3\phi},\\
\mathcal{E}_4&=-6H^2G_4+24H^2{X}\left(G_{4X}+XG_{4XX}\right)-12HX\dot\phi{G_{4\phi{X}}}-6H\dot\phi{G_{4\phi}},\\
\mathcal{E}_5&=2H^3X\dot\phi\left(5G_{5X}+2XG_{5XX}\right)-6H^2X\left(3G_{5\phi}+2XG_{5\phi{X}}\right),
\end{align}
and
\begin{align}
\mathcal{P}_2&=G_2,\\
\mathcal{P}_3&=-2X(G_{3\phi}+\ddot{\phi}G_{3X}),\\
\mathcal{P}_4&=2(3H^2+2\dot{H})G_4-12H^2XG_{4X}-4H\dot{X}G_{4X}-8\dot{H}XG_{4X}-8HX\dot{X}G_{4XX}\notag\\
&\quad\ +2(\ddot{\phi}+2H\dot{\phi})G_{4\phi}+4XG_{4\phi\phi}+4X(\ddot{\phi}-2H\dot{\phi})G_{4\phi{X}},\\
\mathcal{P}_5&=-2X(2H^3\dot{\phi}+2H\dot{H}\dot{\phi}+3H^2\ddot{\phi})G_{5X}-4H^2X^2\ddot{\phi}G_{5XX}+4HX(\dot{X}-HX)G_{5\phi{X}}\notag\\
&\quad\ +2\left[2(HX)^{{\boldsymbol \cdot}}+3H^2X\right]G_{5\phi}+4HX\dot{\phi}G_{5\phi\phi},
\end{align}
with $G_\phi:=\partial_\phi G$ and $G_X:=\partial_X G$. Here, to compute the primordial correlation functions, we do not specify explicit forms of $G_i$. Instead, by following Ref.~\cite{Akama:2019qeh},
we assume the following time scaling behavior of each term of the background equations:
\begin{align}
\mathcal{E}_i, \mathcal{P}_i\propto (-t)^{2\alpha}\propto (-\eta)^{2\alpha/(1-n)},
\end{align}
where $\alpha$ is independent of time. For example, $\mathcal{E}_4, \mathcal{P}_4\supset  \mpl^2H^2\propto(-t)^{-2}$ in a minimally coupled theory such as the k-essence and cubic Galileon theories, and hence $\alpha=-1$ in such theories. See, e.g., Refs.~\cite{Brandenberger:2012zb,Li:2016xjb} for examples of $\alpha=-1$ in the k-essence theory. By extending the gravitational theories, the other cases $\alpha\neq-1$ can be studied. See also Appendix~\ref{app: example} for an example of $\alpha=-2$. In the following section, we show the role of $\alpha$ in spectral indices of primordial power spectra.

\section{Primordial power spectra: two branches of general bounce cosmology}\label{Sec: powerspectra}
In this section, we review two-point correlation functions from the general bounce cosmology studied in Ref.~\cite{Akama:2019qeh}. We also explain two branches of that general model given the properties of long-wavelength perturbations and spacetime anisotropies.

Taking the unitary gauge, $\delta\phi(t,{\bf x})=0$, scalar and tensor perturbations on top of the spatially flat FLRW background are defined by
\begin{align}
\D s^2=-N^2\D t^2+g_{ij}(\D x^i+N^i\D t)(\D x^j+N^j\D t),
\end{align}
where
\begin{align}
N&=1+\delta n,\ N_i=\partial_i\chi,\ g_{ij}=a^2e^{2\zeta}(e^{h})_{ij},
\end{align}
with $(e^h)_{ij}$ being an exponential expansion of the spatial metric including the transverse-traceless tensor, $\partial_i h_{ij}=0=\delta_{ij}h_{ij}$, defined by
\begin{align}
(e^{h})_{ij}:=\delta_{ij}+h_{ij}+\frac{1}{2}h_{ik}h_{kj}+\frac{1}{6}h_{ik}h_{kl}h_{lj}+\cdots.
\end{align}
Once the constraint equations are solved, the solutions of the auxiliary fields $\delta n$ and $\chi$ are expressed by the curvature perturbation $\zeta$ (see Appendix~\ref{Sec: app-perturbed-actions} for their forms). In the Horndeski theory, the quadratic actions of the curvature perturbation $\zeta$ and the tensor perturbations $h_{ij}$ around the spatially flat FLRW background have been found to be of the form~\cite{Kobayashi:2011nu},
\begin{align}
S^{(2)}_s&=\int\D t\D^3xa^3\biggl[\mathcal{G}_S\dot\zeta^2-\frac{\mathcal{F}_S}{a^2}(\partial_i\zeta)^2\biggr],\\
S^{(2)}_t&=\frac{1}{8}\int\D t\D^3x a^3\biggl[\mathcal{G}_T\dot h_{ij}^2-\frac{\mathcal{F}_T}{a^2}(\partial_k h_{ij})^2\biggr].
\end{align}
We write the explicit expressions of $\mathcal{G}_S, \mathcal{F}_S, \mathcal{G}_T$, and $\mathcal{F}_T$ in Appendix~\ref{Sec: app-perturbed-actions}. Note that the positivity of each coefficient, i.e., $\mathcal{G}_S, \mathcal{F}_S, \mathcal{G}_T, \mathcal{F}_T>0$, ensures that both ghost and gradient instabilities are avoided. The linear perturbations obey the following equations of motion,
\begin{align}
E^s=0,\ E^t_{ij}=0,
\end{align}
where
\begin{align}
E^s&:=\frac{\D}{\D t}(a^3\mathcal{G}_S\dot\zeta)-a\mathcal{F}_S\partial^2\zeta,\\
E^t_{ij}&:=\frac{\D}{\D t}(a^3\mathcal{G}_T\dot h_{ij})-a\mathcal{F}_T\partial^2 h_{ij},
\end{align}
with $\partial^2:=\delta^{ij}\partial_i\partial_j$. On the general power-law contracting background, the four coefficients of the quadratic actions have been found to scale as~\cite{Akama:2019qeh} (see also Appendix~\ref{Sec: app-perturbed-actions})
\begin{align}
\mathcal{G}_S, \mathcal{F}_S, \mathcal{G}_T, \mathcal{F}_T\propto (-t)^{2(\alpha+1)}.
\end{align}
The above and Eq.~(\ref{eq: scale-factor}) solve the equations of motion for the linear perturbations. Using mode functions and creation and annihilation operators, one can decompose the quantized fluctuations in Fourier space as
\begin{align}
\zeta(t,{\bf k})&=\zeta_{\bf k}(t)a_{\bf k}+\zeta^*_{\bf -k}(t)a^\dagger_{-{\bf k}},\\
h_{ij}(t,{\bf k})&=\sum_s\left[h^{(s)}_{\bf k}a^{(s)}_{\bf k}e^{(s)}_{ij}({\bf k})+h^{(s)*}_{-\bf k}a^{(s)\dagger}_{-\bf k}e^{(s)*}_{ij}(-{\bf k})\right],
\end{align}
where the curvature and tensor perturbations are Fourier transformed as
\begin{align}
\zeta(t,{\bf x})&=\int\frac{\D^3 k}{(2\pi)^3}\zeta(t,{\bf k})e^{i{\bf k}\cdot{\bf x}},\\
h_{ij}(t,{\bf x})&=\int\frac{\D^3 k}{(2\pi)^3} h_{ij}(t,{\bf k})e^{i{\bf k}\cdot{\bf x}},
\end{align}
and the commutation relations of the creation and annihilation operators are as follows,
\begin{align}
[a_{\bf k},a^\dagger_{{\bf k}'}]&=(2\pi)^3\delta({\bf k}-{\bf k}'),\\
[a^{(s)}_{\bf k},a^{(s')\dagger}_{{\bf k}'}]&=(2\pi)^3\delta_{ss'}\delta({\bf k}-{\bf k}'),\\
{\rm others}&=0,
\end{align}
with the two helicities of gravitational waves $s$ taking $s=\pm$. The transverse and traceless polarization tensor $e^{(s)}_{ij}({\bf k})$ (i.e., $k^i e^{(s)}_{ij}({\bf k})=0=\delta^{ij}e^{(s)}_{ij}({\bf k})$) is normalized by $e^{(s)}_{ij}({\bf k})e^{(s')*}_{ij}({\bf k})=\delta_{ss'}$. 

Introducing canonically normalized perturbations defined by $u:=z_s\zeta$ and $v_{ij}:=z_th_{ij}$ with $z_s:=\sqrt{2}a(\mathcal{G}_S\mathcal{F}_S)^{1/4}$ and $z_t:=(a/2)(\mathcal{G}_T\mathcal{F}_T)^{1/4}$ and varying the quadratic actions with respect to those, the equations of motion for the mode functions of the canonically normalized perturbations $u_{\bf k}$ and $v^{(s)}_{\bf k}$ are found to be
\begin{align}
u_{\bf k}''+\biggl(c_s^2k^2-\frac{z_s''}{z_s}\biggr)u_{\bf k}&=0,\\
v^{(s)''}_{\bf k}+\biggl(c_t^2k^2-\frac{z_t''}{z_t}\biggr)v^{(s)}_{\bf k}&=0,
\end{align}
where the propagation speeds of perturbations $c_s^2:=\mathcal{F}_S/\mathcal{G}_S$ and $c_t^2:=\mathcal{F}_T/\mathcal{G}_T$ are constant during the contracting phase.
Here, we find $c_sk/(aH)=(-c_sk\eta)(n-1)/n$ and $c_tk/(aH)=(-c_tk\eta)(n-1)/n$, and hence the time at which a phase oscillation of each mode ceases is different from that of the so-called sound-horizon crossing for each mode, up to the factor $(n-1)/n$. However, we refer to subhorizon, (sound-)horizon cross, and superhorizon scales as $-ck\eta\gg1$, $-ck\eta=1$, and $-ck\eta\ll1$, respectively, for the curvature perturbations ($c=c_s$) and the tensor ones ($c=c_t$).

The effective mass terms $z_s''/z_s$ and $z_t''/z_t$ in the equations of motion take the same form:
\begin{align}
\frac{z_s''}{z_s}=\frac{z_t''}{z_t}=\frac{1}{\eta^2}\biggl(\nu^2-\frac{1}{4}\biggr),
\end{align}
where
\begin{align}
\nu:=\frac{-1-3n-2\alpha}{2(1-n)}.
\end{align}
Note that the equations of motion for the mode functions of the canonically normalized perturbations are invariant when flipping the sign of $\nu$, while those for the mode functions of the original variables are not, which results in a different time evolution of the original variables for positive and negative values of $\nu$. In solving the equations of motion for the mode functions of the canonically normalized perturbations, we impose the adiabatic vacuum initial condition in the far past when each mode is on the subhorizon scales $-c_sk\eta, -c_tk\eta\gg1$:
\begin{align}
u_{\bf k}(\eta)&\to\frac{1}{\sqrt{2k}}e^{-ic_sk\eta},\\
v^{(s)}_{\bf k}(\eta)&\to\frac{1}{\sqrt{2k}}e^{-ic_tk\eta}.
\end{align}
As a result, using the common $\nu$, the mode functions of the curvature and tensor perturbations are expressed as
\begin{align}
\zeta_{\bf k}(\eta)&=\frac{1}{\sqrt{2}a(\mathcal{G}_S\mathcal{F}_S)^{1/4}}\cdot \frac{\sqrt{\pi}}{2}\sqrt{-c_s\eta}H^{(1)}_\nu(-c_sk\eta),\\
h^{(s)}_{\bf k}(\eta)&=\frac{2}{a(\mathcal{G}_T\mathcal{F}_T)^{1/4}}\cdot \frac{\sqrt{\pi}}{2}\sqrt{-c_t\eta}H^{(1)}_\nu(-c_tk\eta),
\end{align}
with the Hankel function of the first kind $H^{(1)}_\nu$.
Through the two-point correlation functions, dimensionless power spectra $\mathcal{P}_\zeta$ and $\mathcal{P}_h$ are defined by
\begin{align}
\langle\zeta({\bf k})\zeta({\bf k}')\rangle=(2\pi)^3\delta({\bf k}+{\bf k}')\frac{2\pi^2}{k^3}\mathcal{P}_\zeta,\\
\langle\xi^{(s)}({\bf k})\xi^{(s')}({\bf k}')\rangle=(2\pi)^3\delta_{ss'}\delta({\bf k}+{\bf k}')\frac{\pi^2}{k^3}\mathcal{P}_h,
\end{align}
where $\xi^{(s)}({\bf k}):=h_{ij}({\bf k})e^{(s)*}_{ij}({\bf k})$.
From the solutions of the mode functions, general expressions of the power spectra at the end of the contracting phase when the perturbations are on the superhorizon scales, $-c_sk\eta_b\ll1, -c_tk\eta_b\ll1$, can be derived as (see also Ref.~\cite{Nishi:2016ljg} for general expressions of primordial power spectra in power-law universes)
\begin{align}
\mathcal{P}_\zeta(\eta_b)&=\frac{1}{8\pi^2}\frac{1}{c_s\mathcal{F}_{S}}\frac{1}{\eta^2}\biggr|_{\eta=\eta_b}\biggl[2^{|\nu|-3/2}\frac{\Gamma(|\nu|)}{\Gamma(3/2)}\biggr]^2|c_sk\eta_b|^{n_s-1}\notag\\
&=\frac{1}{8\pi^2}\biggl(\frac{1-n}{n}\biggr)^2\frac{H^2}{c_s\mathcal{F}_S}\biggr|_{\eta=\eta_b}\biggl[2^{|\nu|-3/2}\frac{\Gamma(|\nu|)}{\Gamma(3/2)}\biggr]^2|c_sk\eta_b|^{n_s-1},\\
\mathcal{P}_h(\eta_b)&=\frac{2}{\pi^2}\frac{1}{c_t\mathcal{F}_{T}}\frac{1}{\eta^2}\biggr|_{\eta=\eta_b}\biggl[2^{|\nu|-3/2}\frac{\Gamma(|\nu|)}{\Gamma(3/2)}\biggr]^2|c_tk\eta_b|^{n_t}\notag\\
&=\frac{2}{\pi^2}\biggl(\frac{1-n}{n}\biggr)^2\frac{H^2}{c_t\mathcal{F}_T}\biggr|_{\eta=\eta_b}\biggl[2^{|\nu|-3/2}\frac{\Gamma(|\nu|)}{\Gamma(3/2)}\biggr]^2|c_tk\eta_b|^{n_t},
\end{align}
where we used Eq.~(\ref{eq: H-etab}) and parametrized time-dependent quantities as $Q(\eta)=Q(\eta_b)(\eta/\eta_b)^{\bullet}$.
The spectral indices of both scalar and tensor power spectra are of the form, 
\begin{align}
n_s-1=n_t=3-2|\nu|.
\end{align}
The general framework is thus classified into two branches in terms of the scale invariance of the power spectra: $\nu=3/2$ ($\alpha=-2$) and $\nu=-3/2$ ($\alpha=1-3n$). We will understand these branches by invoking a conformal transformation later. In our paper, we focus on the exact scale-invariant cases, but the CMB experiments have clarified that the scalar power spectrum is slightly red-tilted, $n_s-1\simeq -0.04$~\cite{Planck:2018jri}, which can be realized by choosing $n$ and $\alpha$ properly:
\begin{align}
|\nu|=\frac{|-1-3n-2\alpha|}{2(1-n)}\simeq\frac{3}{2}+0.02
\end{align}
Small deviations from the exact flat case such that $|\nu|-3/2=\mathcal{O}(10^{-2})$ give $\mathcal{O}(10^{-2})$ corrections to the mode functions, that is, the corrections to the power spectra would be at most of $\mathcal{O}(10^{-2})$. It would be reasonable to anticipate that the bispectra receive $\mathcal{O}(10^{-2})$ corrections to those computed for $|\nu|=3/2$.\footnote{The bispectra are computed by time integrations for the in-in formalism. In addition to the mode functions, the change of coupling function of cubic interactions due to the small deviations from $\alpha=1-3n$ or $\alpha=-2$ can affect the resultant bispectra. In our model, the coupling function is a function of $\eta$ raised to a positive power, and thus naively the difference of the coupling may be especially large the further back in time one goes (i.e., in a larger $|\eta|$ region). However, the further back in time, the less contribution to the generation of non-Gaussianities due to a rapid oscillation of integrands originating from a product of mode functions. Such a rapidly oscillating regime does not contribute to the resultant non-Gaussianities, and thus it would be reasonable to expect that the impacts of the $\mathcal{O}(10^{-2})$ deviations from the scale invariance to the bispectra are $\mathcal{O}(10^{-2})$. } In the present paper, we ignore these expected corrections and proceed our analysis of primordial correlation functions in the exact scale-invariant cases.

Here, certain aspects of general bounce cosmology need to be mentioned.  
The case of $\alpha=-2$ ($\nu=3/2$) indicates $\mathcal{G}_S, \mathcal{F}_S, \mathcal{G}_T, \mathcal{F}_T\propto(-t)^{-2}\propto(-\eta)^{-2/(1-n)}$, that is, all of the coefficients of the quadratic actions are asymptotic to $0$ in the past infinity. We then have
\begin{align}
\int_{-\infty}^t\D t' a(t')\mathcal{F}_T(t')\propto\left[(-t')^{n-1}\right]^t_{-\infty}=(-t)^{n-1},
\end{align}
which is finite for a finite $t$. Therefore, models in this branch avoid the no-go theorem for the nonsingular cosmologies in the Horndeski theory. Motivated by this advantage, an explicit contracting model (corresponding to $\nu=3/2$ or $\nu\simeq3/2$ in our paper depending on a parameter choice) has been studied~\cite{Ageeva:2022asq,Ageeva:2024knc}. However, at the same time, the perturbations can suffer from a strong coupling issue in the past infinity. Let us inspect this situation in more detail by using the canonically normalized variables. Note that in the present model, up to the overall factors, both quadratic and cubic actions for $\nu=3/2$ take the same form as those in de Sitter inflation, and thus the following argument applies for inflation as well. By using the canonically normalized perturbation, we have
 \begin{align}
S_u&=\int\D y_s\D^3x\biggl[\frac{1}{2}\biggl(\frac{\D u}{\D y_s}\biggr)^2-\frac{1}{2}(\partial_i u)^2+\frac{1}{2}\frac{1}{z_s}\frac{\D^2 z_s}{\D y_s^2}u^2+\frac{1}{\Lambda^2}\biggl(\frac{\D u}{\D y_s}\biggr)^3+\cdots\biggr],
 \end{align}
 where we introduced $y_s$ defined by $\D y_s:=\D\eta/c_s$. Now $\Lambda$ is asymptotic to $0$ in the past infinity in the present framework for $\nu=3/2$ (where $aH\propto(-\eta)^{-1}, a^2\mathcal{G}_S\propto(-\eta)^{-2}, ..$) and de Sitter inflation (where $aH=(-\eta)^{-1}, a^2\mathcal{G}_S\propto(-\eta)^{-2}, ..$) with the same time dependence, $\Lambda\propto(-\eta)^{-1}$. 
 See e.g., Refs.~\cite{Ageeva:2018lko,Ageeva:2020gti,Ageeva:2020buc,Ageeva:2021yik,Ageeva:2022fyq,Ageeva:2022asq,Akama:2022usl,Ageeva:2023nwf,Ageeva:2024knc} where the strong coupling issue in the past infinity was studied, and whether the above asymptotic behavior of $\Lambda$ spoils classical treatments of cosmological models was found to be model-dependent. In the present paper, we will not go into the strong coupling issue, but in light of the asymptotic behavior of the coefficients of the quadratic actions, realistically the models for $\nu=3/2$ can be considered under the condition such that
the strong coupling scale $\Lambda\propto(-\eta)^{-1}$ is higher than a classical energy scale, e.g., $H\propto t^{-1}\propto(-\eta)^{-1/(1-n)}$, which may give rise to constraints on the viable model space, instead of rulling out the branch of $\nu=3/2$ immediately. Indeed, the explicit example in Ref.~\cite{Ageeva:2022asq} allows for a model space
that can justify the classical treatment of a power-law contracting background for $\nu=3/2$ in a subclass of the Horndeski theory, and hence a wider framework used in the present paper would allow such a viable parameter space as well. (See Ref.~\cite{Ageeva:2022asq} for more details on this point studied through perturbative unitarity.)\footnote{If strong coupling shows up in the past infinity depending on model parameters, one of the options to overcome that problem would be to introduce an early phase before the contracting phase for $\nu=3/2$. For the model studied in Ref.~\cite{Ageeva:2022asq,Ageeva:2024knc}, if the power spectrum of the scalar perturbations is slightly red-tilted (i.e., $\nu-3/2=\mathcal{O}(10^{-2})>0$) instead of exactly flat (i.e., $\nu=3/2$), the strong coupling in the past infinity was found to show up. Then, a proposed idea was to introduce an early phase for $\nu-3/2\sim0<0$ in which the strong coupling is absent and then slowly connect to the contracting phase for $\nu-3/2\sim0>0$ to obtain the slightly-red scalar spectrum. As long as such a slow transition happens in the far past, that transition would not affect the primordial correlation functions at a late time.} Instead of the strong coupling issue in the past infinity, we will discuss perturbativity by evaluating ratios of cubic Lagrangians to the quadratic ones at and after horizon-cross scales.

A general expression of the tensor-to-scalar ratio is computed as
\begin{align}
r:=\frac{\mathcal{P}_h}{\mathcal{P}_\zeta}=16\biggl(\frac{c_t}{c_s}\biggr)^{3-2|\nu|}\frac{c_s}{c_t}\frac{\mathcal{F}_{S}}{\mathcal{F}_{T}}. \label{eq: tensor-to-scalar}
\end{align}
Since the power spectrum of $\zeta$ and that of $h_{ij}$ depend on time in the same way, the resultant tensor-to-scalar ratio is constant irrespective of the value of $\nu$, while the powers spectra on the superhorizon scales grow with time for $\nu<0$. Before moving on to the next subsection, we briefly review two differences between the two branches. To do so, let us first perform a conformal transformation $g_{\mu\nu}\to\tilde g_{\mu\nu}=\Omega^2 g_{\mu\nu}$ under which we have
\begin{align}
\D\tilde s^2&\to\Omega^2(\eta)a^2(\eta)(-\D\eta^2+\delta_{ij}\D x^i\D x^j)\notag\\
&= \tilde a^2(\tilde\eta)(-\D\tilde\eta^2+\delta_{ij}\D x^i\D x^j),
\end{align}
where $\tilde a:=\Omega a$ and $\D\tilde\eta:=\D\tilde t/\tilde a=\D t/a=\D\eta$. Here, we choose the time dependence of the conformal factor $\Omega$ in such a way that all of the coefficients of the quadratic actions are constant, $\mathcal{G}_S, \mathcal{F}_S, \mathcal{G}_T, \mathcal{F}_T\to{\rm const.}$, as in the conventional slow-roll inflation (with time variations of the so-called slow-roll parameters being ignored) and matter-dominated contracting models.\footnote{We have $z_s''/z_s=z_t''/z_t=\tilde a''/\tilde a$ and $z_s, z_t\propto\tilde a$ (i.e., $z_s$ and $z_t$ correspond to the scale factor of the conformal metric), and inspecting the linear perturbations in terms of the conformal transformation is equivalent to doing those in terms of $u, z_s, v_{ij}$, and $z_t$.} This can be achieved by choosing $\Omega\propto(-\eta)^{(\alpha+1)/(1-n)}$ which yields
\begin{align}
\tilde a(\tilde\eta)\propto (-\tilde\eta)^{(1+\alpha+n)/(1-n)}=(-\tilde\eta)^{1/2-\nu}.
\end{align}
In particular, we have $\tilde a\propto(-\tilde\eta)^{-1}$ for $\nu=3/2$ and $\tilde a\propto(-\tilde\eta)^2$ for $\nu=-3/2$, which means that the power spectra are scale invariant when the scale factor of the conformal metric $\tilde a$ is of de Sitter or a matter-dominated contracting universe, even if the scale factor of the background metric $a$ is neither. See Refs.~\cite{Nandi:2019xag,Nandi:2020sif,Nandi:2020szp,Nandi:2022twa}
where the authors have studied bounce models that are related to inflation via a conformal transformation in a subclass of the Horndeski theory.

The other difference is a property of spacetime anisotropies. In general, contracting models suffer from the growth of anisotropies~\cite{Belinsky:1970ew}. 
In a Bianchi-type I anisotropic universe with the following metric,
\begin{align}
\D s^2&=-\D t^2+a^2[e^{2(\beta_++\sqrt{3}\beta_-)}\D x^2+e^{2(\beta_+-\sqrt{3}\beta_-)}\D y^2+e^{-4\beta_+}\D z^2].
\end{align}
the evolution equations for the anisotropies $\beta_{(+,-)}$ have been obtained in the Horndeski theory as~\cite{Nishi:2015pta,Tahara:2018orv},
\begin{align}
\frac{\D}{\D t}\{a^3[\mathcal{G}_T\dot\beta_+-2\mu(\dot\beta_+^2-\dot\beta_-^2)]\}&=0,\\
\frac{\D}{\D t}[a^3(\mathcal{G}_T\dot\beta_-+4\mu\dot\beta_+\dot\beta_-)]&=0.
\end{align}
Then, let us study the case of initially small anisotropies such that the first terms in the above two equations dominate the others, which yields
\begin{align}
\dot\beta_{(+,-)}\propto\frac{1}{a^3\mathcal{G}_T}\sim\dot h_{ij}\ {\rm on}\ {\rm the\ superhorizon\ scales}. \label{eq: aniso}
\end{align}
Since the tensor perturbations on the superhorizon scales are constant for $\nu>0$ and grow for $\nu<0$, the decay of the initially small anisotropies indicates the conservation of the superhorizon modes. 
In particular, when requiring the decay of the spacetime anisotropies and the scale invariance of the scalar power spectrum simultaneously in the present framework, $\nu=3/2$ is chosen. Throughout our paper, we study the case of $\nu=-3/2$ under the assumption that the contracting phase successfully connects to the subsequent expanding phase before the isotropic contracting background is spoiled by the spacetime anisotropies, and we will not discuss a concrete realization of such a successful bounce. See Ref.~\cite{Khoury:2001wf} for an example of avoiding the growth of the anisotropies with the so-called Ekpyrotic contracting phase during which the anisotropies are diluted away due to a matter field whose energy density grows faster than that of anisotropies.

\subsection{$\nu=3/2$}\label{SubSec: power-const}
In this subsection, we show the explicit forms of the mode functions and primordial power spectra for $\nu=3/2$, i.e., $\alpha=-2$. Since $\alpha=-1$ in minimally coupled theories, the case of $\alpha=-2$ can be studied only in modified gravity theories. First, the mode functions are of the form,
\begin{align}
\zeta_{\bf k}(\eta)&=-\frac{i}{2\sqrt{\mathcal{F}_Sc_s}}\biggr|_{\eta=\eta_b}k^{-3/2}\frac{1}{\eta_b}(1+ic_sk\eta)e^{-ic_sk\eta},\\
h^{(s)}_{\bf k}(\eta)&=-\frac{i\sqrt{2}}{\sqrt{\mathcal{F}_Tc_t}}\biggr|_{\eta=\eta_b}k^{-3/2}\frac{1}{\eta_b}(1+ic_tk\eta)e^{-ic_tk\eta},
\end{align}
from which the power spectra on the superhorizon scales read
\begin{align}
\mathcal{P}_\zeta(\eta)&=\mathcal{P}_\zeta(\eta_b)=\frac{1}{8\pi^2}\biggl(\frac{1-n}{n}\biggr)^2\frac{H^2}{c_s\mathcal{F}_S},\\
\mathcal{P}_h(\eta)&=\mathcal{P}_h(\eta_b)=\frac{2}{\pi^2}\biggl(\frac{1-n}{n}\biggr)^2\frac{H^2}{c_t\mathcal{F}_T}.
\end{align}
Note that $H^2/(c_s\mathcal{F}_S)$ and $H^2/(c_t\mathcal{F}_T)$ are constant for $\nu=3/2$. Here, by evaluating the power spectra at the sound-horizon cross scales, $-c_sk\eta=1, -c_tk\eta=1$, one can reproduce $n_s-1=n_t=3-2\nu$ from $n_s-1=\D\ln\mathcal{P}_\zeta/(\D\ln k)|_{k=1/(-c_s\eta)}$ and $n_t=\D\ln\mathcal{P}_t/(\D\ln k)|_{k=1/(-c_t\eta)}$.

Then, by substituting $\nu=3/2$ into Eq.~(\ref{eq: tensor-to-scalar}), the tensor-to-scalar ratio reads
\begin{align}
r=16\frac{c_s}{c_t}\frac{\mathcal{F}_S}{\mathcal{F}_T}. \label{eq: r-nu3/2}
\end{align}
For a canonical potential-driven slow-roll inflation that resides within the k-essence theory, one has $c_t=1$ and $\mathcal{F}_S/\mathcal{F}_T\simeq \epsilon\simeq -n_t/2$ where $\epsilon:=-\dot H/H^2$, which results in a consistency relation $r=-8c_sn_T$. (See also e.g., Refs.~\cite{Kobayashi:2010cm,Kamada:2010qe,Kobayashi:2011nu,Ohashi:2012wf,Unnikrishnan:2013rka} for the violation of this consistency relation caused by $G_i$ ($i=3,4,5$) terms.) 
In the present case, $c_t$ is not necessarily unity and the relation between $\mathcal{F}_S$ and $\mathcal{F}_T$ does not necessarily hold, and the consistency relation $r=-8c_sn_t$ is generally violated. See also Appendix~\ref{app: example} for a concrete example with that violation.

\subsection{$\nu=-3/2$}\label{SubSec: power-grow}

We then study the other branch, $\alpha=1-3n$. In this branch, the mode functions of both scalar and tensor perturbations have been obtained as~\cite{Akama:2019qeh}
\begin{align}
\zeta_{\bf k}(\eta)&=-\frac{i}{2\sqrt{\mathcal{F}_Sc_s}}\biggr|_{\eta=\eta_b}k^{-3/2}\frac{\eta_b^2}{\eta^3}(1+ic_sk\eta)e^{-ic_sk\eta},\\
h^{(s)}_{\bf k}(\eta)&=-\frac{i\sqrt{2}}{\sqrt{\mathcal{F}_Tc_t}}\biggr|_{\eta=\eta_b}k^{-3/2}\frac{\eta_b^2}{\eta^3}(1+ic_tk\eta)e^{-ic_tk\eta}.
\end{align}
On the superhorizon scales, the concrete forms of the power spectra are obtained as
\begin{align}
\mathcal{P}_\zeta(\eta)&=\frac{1}{8\pi^2}\frac{1}{\mathcal{F}_Sc_s}\biggr|_{\eta=\eta_b}\frac{\eta_b^4}{\eta^6},\\
\mathcal{P}_h(\eta)&=\frac{2}{\pi^2}\frac{1}{\mathcal{F}_Tc_t}\biggr|_{\eta=\eta_b}\frac{\eta_b^4}{\eta^6},
\end{align}
which grow in proportion to $1/\eta^6$. By evaluating the above two power spectra at the end of the contracting phase $\eta=\eta_b$, one obtains~\cite{Akama:2019qeh}
\begin{align}
\mathcal{P}_\zeta(\eta_b)=\frac{1}{8\pi^2}\biggl(\frac{1-n}{n}\biggr)^2\frac{H^2}{c_s\mathcal{F}_S}\biggr|_{\eta=\eta_b},\\
\mathcal{P}_h(\eta_b)=\frac{2}{\pi^2}\biggl(\frac{1-n}{n}\biggr)^2\frac{H^2}{c_t\mathcal{F}_T}\biggr|_{\eta=\eta_b}.
\end{align}
The tensor-to-scalar ratio in this branch takes the same form as for $\nu=3/2$ in Eq.~(\ref{eq: r-nu3/2}).

\section{Primordial full bispectra}\label{Sec: bispectra}
In this section, let us study the full bispectra originating from the curvature and tensor perturbations on the power-law contracting background. Let us define those bispectra by
\begin{align}
\langle \zeta({\bf k}_1)\zeta({\bf k}_2)\zeta({\bf k}_3)\rangle&=:(2\pi)^3\delta({\bf k}_1+{\bf k}_2+{\bf k}_3)\mathcal{B}_{s}, \label{eq: def-sss}\\
\langle \zeta({\bf k}_1)\zeta({\bf k}_2)\xi^{(s_3)}({\bf k}_3)\rangle&=:(2\pi)^3\delta({\bf k}_1+{\bf k}_2+{\bf k}_3)\mathcal{B}^{s_3}_{sst},  \label{eq: def-st}\\
\langle \zeta({\bf k}_1)\xi^{(s_2)}({\bf k}_2)\xi^{(s_3)}({\bf k}_3)\rangle&=:(2\pi)^3\delta({\bf k}_1+{\bf k}_2+{\bf k}_3)\mathcal{B}^{s_2s_3}_{stt},  \label{eq: def-stt}\\
\langle \xi^{(s_1)}({\bf k}_1)\xi^{(s_2)}({\bf k}_2)\xi^{(s_3)}({\bf k}_3)\rangle&=:(2\pi)^3\delta({\bf k}_1+{\bf k}_2+{\bf k}_3)\mathcal{B}^{s_1s_2s_3}_{t}.  \label{eq: def-ttt}
\end{align}
Utilizing the in-in formalism~\cite{Maldacena:2002vr}, we compute the three-point correlation functions from self and cross interactions
as
\begin{align}
\langle \zeta({\bf k}_1)\zeta({\bf k}_2)\zeta({\bf k}_3)\rangle&=-i\int_{\eta_i}^{\eta_b}\D\eta' a(\eta')\langle[\zeta(\eta,{\bf k}_1)\zeta(\eta,{\bf k}_2)\zeta(\eta,{\bf k}_3),H^{s}_{\rm int}(\eta')]\rangle,\\
\langle \zeta({\bf k}_1)\zeta({\bf k}_2)\xi^{(s_3)}({\bf k}_3)\rangle&=-i\int_{\eta_i}^{\eta_b}\D\eta' a(\eta')\langle[\zeta(\eta,{\bf k}_1)\zeta(\eta,{\bf k}_2)\xi^{(s_3)}(\eta,{\bf k}_3),H^{sst}_{\rm int}(\eta')]\rangle,\\
\langle \zeta({\bf k}_1)\xi^{(s_2)}({\bf k}_2)\xi^{(s_3)}({\bf k}_3)\rangle&=-i\int_{\eta_i}^{\eta_b}\D\eta' a(\eta')\langle[\zeta(\eta,{\bf k}_1)\xi^{(s_2)}(\eta,{\bf k}_2)\xi^{(s_3)}(\eta,{\bf k}_3),H^{stt}_{\rm int}(\eta')]\rangle,\\
\langle \xi^{(s_1)}({\bf k}_1)\xi^{(s_2)}({\bf k}_2)\xi^{(s_3)}({\bf k}_3)\rangle&=-i\int_{\eta_i}^{\eta_b}\D\eta' a(\eta')\langle[\xi^{(s_1)}(\eta,{\bf k}_1)\xi^{(s_2)}(\eta,{\bf k}_2)\xi^{(s_3)}(\eta,{\bf k}_3),H^{t}_{\rm int}(\eta')]\rangle,
\end{align}
where $\eta_i$ denotes the conformal time when the perturbation modes are on the subhorizon scales, and each interaction Hamiltonian is derived from the cubic Lagrangians:
\begin{align}
H^s_{\rm int}&:=-\int\D^3x\mathcal{L}^{(3)}_s,\ H^{sst}_{\rm int}:=-\int\D^3x\mathcal{L}^{(3)}_{sst}, H^{stt}_{\rm int}:=-\int\D^3x\mathcal{L}^{(3)}_{stt}, H^t_{\rm int}:=-\int\D^3x\mathcal{L}^{(3)}_{t}.
\end{align}
See Appendix~\ref{Sec: app-perturbed-actions} for the explicit forms of the cubic Lagrangians, $\mathcal{L}^{(3)}_s, \mathcal{L}^{(3)}_{sst}, \mathcal{L}^{(3)}_{stt}, \mathcal{L}^{(3)}_{t}$. Then, we parametrize the bispectra normalized by the square of primordial power spectra as
\begin{align}
\mathcal{B}_s=\frac{(2\pi)^4\mathcal{P}_\zeta^2}{k_1^3k_2^3k_3^3}\mathcal{A}_s,\ 
\mathcal{B}^{s_3}_{sst}=\frac{(2\pi)^4\mathcal{P}_\zeta\mathcal{P}_h}{k_1^3k_2^3k_3^3}\mathcal{A}^{s_3}_{sst},\ 
\mathcal{B}^{s_2s_3}_{stt}=\frac{(2\pi)^4\mathcal{P}_\zeta\mathcal{P}_h}{k_1^3k_2^3k_3^3}\mathcal{A}^{s_2s_3}_{stt},\
\mathcal{B}^{s_1s_2s_3}_{t}=\frac{(2\pi)^4\mathcal{P}_h^2}{k_1^3k_2^3k_3^3}\mathcal{A}^{s_1s_2s_3}_{t}.
\end{align}
Before showing the explicit forms of the bispectra, let us mention the expected relations of the bispectra between inflation, matter bounce, and the present contracting model. For simplicity, we focus on the scalar-scalar-scalar bispectrum. The cubic action and the interaction Hamiltonian of the scalar-scalar-scalar interaction are of the form,
\begin{align}
S^{(3)}_s=\int\D \eta\D^3x\mathcal{L}^{(3)}_s=-\int\D^3x H^s_{\rm int},
\end{align}
where
\begin{align}
{\cal L}_s^{(3)}&={\mathcal G}_S
\biggl\{a\frac{\Lambda_1}{H}
\zeta'^3+a^2\Lambda_2\zeta\zeta'^2+a^2\Lambda_3\zeta\left(\partial_i\zeta\right)^2
+\frac{\Lambda_4}{H^2}
\zeta'^2\partial^2\zeta+a^2\Lambda_5\zeta'\partial_i\zeta\partial_i
(a\psi)+\Lambda_6\partial^2\zeta\left(\partial_i\psi\right)^2\notag\\
&\quad\quad\quad
+\frac{\Lambda_7}{H^2}
\left[\partial^2\zeta\left(\partial_i\zeta\right)^2-\zeta\partial_i\partial_j\left(\partial_i\zeta\partial_j\zeta\right)\right]+\frac{a\Lambda_8}{H}\left[\partial^2\zeta\partial_i\zeta\partial_i(a\psi)-\zeta\partial_i\partial_j\left(\partial_i\zeta\partial_j(a\psi)\right)\right]\biggr\}\notag\\
&\quad +aF(\zeta)E_S, \label{eq: cubic}
\end{align}
where $\Lambda_i$ and $F(\zeta)$ are given in Appendix~\ref{Sec: app-perturbed-actions}.
Let us focus on the interaction Hamiltonian originating from the first interaction term and show its time dependence in the cases of a quasi-de Sitter inflation where $a\simeq-1/(H\eta), \mathcal{G}_S\simeq{\rm const.}, H\simeq{\rm const.}$, and $\Lambda_1\simeq{\rm const.}$ and a matter-dominated contracting model where $a\propto(-\eta)^2, \mathcal{G}_S={\rm const.}, H\propto(-\eta)^{-3}$, and $\Lambda_1={\rm const.}$, which can be confirmed by taking $n=2/3$ and $\alpha=-1$ in the present model (see also Ref~\cite{Cai:2009fn} for a matter bounce model having such time dependence). In the former case, one has
\begin{align}
H^s_{{\rm int}}\supset-\int\D\eta a\mathcal{G}_S\frac{\Lambda_1}{H}\zeta'^3=({\rm const.})\int\D\eta(-\eta)^{-1}\zeta'^3, \label{eq: cubic-timedependence-inf}
\end{align}
and in the latter case, one has
\begin{align}
H^s_{{\rm int}}\supset-\int\D\eta a\mathcal{G}_S\frac{\Lambda_1}{H}\zeta'^3=({\rm const.})\int\D\eta(-\eta)^{5}\zeta'^3, \label{eq: cubic-timedependence-md}
\end{align}
On the general contracting background, we have
\begin{align}
H^s_{\rm int}&\supset -\int\D\eta a\mathcal{G}_S\frac{\Lambda_1}{H}\zeta'^3= ({\rm const.})\int\D\eta(-\eta)^{2(1-\nu)}\zeta'^3
\label{eq: sss-conformal}
\end{align}
By taking $\nu=3/2$ and $\nu=-3/2$ in the above, one can find that, up to the overall factor, Eq.~(\ref{eq: sss-conformal}) for $\nu=3/2$ and that for $\nu=-3/2$ are identical to Eq.~(\ref{eq: cubic-timedependence-inf}) and Eq.~(\ref{eq: cubic-timedependence-md}), respectively.\footnote{The relation between inflation, matter bounce, and the present model can also be understood by invoking the conformal transformation performed in the previous section. Under this transformation, it can be shown that up to overall factors, the coefficients of the integrands appearing in the in-in formalism can be written by the quantities which characterize the time evolution of the spacetime, i.e., $\tilde a$ and $\tilde H(:=(1/\tilde a^2)\D\tilde a/\D\tilde\eta)$ just as the time dependence of the coefficients of the quadratic actions was solely determined by $\tilde a$. This indicates that the momentum dependence of bispectra for $\nu=3/2$ and $\nu=-3/2$ is, respectively, identical to that found in de Sitter inflation and matter-dominated contracting models, where the time dependence of the coefficients of cubic interactions is determined by $a$ and $H$.} In particular, the momentum dependence of the bispectrum is determined by the time integrations in the in-in formalism, and as will be shown later, the momentum dependence for $\nu=3/2$ and $\nu=-3/2$ is identical to that for inflation and matter bounce, respectively. The full bispectra with the full Horndeski action have been studied in the context of inflation in Ref.~\cite{Gao:2012ib}, and hence one can confirm that the above equivalence holds for the other cubic interactions in the full Horndeski action. On the other hand, only the scalar and tensor auto-bispectra from a matter-dominated contracting model have been studied in a subclass of the Horndeski theory~\cite{Cai:2009fn,Chowdhury:2015cma,Li:2016xjb}. One can thus partly verify the above equivalence between matter bounce and the present model by comparing the time dependence of the coefficients in the interaction Hamiltonians for $n=2/3$ and $\alpha=-1$ (i.e., matter-dominated contracting models in minimally coupled theories) and that for $\alpha=1-3n$ (or equivalently $\nu=-3/2$) without specifying the values of $n$ and $\alpha$. However, the important point here is that the power of time of the coefficients of the interaction Hamiltonians is solely written by $\nu$ (i.e., the parameters $n$ and $\alpha$ appear only via $\nu$), which makes it easier to obtain analytical expressions of the full bispectra for $\nu=\pm3/2$. (See also Appendix~\ref{Sec: app-perturbed-actions} for the time dependence of the other coefficients of the interaction Hamiltonians.) 

Before moving on to the next subsection, let us highlight the fact that the assumptions on the time scaling of the background equations make it possible to compute the full bispectra without specifying a concrete Lagrangian (i.e., explicit forms of the arbitrary functions $G_i$ in the Lagrangian). Therefore, the expressions of the full bispectra in the following sections are applicable to a wide variety of bouncing cosmologies studied within the Horndeski theory.

\subsection{$\nu=3/2$}\label{SubSec: bispectra-const}
In this subsection, we show a complete set of bispectra for $\nu=3/2$ where the perturbations become constant on the superhorizon scales. 

\subsubsection{Scalar-scalar-scalar bispectrum}\label{SubSubSec: sss-bispectra-const}
Using the in-in formalism, the scalar auto-bispectrum is obtained as
\begin{align}
\mathcal{A}_s&=\biggl[\frac{3\Lambda_1}{2}\frac{n-1}{n}+\frac{3\Lambda_4}{c_s^2}\biggl(\frac{n-1}{n}\biggr)^2\biggr]\frac{(k_1k_2k_3)^2}{K^3}+\frac{\Lambda_2}{4K}\biggl(2\sum_{i>j}k_i^2k_j^2-\frac{1}{K}\sum_{i\neq j}k_i^2k_j^3\biggr)\notag\\
&\quad+\frac{\Lambda_3}{8c_s^2}\biggl(\sum_i k_i^3+\frac{4}{K}\sum_{i>j}k_i^2k_j^2-\frac{2}{K^2}\sum_{i\neq j}k_i^2k_j^3\biggr)+\frac{\Lambda_5}{8}\biggl(\sum_i k_i^3-\frac{1}{2}\sum_{i\neq j}k_i k_j^2-\frac{2}{K^2}\sum_{i\neq j}k_i^2k_j^3\biggr)\notag\\
&\quad +\frac{\Lambda_6}{8}\frac{1}{K^2}\biggl(2\sum_i k_i^5+\sum_{i\neq j}k_i k_j^4-3\sum_{i\neq j}k_i^2k_j^3-2k_1k_2k_3\sum_{i>j}k_1k_j\biggr)\notag\\
&\quad +\frac{3\Lambda_7}{8c_s^4}\biggl(\frac{n-1}{n}\biggr)^2\frac{1}{K}\biggl(\sum_i k_i^4-2\sum_{i<j}k_i^2k_j^2\biggr)\biggl(1+\frac{1}{K^2}\sum_{i>j}k_ik_j+\frac{3k_1k_2k_3}{K^3}\biggr)\notag\\
&\quad +\frac{\Lambda_8}{16c_s^2}\frac{n-1}{n}\frac{1}{K^2}\biggl(7K\sum_i k_i^4+3k_1k_2k_3\sum_i k_i^2-2\sum_i k_i^5-5k_1k_2k_3K^2-12\sum_{i\neq j}k_i^2k_j^3\biggr), \label{eq: 3pt-sss-const}
\end{align}
where $K:=k_1+k_2+k_3$. Here, for the purpose of studying the contracting phase, we have imposed $0<n<1$. However, this condition does not affect the time integrations in the in-in formalism, and the above expression is valid not only for power-law contracting models ($0<n<1$ with $\eta$ ranging from $-\infty$) but also for power-law expanding models ($n>1$ with $\eta$ ranging from $-\infty$). Especially, by taking the limit $n\to\infty$ schematically (which yields e.g., $aH\to-\eta^{-1}$ as in de Sitter), one can check that the above reproduces the expression in the generalized G-inflation~\cite{Gao:2011qe,DeFelice:2011uc} which is a unified framework of inflation with the full Horndeski action~\cite{Kobayashi:2011nu}.

\subsubsection{Scalar-scalar-tensor bispectrum}\label{SubSubSec: sst-bispectra-const}
The scalar-scalar-tensor cross-bispectrum is computed as
\begin{align}
\mathcal{A}^{s_3}_{sst}&=\frac{\pi^2}{2}\biggl(\frac{n}{n-1}\biggr)^2\mathcal{P}_\zeta\sum_{q=1}^6\mathcal{J}^{(q)}(k_1,k_2,k_3)\mathcal{V}^{(q)}_{s_3}({\bf k}_1,{\bf k}_2,{\bf k}_3)+({\bf k}_1\leftrightarrow{\bf k}_2),\label{eq: 3pt-sst-const}
\end{align}
where
\begin{align}
\mathcal{V}^{(1)}_{s_3}&=k_{1i}k_{2j}e^{(s_3)*}_{ij}({\bf k}_3)=\mathcal{V}^{(2)}_{s_3},\\
\mathcal{V}^{(3)}_{s_3}&=\frac{1}{k_2^2}\mathcal{V}^{(1)}_{s_3},\ \mathcal{V}^{(4)}_{s_3}=\frac{k_3^2}{k_2^2}\mathcal{V}^{(1)}_{s_3},\ \mathcal{V}^{(5)}_{s_3}=k_3^2\mathcal{V}^{(1)}_{s_3},\ \mathcal{V}^{(6)}_{s_3}=\frac{k_3^2}{k_1^2k_2^2}\mathcal{V}^{(1)}_{s_3}.
\end{align}
and
\begin{align}
\mathcal{J}^{(1)}&=-\frac{c_1}{H^2}\frac{c_s^3(k_1+k_2)(k_1^2+k_1k_2+k_2^2)+2c_s^2c_t(k_1^2+k_1k_2+k_2^2)k_3+2c_sc_t^2(k_1+k_2)k_3^2+c_t^3k_3^3}{K''^2},\\
\mathcal{J}^{(2)}&=\frac{c_2}{H}\frac{n-1}{n}\frac{c_t^2k_3^2\left[2c_s^2(k_1^2+3k_1k_2+k_2^2)+3c_sc_t(k_1+k_2)k_3+c_t^2k_3^2\right]}{K''^3},\\
\mathcal{J}^{(3)}&=\frac{c_3}{H^2}\frac{c_s^2c_t^2k_2^2k_3^2(c_sk_1+K'')}{K''^2},\\
\mathcal{J}^{(4)}&=\frac{c_4}{H}\frac{n-1}{n}\frac{c_s^2k_2^2\left[c_s^2(k_1+k_2)(2k_1+k_2)+3c_sc_t(2k_1+k_2)k_3+2c_t^2k_3^2\right]}{K''^3},\\
\mathcal{J}^{(5)}&=c_5\biggl(\frac{n-1}{n}\biggr)^2\frac{2}{K''^2}\notag\\
&\quad\ \times\left[c_s^3(k_1+k_2)(k_1^2+3k_1k_2+k_2^2)+4c_s^2c_t(k_1^2+3k_1k_2+k_2^2)k_3+4c_sc_t^2(k_1+k_2)k_3^2+c_t^3k_3^3\right],\\
\mathcal{J}^{(6)}&=\frac{c_6}{H^2}\frac{c_s^4k_1^2k_2^2(K''+c_tk_3)}{K''^2},
\end{align}
with $K'':=c_s(k_1+k_2)+c_tk_3$. The details of the derivation of the products of the polarization tensors $\mathcal{V}^{(i)}_{s_2s_3}$ are summarized in Appendix~\ref{Sec: app-polarization-tensor}. Note that $c_1/H^2, c_2/H, c_3/H^2,c_4/H,c_5$, and $c_6/H^2$ are constant.

Similarly to the scalar auto-bispectrum, by taking $n\to\infty$, one can check that the above reproduces the expression in the generalized G-inflation~\cite{Gao:2011vs}.\footnote{Regarding the normalization factor, we have used the following relation:
\begin{align}
\mathcal{P}_\zeta^2\frac{H^2}{c_t\mathcal{F}_T}=\mathcal{P}_\zeta\mathcal{P}_h \frac{H^2}{rc_t\mathcal{F}_T}=\mathcal{P}_\zeta\mathcal{P}_h\cdot\mathcal{P}_\zeta\frac{\pi^2}{2}\biggl(\frac{n}{1-n}\biggr)^2.
\end{align}
The authors of Ref.~\cite{Gao:2012ib} have normalized the bispectrum by $\mathcal{B}^{s_3}_{sst}\propto\mathcal{P}_\zeta^2\mathcal{A}^{s_3}_{sst}$, whereas we have done by $\mathcal{B}^{s_3}_{sst}\propto\mathcal{P}_\zeta\mathcal{P}_h\mathcal{A}^{s_3}_{sst}$, and hence the overall factors of $\mathcal{A}^{s_3}_{sst}$ between Ref.~\cite{Gao:2012ib} and the present paper are different even under $n\to\infty$. However, one can check the consistency by including the normalization factors. 
}

\subsubsection{Scalar-tensor-tensor bispectrum}\label{SubSubSec: stt-bispectra-const}
The scalar-tensor-tensor cross-bispectrum is computed as
\begin{align}
\mathcal{A}^{s_2s_3}_{stt}&=\frac{\pi^2}{4}\biggl(\frac{n}{n-1}\biggr)^2\mathcal{P}_h\sum_{q=1}^7\mathcal{I}^{(q)}(k_1,k_2,k_3)\mathcal{V}^{(q)}_{s_2s_3}({\bf k}_1,{\bf k}_2,{\bf k}_3)+({\bf k}_2,s_2\leftrightarrow{\bf k}_3,s_3),\label{eq: 3pt-stt-const}
\end{align}
where
\begin{align}
\mathcal{V}^{(1)}_{s_2s_3}&=e^{(s_2)*}_{ij}({\bf k}_2)e^{(s_3)*}_{ij}({\bf k}_3)=\mathcal{V}^{(4)}_{s_2s_3},\\
\mathcal{V}^{(2)}_{s_2s_3}&=\frac{k_1^2-k_2^2-k_3^2}{2}\mathcal{V}^{(1)}_{s_2s_3},\mathcal{V}^{(3)}_{s_2s_3}=-\frac{k_1^2-k_2^2+k_3^2}{2k_1^2}\mathcal{V}^{(1)}_{s_2s_3},\ \mathcal{V}^{(5)}_{s_2s_3}=k_1^2\mathcal{V}^{(1)}_{s_2s_3},\\
\mathcal{V}^{(6)}_{s_2s_3}&=\hat k_{1m}\hat k_{1n}e^{(s_2)*}_{ij}({\bf k}_2)e^{(s_3)*}_{ij}({\bf k}_3),\ \mathcal{V}^{(7)}_{s_2s_3}=k_1^2\mathcal{V}^{(6)}_{s_2s_3},
\end{align}
and
\begin{align}
\mathcal{I}^{(1)}&=\frac{b_1}{H^2}\frac{c_t^4k_2^2k_3^2(c_sk_1+K')}{K'^2},\\
\mathcal{I}^{(2)}&=-\frac{b_2}{H^2}\frac{c_s^3k_1^3+2c_s^2c_tk_1^2(k_2+k_3)+2c_sc_t^2k_1(k_2^2+k_2k_3+k_3^2)+c_t^3(k_2+k_3)(k_2^2+k_2k_3+k_3^2)}{K'^2},\\
\mathcal{I}^{(3)}&=\frac{b_3}{H^2}\frac{c_s^2c_t^2k_1^2k_2^2(K'+c_tk_3)}{K'^2},\\
\mathcal{I}^{(4)}&=\frac{b_4}{H}\frac{n-1}{n}\frac{2c_s^2c_t^4k_1^2k_2^2k_3^2}{K'^3},\\
\mathcal{I}^{(5)}&=b_5\biggl(\frac{n-1}{n}\biggr)^2\frac{2c_t^4k_2^2k_3^2(3c_sk_1+K')}{K'^4},\\
\mathcal{I}^{(6)}&=\frac{b_6}{H}\frac{n-1}{n}\frac{2c_s^2c_t^4k_1^2k_2^2k_3^2}{K'^3},\\
\mathcal{I}^{(7)}&=b_7\biggl(\frac{n-1}{n}\biggr)^2\frac{2c_t^4k_2^2k_3^2(3c_sk_1+K')}{K'^4},
\end{align}
with $K':=c_sk_1+c_t(k_2+k_3)$. Note that $b_1/H^2, b_2/H^2, b_3/H^2,b_4/H,b_5, b_6/H$, and $b_7$ are constant.

Similarly to the case of the scalar-scalar-tensor bispectrum, the derivation of the products of the polarization tensors is written in Appendix~\ref{Sec: app-polarization-tensor}. Taking $n\to\infty$, one can check that the above reproduces the expression in the generalized G-inflation~\cite{Gao:2011vs}.\footnote{Regarding the normalization factor, we have used the following relation:
\begin{align}
\mathcal{P}_\zeta^2\frac{8H^2\mathcal{F}_Sc_s}{\mathcal{F}_T^2c_t^2}=\mathcal{P}_\zeta\mathcal{P}_h\cdot\mathcal{P}_h\frac{\pi^2}{4}\biggl(\frac{n}{1-n}\biggr)^2.
\end{align}
The authors of Ref.~\cite{Gao:2012ib} have normalized the bispectrum by $\mathcal{B}^{s_2s_3}_{stt}\propto\mathcal{P}_\zeta^2\mathcal{A}^{s_2s_3}_{stt}$, whereas we have done by $\mathcal{B}^{s_2s_3}_{stt}\propto\mathcal{P}_\zeta\mathcal{P}_h\mathcal{A}^{s_2s_3}_{stt}$. Similarly to the scalar-scalar-tensor bispectrum, one can check the consistency under $n\to\infty$ by including the normalization factors.
}

\subsubsection{Tensor-tensor-tensor bispectrum}\label{SubSubSec: ttt-bispectra-const}
The tensor auto-bispectrum is computed as
\begin{align}
\mathcal{A}^{s_1s_2s_3}_t&=-\frac{K}{16}\biggl[1-\frac{1}{K^3}\sum_{i\neq j}k_i^2 k_j-4\frac{k_1k_2k_3}{K^3}\biggr]F_{\rm GR}+\frac{\mu H}{4\mathcal{G}_T}\frac{n-1}{n}\frac{k_1^2k_2^2k_3^2}{K^3}F_{\rm New}, \label{eq: ttt-bispectrum-nu3/2}
\end{align}
where
\begin{align}
F_{\rm GR}&:=e^{(s_1)*}_{ik}({\bf k}_1)e^{(s_2)*}_{jl}({\bf k}_2)\biggl[k_{3k}k_{3l}e^{(s_3)*}_{ij}({\bf k}_3)-\frac{1}{2}k_{3i}k_{3k}e^{(s_3)*}_{jl}({\bf k}_3)\biggr]+(5\ {\rm perm.}),\\
F_{\rm New}&:=e^{(s_1)*}_{ij}({\bf k}_1)e^{(s_2)*}_{jk}({\bf k}_2)e^{(s_3)*}_{ki}({\bf k}_3),
\end{align}
with $\mu:=\dot\phi X G_{5X}$. Here, the cubic interaction terms are of the form $h^2\partial^2 h$ and $\dot h^3$. The former interaction term exists in general relativity (or minimally coupled theories), while the latter one appears anew by introducing the $G_{5X}$ term. We thus labeled the products of the polarization tensors originating from the former and the latter as ``GR'' and ``New'', respectively. Also, by taking $n\to\infty$, one can check that the above reproduces the expression in the generalized G-inflation~\cite{Gao:2011vs,Gao:2012ib}. 

One can find from Eq.~(\ref{eq: ttt-bispectrum-nu3/2}) that the GR-type cubic operator yields the non-Gaussian amplitude which is independent of any parameters in the theory. We will mention this point in more detail in the following subsection.

\subsubsection{Non-Gaussianity consistency relations}\label{SubSubSec: consistency-const}
In this subsection, let us comment on the forms of the bispectra for several squeezed momentum triangle configurations to discuss the so-called Maldacena's consistency relation known in the context of inflation~\cite{Maldacena:2002vr,Creminelli:2004yq}.  The explicit forms of the non-Gaussianity consistency relations for the three-point functions in Eqs.~(\ref{eq: def-sss})--(\ref{eq: def-ttt}) are as follows (see Appendix~\ref{Sec: app-consistency} for a review of the derivation),
\begin{align}
\langle \zeta({\bf k}_1)\zeta({\bf k}_2)\zeta({\bf k}_3)\rangle|_{k_1\simeq k_2\gg k_3}&=(2\pi)^3\delta({\bf k}_1+{\bf k}_2+{\bf k}_3)\frac{(2\pi)^4\mathcal{P}_\zeta(k_1)\mathcal{P}_\zeta(k_3)}{k_1^3k_2^3k_3^3}\notag\\
&\quad\ \times\frac{1}{4}(1-n_s)k_2^3,\label{eq: sss-consistency-relation}\\
\langle \zeta({\bf k}_1)\zeta({\bf k}_2)\xi^{(s_3)}({\bf k}_3)\rangle|_{k_1\simeq k_2\gg k_3}&=(2\pi)^3\delta({\bf k}_1+{\bf k}_2+{\bf k}_3)\frac{(2\pi)^4\mathcal{P}_\zeta(k_1)\mathcal{P}_h(k_3)}{k_1^3k_2^3k_3^3}\notag\\
&\quad\ \times\frac{1}{8}\biggl(2-\frac{n_s}{2}\biggr)k_2^3\frac{k_{1i}k_{1j}}{k_1^2}e^{(s_3)*}_{ij}({\bf k}_3),\label{eq: sst-consistency-relation}\\
\langle \zeta({\bf k}_1)\xi^{(s_2)}({\bf k}_2)\xi^{(s_3)}({\bf k}_3)\rangle|_{k_1\ll k_2\simeq k_3}&=(2\pi)^3\delta({\bf k}_1+{\bf k}_2+{\bf k}_3)\frac{(2\pi)^4\mathcal{P}_\zeta(k_1)\mathcal{P}_h(k_3)}{k_1^3k_2^3k_3^3}\notag\\
&\quad\ \times\frac{1}{8}(-n_t)k_2^3\delta_{s_2s_3},\label{eq: stt-consistency-relation}\\
\langle \xi^{(s_1)}({\bf k}_1)\xi^{(s_2)}({\bf k}_2)\xi^{(s_3)}({\bf k}_3)\rangle|_{k_1\simeq k_2\gg k_3}&=(2\pi)^3\delta({\bf k}_1+{\bf k}_2+{\bf k}_3)\frac{(2\pi)^4\mathcal{P}_h(k_1)\mathcal{P}_h(k_3)}{k_1^3k_2^3k_3^3}\notag\\
&\quad\ \times\frac{1}{16}\biggl(\frac{3-n_t}{2}\biggr)\delta_{s_1s_2}k_2^3\frac{k_{1i}k_{1j}}{k_1^2}e^{(s_3)*}_{ij}({\bf k}_3).\label{eq: ttt-consistency-relation}
\end{align}
Since the general bounce model for $\nu=3/2$ is conformally equivalent to a single-clock inflation, we discuss the above consistency relations. To do so, instead of using the above forms of squeezed bispectra, let us define the non-linearity parameters by
\begin{align}
f^s_{\rm NL}&:=\frac{10}{3}\frac{\mathcal{A}_s}{\sum_i k_i^3},\ f^{sst}_{\rm NL}:=\frac{10}{3}\frac{\mathcal{A}^{s_3}_{sst}}{\sum_i k_i^3},\notag\\
f^{stt}_{\rm NL}&:=\frac{10}{3}\frac{\mathcal{A}^{s_2s_3}_{stt}}{\sum_i k_i^3},\ f^{t}_{\rm NL}:=\frac{10}{3}\frac{\mathcal{A}^{s_1s_2s_3}_{t}}{\sum_i k_i^3}. \label{eq: def-fnl}
\end{align}
By taking $n_s\to1$ and $n_t\to0$ to the above parameters derived from Eqs.~(\ref{eq: sss-consistency-relation})--(\ref{eq: ttt-consistency-relation}), we have
\begin{align}
{\bf f}^{\rm sq}_{\rm NL}&:=
\begin{pmatrix}
f^{s}_{\rm NL} & f^{sst}_{\rm NL} \\
f^{stt}_{\rm NL} & f^t_{\rm NL}
\end{pmatrix}\biggr|_{{\rm squeezed}}\notag\\
&=\frac{5}{12}\begin{pmatrix}
1-n_s & \frac{1}{8}(4-n_s) \\
-\frac{n_t}{2} & \frac{1}{16}(3-n_t)\delta_{s_1s_2}
\end{pmatrix}\\
&\to\begin{pmatrix}
0 & \frac{5}{32} \\
0 & \frac{5}{64}\delta_{s_1s_2}
\end{pmatrix}, \label{eq: fnl-consistency}
\end{align}
where the subscript ``sq'' indicates a set of the non-linearity parameter evaluated under the squeezed limits written in Eqs.~(\ref{eq: sss-consistency-relation})--(\ref{eq: ttt-consistency-relation}).
To the non-linearity parameters derived from the bispectra for $\nu=3/2$, we take the same squeezed limits under which we have
\begin{align}
{\bf f}^{\rm sq}_{\rm NL}&\to
\begin{pmatrix}
\frac{5}{24}\left(\Lambda_2+3\frac{\Lambda_3}{c_s^2}\right)& \frac{5}{32} \\
\frac{5}{6\mathcal{F}_{T}}(3b_2+c_t^2b_1)\delta_{s_2s_3} & \frac{5}{64}\delta_{s_1s_2}
\end{pmatrix}.
\end{align}
By using the explicit forms of $\Lambda_2, \Lambda_3, b_2$, and $b_4$, we can find
\begin{align}
\Lambda_2+\frac{3\Lambda_3}{c_s^2}&=\frac{4H\mathcal{G}_T}{nc_s^2\Theta}(\alpha+2),\\
\frac{1}{\mathcal{F}_T}(3b_2+c_t^2b_1)&=\frac{H\mathcal{G}_T}{2nc_t^2\Theta}(\alpha+2),
\end{align}
both of which vanish for $\alpha=-2$ (i.e., $\nu=3/2$). Therefore, the present bounce model for $\nu=3/2$ recovers Eq.~(\ref{eq: fnl-consistency}). Here, in our computations of the three-point functions, we have imposed $n_s-1=0=n_t$, and hence it would be interesting to see whether the explicit forms of Eqs.~(\ref{eq: sss-consistency-relation})--(\ref{eq: ttt-consistency-relation}) hold by taking into account the deviations from the scale invariance (i.e., $n_s-1\ll1$ and $n_t\ll1$) in the computations of the three-point functions, which is beyond the scope of the present paper.\footnote{See Ref.~\cite{Ageeva:2024knc} which reported an enhancement of a squeezed-type scalar non-Gaussianity for $c_s\ll1$ in a contracting model corresponding to $\nu-3/2\simeq0>0$ (i.e., the case of not an exactly-flat scalar power spectrum but a slightly-red one). However, the authors of that paper showed their numerical results of the non-linearity parameters of scalar non-Gaussianity for $\nu-3/2=\mathcal{O}(10^{-2})$, and it is unknown whether those enhancements originated from the small deviations or not. It is therefore important to compute the bispectra in the slightly-red case.}
Here, overall amplitudes of the cubic interactions are arbitrary (i.e., the time variations of time-dependent quantities are not suppressed by the conventional slow-roll parameters), and hence each coefficient can, in principle, be of $\mathcal{O}(1)$. The non-squeezed non-Gaussianities are therefore not necessarily much smaller than unity. 

\subsubsection{Shapes}
In this subsection, we study the shapes of the bispectra that we computed in the previous subsections. As we have mentioned before, the momentum dependence in the present model for $\nu=3/2$ is identical to that in the generalized G-inflation. Therefore, the analysis of the shape for $\nu=3/2$ is mostly overlapped with that in the literature. However, the impact of the non-Gaussianity consistency relation on the shape of the cross-bispectrum has not been discussed. Furthermore, the consequence of the consistency relation is important when we discuss the differences in the shapes between $\nu=3/2$ and $\nu=-3/2$. We thus highlight several points related to the consistency relation in this subsection and leave the other detailed analysis to Appendix~\ref{Appendix: shape} in which we newly study the shape of the scalar-tensor-tensor cross-bispectrum in the mixed-helicity case, $s_2=-s_3=+1$. A summary of the following analysis is given in Table~\ref{table: auto-nu3/2} for the auto-bispectra and in Table~\ref{table: cross-nu3/2} for the cross-bispectra.

\begin{figure}\centering
\subfloat[$\mathcal{A}_{s,2}/(k_1k_2k_3)$.]{\label{a}\includegraphics[width=.5\linewidth]{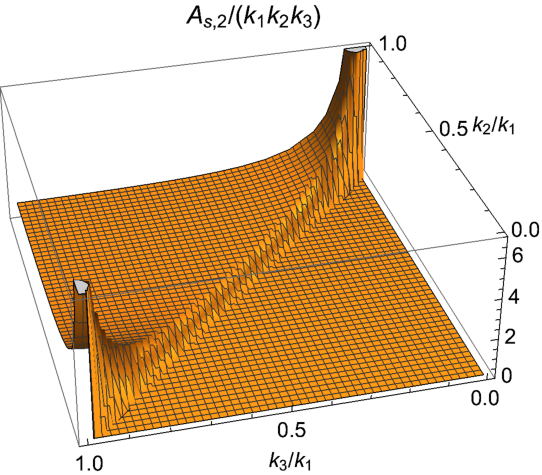}}\hfill
\subfloat[$\mathcal{A}_{s,3}/(k_1k_2k_3)$.]{\label{b}\includegraphics[width=.5\linewidth]{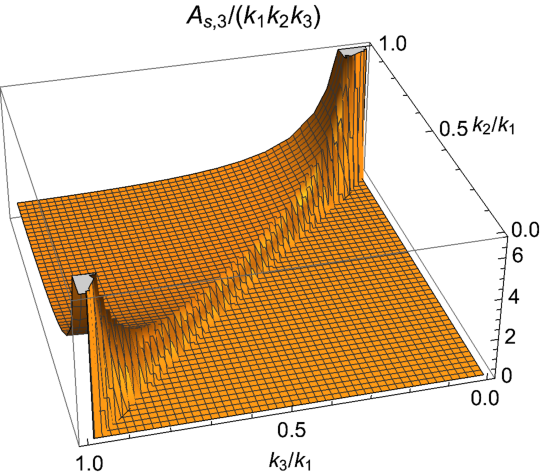}}\par 
\subfloat[$(\mathcal{A}_{s,2}+\mathcal{A}_{s,3})/(k_1k_2k_3)$.]{\label{c}\includegraphics[width=.5\linewidth]{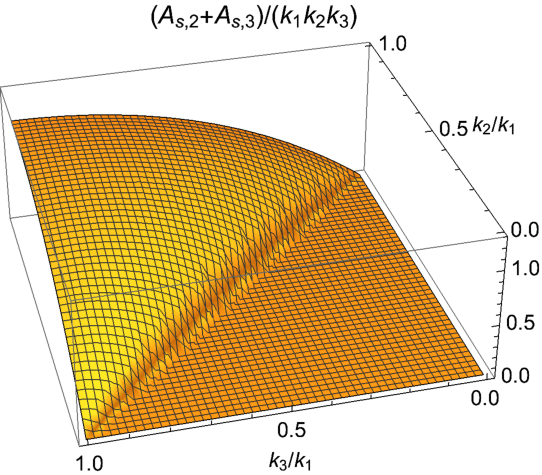}}
\caption{The plot of the bispectrum $\mathcal{A}_s$ from $\Lambda_2$ (a), $\Lambda_3$ (b) and both of them $(c)$ as a function of $k_2/k_1$ and $k_3/k_1$. We set $\mathcal{A}_/(k_1k_2k_3)$ to $1$ for the equilateral configuration.}
\label{Fig: Lambda23}
\end{figure}

Hereafter, we express $\mathcal{A}$ coming from each cubic interaction as $\mathcal{A}_\bullet$; for instance $\mathcal{A}_{s,1}$ stands for the $\Lambda_1$ term of $\mathcal{A}_s$. If a bispectrum has a local shape, i.e.,  a divergent peak at a squeezed limit $k_1\ll k_2\simeq k_3$, the bispectrum at that limit is like
\begin{align}
\frac{\mathcal{A}}{k_1k_2k_3}\propto \frac{k_2}{k_1}. \label{eq: expression-local}
\end{align}
Among the bispectra, the scalar auto-bispectrum and the scalar-tensor-tensor cross-bispectrum are suppressed at the squeezed limit, $k_1\ll k_2\simeq k_3$. When we study each term of these bispectra, some of them are found to peak at that squeezed limit as Eq.~(\ref{eq: expression-local}).
First examples that we study are the $\mathcal{A}_{s,2}$ and $\mathcal{A}_{s,3}$ terms, each of which has a divergent behavior at the squeezed limit as
\begin{align}
\frac{\mathcal{A}_{s,2}}{k_1k_2k_3}&\simeq \frac{k_2}{8k_1}\Lambda_2,\ \frac{\mathcal{A}_{s,3}}{k_1k_2k_3}\simeq \frac{3k_2}{8c_s^2k_1}\Lambda_3. \label{eq: sss23-squeezed}
\end{align}
Here, to show the consistency relation, we have used $\Lambda_2+3\Lambda_3/c_s^2=0$ for $\alpha=-2$. Under this relation, the divergent behavior in Eq.~(\ref{eq: sss23-squeezed}) cancels out, and $(\mathcal{A}_{s,2}+\mathcal{A}_{s,3})/(k_1k_2k_3)$ has a peak at the equilateral limit as shown in Figure~\ref{Fig: Lambda23}.
The other terms i.e., $\Lambda_i$ ($i\neq2,3$), peak at the equilateral limit. As a result, the scalar auto-bispectrum is found to be close to the equilateral shape, which is consistent with the analysis in the generalized G-inflation~\cite{Gao:2011qe,DeFelice:2011uc,DeFelice:2013ar}.

We next study the shapes of the scalar-tensor-tensor bispectrum. As has been studied in Ref.~\cite{Gao:2012ib}, the cross-bispectrum has $c_t/c_s$ dependence, and hence we consider the three cases, $c_t/c_s=1, 100, 0.01$. The shape of each term for $c_t/c_s=1$ is similar to that for $c_t/c_s=100$.

\begin{figure}\centering
\subfloat[$c_t/c_s=1$.]{\label{a}\includegraphics[width=.5\linewidth]{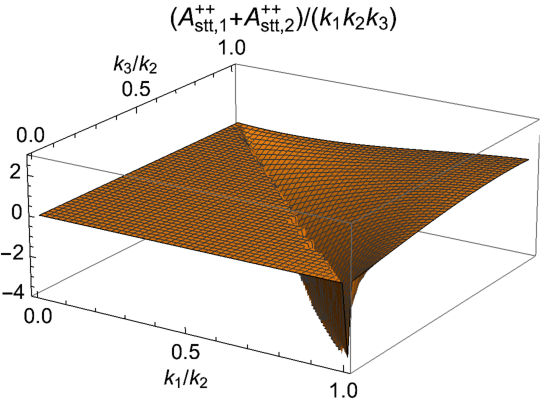}}\hfill
\subfloat[$c_t/c_s=100$.]{\label{b}\includegraphics[width=.5\linewidth]{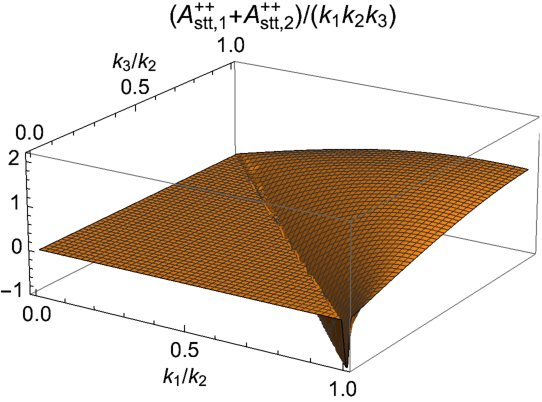}}\par 
\subfloat[$c_t/c_s=0.01$]{\label{c}\includegraphics[width=.5\linewidth]{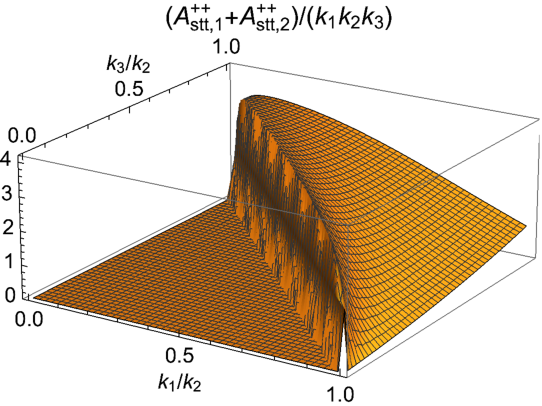}}
\caption{ $(\mathcal{A}^{++}_{stt,1}+\mathcal{A}^{++}_{stt,2})/(k_1k_2k_3)$ as a function of $k_1/k_2$ and $k_3/k_2$. We set $\mathcal{A}_/(k_1k_2k_3)$ to $1$ for the equilateral configuration. The plots of (a), (b), and (c) correspond to the ones for $c_t/c_s=1, c_t/c_s=100$, and $c_t/c_s=0.01$, respectively.}
\label{Fig: b12}
\end{figure}

\begin{figure} [htb]
     \begin{tabular}{cc}
        \begin{minipage}{0.45\hsize}
            \centering
            \includegraphics[width=7.cm]{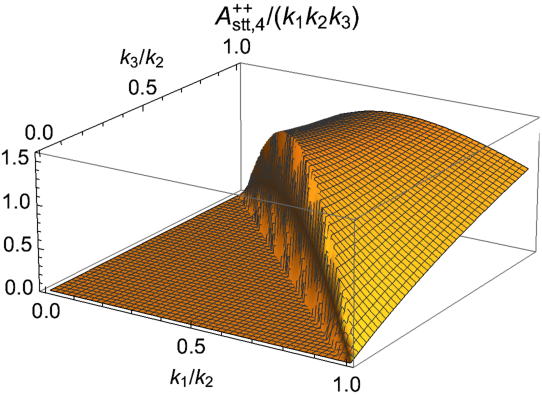}
        \end{minipage} &
        \begin{minipage}{0.45\hsize}
            \centering
            \includegraphics[width=7.cm]{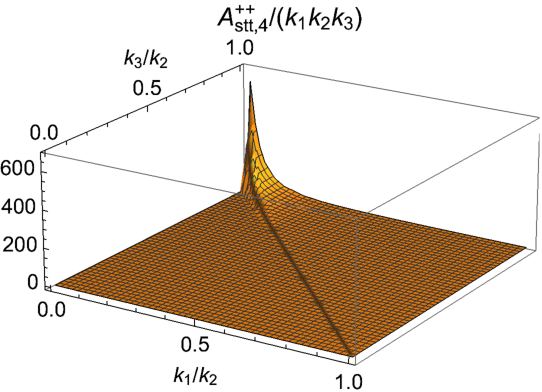}
        \end{minipage} 
    \end{tabular}
\caption{\emph{Left}:$\mathcal{A}^{++}_{stt,4}/(k_1k_2k_3)$ as a function of $k_1/k_2$. We have taken $c_t/c_s=1$ and normalized the plot to $1$ for the equilateral configuration.\emph{Right}: $\mathcal{A}^{++}_{stt,4}/(k_1k_2k_3)$ as a function of $k_1/k_2$ and $k_3/k_2$. We have taken $c_t/c_s=0.01$ and normalized the plot to $1$ for the equilateral configuration.}\label{Fig: b4}
\end{figure}

Among the cross-bispectra, $\mathcal{A}^{++}_{stt,1}/(k_1k_2k_3)$ and $\mathcal{A}^{++}_{stt,2}/(k_1k_2k_3)$ behave at the squeezed limit, $k_1\ll k_2\simeq k_3$, as 
\begin{align}
\frac{\mathcal{A}^{++}_{stt,1}}{k_1k_2k_3}\simeq \frac{c_t^2b_1}{2\mathcal{F}_T}\frac{k_2}{k_1},\ \frac{\mathcal{A}^{++}_{stt,2}}{k_1k_2k_3}\simeq \frac{3b_2}{2\mathcal{F}_T}\frac{k_2}{k_1}, \label{eq: stt12-squeezed}
\end{align}
each of which has a divergent peak at the squeezed limit. Similarly to the scalar auto-bispectrum, this divergent behavior cancels out by using the condition that we have used in Sec.~\ref{Sec: bispectra}, i.e., $c_t^2b_1+3b_2=0$ which holds for $\alpha=-2$. Under this relation, one can find that each of the two terms in Eq.~(\ref{eq: stt12-squeezed}) indeed cancels out. The same cancellation has occurred for the scalar auto-bispectrum, and hence we show only the plot of $(\mathcal{A}^{s_2s_3}_{stt,1}+\mathcal{A}^{s_2s_3}_{stt,2})/(k_1k_2k_3)$.
Figure~\ref{Fig: b12} shows that the aforementioned sharp peak at the squeezed limit cancels out, and the resultant bispectrum has a $c_t/c_s$-dependent shape that is different from the local shape.\footnote{The amplitude for $c_t/c_s=0.01$ is slightly enhanced than those for $c_t/c_s=1, 100$ at $k_1/k_2\ll1$. $(\mathcal{A}^{s_2s_3}_{stt,1}+\mathcal{A}^{s_2s_3}_{stt,2})/(k_1k_2k_3)$ behaves at $k_3\simeq k_2\gg k_1$ as
\begin{align}
\frac{\mathcal{A}^{++}_{stt,1}+\mathcal{A}^{++}_{stt,2}}{k_1k_2k_3}&\propto \frac{k_1}{k_2^2}[k_1^3+4(c_t/c_s)k_1^2k_2+2(3c_t^2/c_s^2-1)k_1k_2^2+2(c_t/c_s)(3c_t^2/c_s^2-4)k_2^3]\notag\\
&\quad\ \times\frac{1}{[k_1+2(c_t/c_s)k_2]^2}.
\end{align}
For $c_t/c_s\ll1$, the above takes its maximum value for $k_1/k_2\sim\sqrt{c_t/c_s}\ll1$. For $c_t/c_s\geq1$, the $k_2$ term dominates the $k_1$ term in the denominator of the second line, and hence any enhancements around $k_1\ll k_2$ did not appear. This has resulted in the difference in the shapes between $c_t/c_s\geq1$ and $c_t/c_s\ll1$.\label{foot:b12-const}} 
The other bispectra do not show any divergent peaks at the squeezed limit, but the $b_4, b_5, b_6$, and $b_7$ terms can peak at that limit only for $c_t/c_s\ll1$. Those have similar shapes, and thus we focus on the $b_4$ term. Figure~\ref{Fig: b4} shows that $\mathcal{A}^{++}_{stt,4}/(k_1k_2k_3)$ for $c_t/c_s=1$ does not have a sharp peak, while that for $c_t/c_s\ll 1$ has a peak around $k_1/k_2\ll1, k_3/k_2\simeq1$. 
However, we emphasize that this $c_t/c_s$-dependent peak appears for a momentum triangle configuration away from the exact squeezed one, $k_1\to0$.\footnote{$\mathcal{A}^{++}_{stt,4}/(k_1k_2k_3)$ can be written as
\begin{align}
\frac{\mathcal{A}^{++}_{stt,4}}{k_1k_2k_3}&\propto \frac{k_1[k_1^2-(k_2+k_3)^2]^2}{k_2k_3[k_1+(c_t/c_s)(k_2+k_3)]^3}.
\end{align}
For $k_2\simeq k_3$ and $c_t/c_s\ll1$, the above takes the maximum value for $k_1\simeq (c_t/c_s)k_2$. For 
$k_2\simeq k_3$ and $c_t/c_s\geq1$, the $k_2$ and $k_3$ terms dominate the $k_1$ term in the denominator of the above, and hence any enhancements around $k_1/k_2\ll1$ did not appear. Note that the above is not enhanced at the exact squeezed limit $k_1\to0$, and the sharpness and location of the peak are different from those of the local shape.}

A summary of the shapes for $\nu=3/2$ is given in Table~\ref{table: auto-nu3/2} for the auto-bispectra and Table~\ref{table: cross-nu3/2} for the cross-bispectra. See Appendix~\ref{Appendix: shape} for the detailed analysis of the shapes of the other terms.

\begin{table}[htb]
  \centering
  \begin{tabular}{|c|c|c|c|} \hline
    \diagbox{\quad\quad\quad\quad\quad\ }{\quad\quad\quad\quad\quad\ } & \\ \hline
    $\Lambda_i\ (i\neq2,3)$ & equilateral \\ \hline
    $\Lambda_{2+3}$ & equilateral \\ \hline
    ${\rm GR}$\ ($s_1=s_2=s_3=+1$)& squeezed\\ \hline
    ${\rm New}$\ ($s_1=s_2=s_3=+1$)& equilateral\\ \hline
    ${\rm GR}$\ ($s_1=s_2=-s_3=+1$)& squeezed\\ \hline
    ${\rm New}$\ ($s_1=s_2=-s_3=+1$)& other\\ \hline
  \end{tabular}
  \caption{Summary of the shapes of the auto-bispectra for $\nu=3/2$. The shape that appears to be away from the squeezed or equilateral ones is denoted by ``other''.}
  \label{table: auto-nu3/2}
\end{table}

\begin{table}[htb]
  \centering
  \begin{tabular}{|c|c|c|c|} \hline
    \diagbox{\quad\quad\quad\quad\quad }{\quad\quad\quad\quad\quad } &  $c_t/c_s=1$ &  $c_t/c_s=0.01$ &  $c_t/c_s=100$ \\ \hline
    $c_1$ & squeezed & squeezed & other \\ \hline
    $c_i\ (i\neq1)$ & equilateral & equilateral & other \\ \hline
    $b_{(1+2,3)}$\ ($s_2=s_3=+1$) & other & other & other \\ \hline
    $b_{(4,6,7)}$\ ($s_2=s_3=+1$) & other & squeezed & other \\ \hline
    $b_5$\ ($s_2=s_3=+1$) & $-$ & squeezed & other \\ \hline
    $b_{(1+2,3,4)}$\ ($s_2=-s_3=+1$) & other & other & other \\ \hline
    $b_{5}$\ ($s_2=-s_3=+1$) & $-$ & other & other \\ \hline
    $b_{(6,7)}$\ ($s_2=-s_3=+1$) & equilateral & equilateral & equilateral \\ \hline
  \end{tabular}
  \caption{Summary of the shapes of the cross-bispectra for $\nu=3/2$. The shapes that appear to be away from the squeezed or equilateral ones are denoted by ``other''. Note that $b_5\propto(c_s^2/c_t^2-1)=0$ for $c_t/c_s=1$.}
  \label{table: cross-nu3/2}
\end{table}

\subsection{$\nu=-3/2$}\label{SubSec: bispectra-grow}
In this subsection, we show the bispectra for $\nu=-3/2$ where the field redefinitions yield non-negligible contributions due to the growth of the perturbations. 
\subsubsection{Scalar-scalar-scalar bispectrum}\label{SubSubSec: sss-bispectra-grow}
On the present contracting background, the scalar auto-bispectrum has been computed as~\cite{Akama:2019qeh},
\begin{align}
\mathcal{A}_s&=\frac{1}{8}\biggl[\biggl(\frac{9(1-n)}{n}\Lambda_1-\Lambda_2+\frac{\Lambda_5}{2}\biggr)\sum_i k_i^3+\frac{\Lambda_6}{2}\sum_{i\neq j}k_i^2k_j\notag\\
&\quad +\frac{1}{2k_1^2k_2^2k_3^2}\biggl(\Lambda_6\sum_i k_i^9-(\Lambda_5+\Lambda_6)\sum_{i\neq j}k_i^7k_j^2-\Lambda_6\sum_{i\neq j}k_i^6k_j^3+(\Lambda_5+\Lambda_6)\sum_{i\neq j}k_i^5k_j^4\biggr)\biggr]\notag\\
&\quad +\frac{3}{8}\frac{1-n}{n}\biggl[(\Lambda_{{\rm red}_1}-4\Lambda_{{\rm red}_2})\sum_i k_i^3+\frac{\Lambda_{{\rm red}_1}}{4}\sum_{i\neq j}k_i^2k_j\notag\\
&\quad\ -\frac{\Lambda_{{\rm red}_1}}{4}\frac{1}{k_1^2k_2^2k_3^2}\biggl(\sum_{i\neq j}k_i^7k_j^2+\sum_{i\neq j}k_i^6k_j^3-2\sum_{i\neq j}k_i^5k_j^4\biggr)\biggr],\label{eq: 3pt-sss-grow}
\end{align}
where the $\Lambda_3, \Lambda_4, \Lambda_7$, and $\Lambda_8$ terms were suppressed on the superhorizon scales, and hence those were ignored. In the models having the growing superhorizon modes, the non-Gaussianities are mainly generated on the superhorizon scales, which indicates that the spatial derivatives suppress the generation of non-Gaussianities. Actually, by means of power counting, one can find that the $\Lambda_3, \Lambda_4, \Lambda_7$, and $\Lambda_8$ terms involve spatial derivatives e.g., $\Lambda_3(\partial_i)^2\zeta^3$, while the others do not, e.g., $\Lambda_1(\partial_i)^0(\partial_t)^3\zeta^3$. As a result, the non-Gaussianities from those four terms were suppressed on the superhorizon scales.\footnote{Whether this interpretation holds or not depends on the time dependence of coupling of cubic interactions. In fact, there is an exception to this suppression, as has been studied in the context of cubic Weyl theories~\cite{Akama:2024bav}. The non-Gaussianities from the cubic interactions involving spatial derivatives can be enhanced with nontrivial time dependence of the coupling such that the magnitude of the coupling is enhanced at horizon-cross scales. However, this was not the case for the present model.} Unlike in the case of $\nu=3/2$, the field redefinition yields non-negligible contributions to the three-point functions. See also Appendix~\ref{Sec: app-boundary} for more details. A concrete example realizing the small tensor-to-scalar ratio and the small scalar non-Gaussianity at the same time has been presented in Ref.~\cite{Akama:2019qeh}.

\subsubsection{Scalar-scalar-tensor bispectrum}\label{SubSubSec: sst-bispectra-grow}
Similarly to the scalar auto-bispectrum, the growth of perturbations results in non-vanishing contributions from the field redefinitions. As a result,  $\mathcal{A}^{s_3}_{sst}$ takes the following form,
\begin{align}
\mathcal{A}^{s_3}_{sst}=\frac{\pi^2}{2}\biggl(\frac{n}{n-1}\biggr)^2\mathcal{P}_\zeta(\eta_b)\biggl[\mathcal{J}^{(3)}|_{\eta=\eta_b}
\mathcal{V}^{(3)}_{s_3}+\mathcal{J}^{(6)}|_{\eta=\eta_b}
\mathcal{V}^{(6)}_{s_3}+\sum_{q=1}^2\mathcal{J}^{({\rm red},q)}|_{\eta=\eta_b}
\mathcal{V}^{({\rm red},q)}_{s_3}\biggr]+({\bf k}_1\leftrightarrow {\bf k}_2), \label{eq: 3pt-sst-grow}
\end{align}
where
\begin{align}
\mathcal{V}^{({\rm red},1)}_{s_3}&=-\frac{1}{k_2^2}\mathcal{V}^{(1)}_{s_3}, \mathcal{V}^{({\rm red},2)}_{s_3}=\frac{k_3^2}{k_1^2}\mathcal{V}^{({\rm red},1)}_{s_3},
\end{align}
and
\begin{align}
\mathcal{J}^{(3)}&=\frac{c_3}{2H^2}[c_s^3(k_1^3-k_2^3)-c_t^3k_3^3],\\
\mathcal{J}^{(6)}&=-\frac{c_6}{2H^2}[c_s^3(k_1^3+k_2^3)-c_t^3k_3^3],\\
\mathcal{J}^{({\rm red},1)}&:=c_{\rm red_1}(c_s^3k_2^3+c_t^3k_3^3),\\
\mathcal{J}^{({\rm red},2)}&:=c_{\rm red_2}c_s^3(k_1^3+k_2^3).
\end{align}
The explicit forms of $c_3, c_6, c_{\rm red_1}$, and $c_{\rm red_2}$ are summarized in Appendix~\ref{Sec: app-perturbed-actions}, and the derivation of the contributions originating from the field redefinitions is summarized in Appendix~\ref{Sec: app-boundary}. The $c_1, c_2, c_4$, and $c_5$ terms are suppressed on the superhorizon scales in proportion to a power of $-k_i\eta\ll1$, and the leading-order bispectrum is determined by the $c_3$, $c_6$, $c_{{\rm red}_1}$, and $c_{\rm red_2}$ terms. One can confirm that, by means of power counting, the cubic operators of the $c_1, c_2, c_4$, and $c_5$ terms involve the spatial derivatives, while the others do not. Note that the $c_3$, $c_6$, and $c_{{\rm red}_1}$ terms are present in a minimally coupled theory, while the $c_{\rm red_2}$ one is only in a non-minimally coupled theory with $G_{5X}\neq0$.

\subsubsection{Scalar-tensor-tensor bispectrum}\label{SubSubSec: stt-bispectra-grow}
We can compute $\mathcal{A}^{s_2s_3}_{stt}$ as
\begin{align}
\mathcal{A}^{s_2s_3}_{stt}&=\frac{\pi^2}{4}\biggl(\frac{n}{n-1}\biggr)^2\mathcal{P}_h(\eta_b)\biggl[\mathcal{I}^{(1)}|_{\eta=\eta_b}\mathcal{V}^{(1)}_{s_2,s_3}+\mathcal{I}^{(3)}|_{\eta=\eta_b}\mathcal{V}^{(3)}_{s_2,s_3}+\mathcal{I}^{(4)}|_{\eta=\eta_b}\mathcal{V}^{(4)}_{s_2,s_3}+\mathcal{I}^{(6)}|_{\eta=\eta_b}\mathcal{V}^{(6)}_{s_2,s_3}\notag\\
&\quad\ +\sum_{q=1}^2\mathcal{I}^{({\rm red},q)}|_{\eta=\eta_b}\mathcal{V}^{({\rm red},q)}_{s_2,s_3}\biggr]+({\bf k}_2,s_2\leftrightarrow {\bf k}_3,s_3),\label{eq: 3pt-stt-grow}
\end{align}
where
\begin{align}
\mathcal{V}^{({\rm red},1)}_{s_2,s_3}=
\mathcal{V}^{({\rm red},2)}_{s_2,s_3}=\mathcal{V}^{(1)}_{s_2s_3},
\end{align}
and
\begin{align}
\mathcal{I}^{(1)}&=\frac{b_1}{2H^2}[c_s^3k_1^2-c_t^3(k_2^3+k_3^3)],\\
\mathcal{I}^{(3)}&=-\frac{b_3}{2H^2}[c_s^3k_1^3+c_t^3(k_2^3-k_3^3)],\\
\mathcal{I}^{(4)}&=\frac{3b_4}{2H}\frac{1-n}{n}[c_s^3k_1^3+c_t^3(k_2^3+k_3^3)],\\
\mathcal{I}^{(6)}&=\frac{3b_6}{2H}\frac{1-n}{n}[c_s^3k_1^3+c_t^3(k_2^3+k_3^3)],\\
\mathcal{I}^{({\rm red}_1)}&=b_{\rm red_1}[c_s^3k_1^3+c_t^3(k_2^3+k_3^3)],\\
\mathcal{I}^{({\rm red}_2)}&=b_{\rm red_2}c_t^3(k_2^3+k_3^3).
\end{align}
The explicit form of each coefficient and the derivation of the $b_{\rm red}$ terms originating from the field redefinitions are summarized in Appendix~\ref{Sec: app-perturbed-actions} and Appendix~\ref{Sec: app-boundary}, respectively. The $b_2$, $b_5$, and $b_7$ terms are suppressed on the superhorizon scales, and hence the leading-order bispectrum is determined by the $b_1$, $b_3$, $b_4$, $b_6$, $b_{\rm red_1}$, and $b_{\rm red_2}$ terms. One can check that, by means of power counting, the cubic operators of the $b_2, b_5$, and $b_7$ terms involve the spatial derivatives, while the others do not. Note that, among the leading-order terms the $b_1$, $b_3$, and $b_{\rm red_2}$ terms are present in a minimally coupled theory, while the $b_4$, $b_6$, $b_{\rm red_1}$ ones are only in a non-minimally coupled theory with $G_{5X}\neq0$.

\subsubsection{Tensor-tensor-tensor bispectrum}\label{SubSubSec: ttt-bispectra-grow}
The full expression of $\mathcal{A}^{s_1s_2s_3}_t$ has been obtained as~\cite{Akama:2019qeh}
\begin{align}
\mathcal{A}^{s_1s_2s_3}_t&=\biggl[-\frac{1}{64}c_t^2\eta_b^2F_{\rm GR}+\frac{3}{16}\frac{1-n}{n}\frac{\mu H}{\mathcal{G}_T}F_{\rm New}\biggr]\biggl(\sum_i k_i^3\biggr).\label{eq: 3pt-ttt-grow}
\end{align}
The first term in $[\cdots]$ is present in a minimally coupled theory, while the second one is in a non-minimally coupled theory with $G_{5X}\neq0$. The first term comes from the cubic operators with only the spatial derivatives and is suppressed on the superhorizon scales, and hence the tensor auto-bispectrum can be non-negligible only in a theory with $G_{5X}\neq0$.

Before closing this section, let us briefly mention the amplitudes of the non-linearity parameters at the squeezed limits originating from the full bispectra. For the models for $\nu=3/2$, the non-Gaussianities at several squeezed limits are suppressed similarly to the squeezed non-Gaussianities from inflation where the Maldacena's consistency relation holds. However, if the perturbations grow on the superhorizon scales and the non-Gaussianities are generated on the superhorizon scales, such suppression does not occur. For example, in matter-dominated contracting models within the k-essence theory (corresponding to $\nu=-3/2$ in the present model), a small scalar propagation speed (i.e., $c_s\ll1$) yields enhanced scalar non-Gaussianities at both squeezed and non-squeezed limits~\cite{Li:2016xjb,Akama:2019qeh}. Thus, the non-suppressed squeezed non-Gaussianity is a characteristic feature of the present model for $\nu=-3/2$. As we will show in the following section, the presence or absence of peaks at squeezed limits will also be a difference in the shapes of bispectra for $\nu=3/2$ and $\nu=-3/2$. 

\subsubsection{Shapes}
In this subsection, we study the shapes of full bispectra for $\nu=-3/2$. As we will show below, the full bispectra for $\nu=-3/2$ peak at squeezed limits over a wider range of parameters than those for $\nu=3/2$. This is one of the consequences of the violation of the non-Gaussianity consistency relations. A summary of the following analysis is given in Table~\ref{table: auto-nu-3/2} for the auto-bispectra and in Table~\ref{table: cross-nu-3/2} for the cross-bispectra.

First, the $\Lambda_1$ term is exactly the same as the so-called local template (i.e., $\mathcal{A}_{s,1}/(k_1k_2k_3)\propto(\sum_i k_i^3)/(k_1k_2k_3)$), and hence this term has a divergent peak at the squeezed limit, $k_i\to0$, which is in contrast to the scalar auto-bispectrum for $\nu=3/2$ which peaks at the equilateral limit. The other terms have similar shapes peaking at the squeezed limit, and hence we do not show those. Our results are consistent with those obtained in Ref.~\cite{Cai:2009fn,Li:2016xjb} where a shape of a scalar auto-bispectrum from matter bounce has been studied within the k-essence theory.\footnote{As has been studied in Ref.~\cite{Li:2016xjb}, the divergent peak can vanish in a certain parameter choice (e.g., $c_s=\sqrt{26/33}$ in the case of that paper). However, this is model-dependent, which is in contrast to the model for $\nu=3/2$ where the divergent peak disappeared regardless of a parameter choice. We do not discuss such a model-dependent feature in the present paper.}

\begin{figure} [htb]
     \begin{tabular}{cc}
        \begin{minipage}{0.45\hsize}
            \centering
            \includegraphics[width=7.cm]{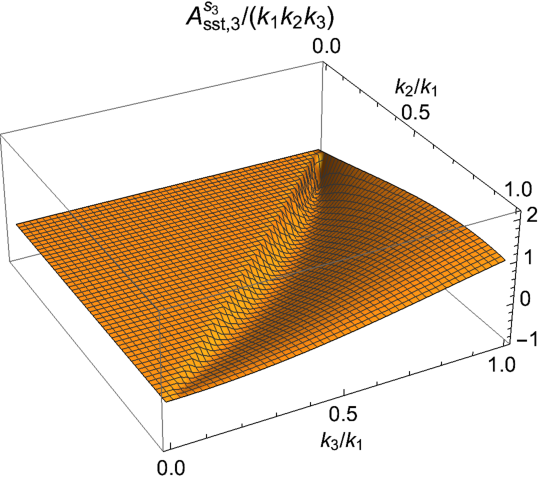}
        \end{minipage} &
        \begin{minipage}{0.45\hsize}
            \centering
            \includegraphics[width=7.cm]{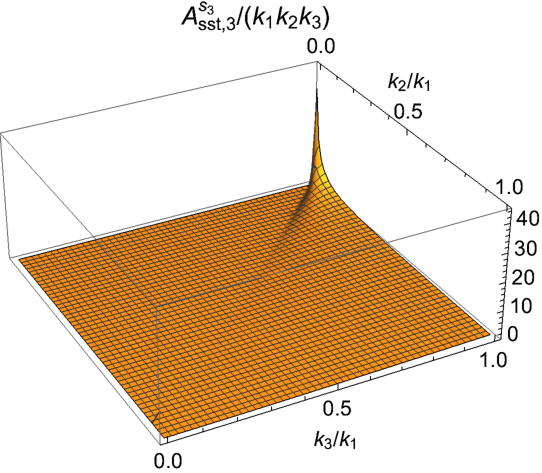}
        \end{minipage} 
    \end{tabular}
\caption{\emph{Left}:$\mathcal{A}^{s_3}_{sst,3}/(k_1k_2k_3)$ as a function of $k_2/k_1$ and $k_3/k_1$. We took $c_t/c_s=1$ and set $\mathcal{A}/(k_1k_2k_3)$ to $1$ for the equilateral configuration.\emph{Right}: $\mathcal{A}^{s_3}_{sst,3}/(k_1k_2k_3)$ as a function of $k_2/k_1$ and $k_3/k_1$. We took $c_t/c_s=100$ and set $\mathcal{A}/(k_1k_2k_3)$ to $1$ for the equilateral configuration.}\label{Fig: C3}
\end{figure}

\begin{figure}[h]
\centering
\includegraphics[width=70mm]{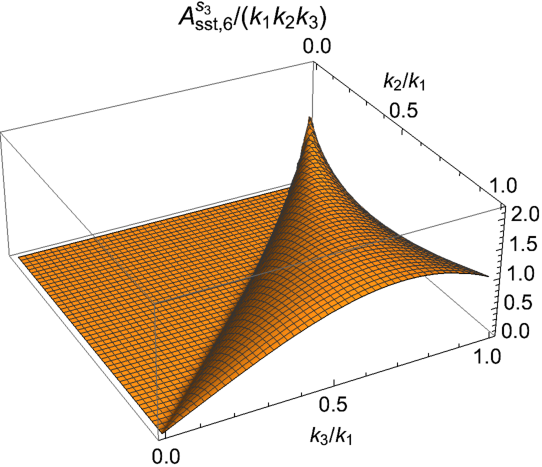}
\captionsetup{justification=raggedright}
\caption{$\mathcal{A}^{s_3}_{sst,6}/(k_1k_2k_3)$ as a function of $k_2/k_1$ and $k_3/k_1$. We took $c_t/c_s=1$ and set $\mathcal{A}/(k_1k_2k_3)$ to $1$ for the equilateral configuration.}\label{Fig: C6}
\end{figure}

\begin{figure} [htb]
     \begin{tabular}{cc}
        \begin{minipage}{0.45\hsize}
            \centering
            \includegraphics[width=7.cm]{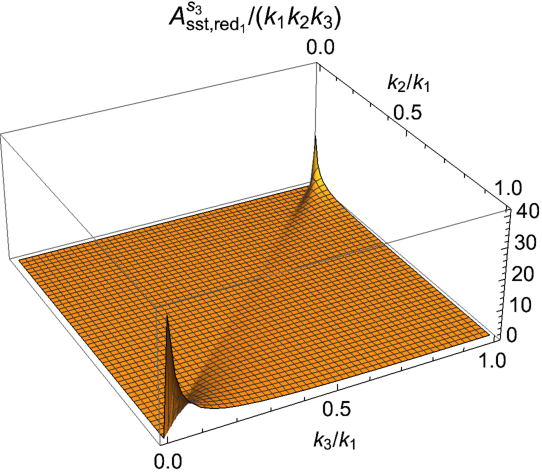}
        \end{minipage} &
        \begin{minipage}{0.45\hsize}
            \centering
            \includegraphics[width=7.cm]{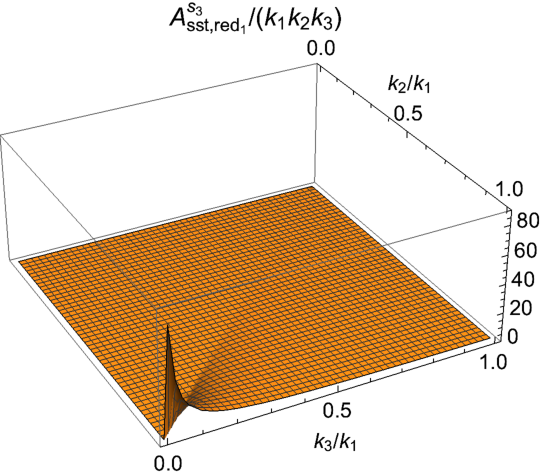}
        \end{minipage} 
    \end{tabular}
\caption{\emph{Left}:$\mathcal{A}^{s_3}_{sst,{\rm red}_1}/(k_1k_2k_3)$ as a function of $k_2/k_1$ and $k_3/k_1$. We took $c_t/c_s=1$ and set $\mathcal{A}/(k_1k_2k_3)$ to $1$ for the equilateral configuration.\emph{Right}: $\mathcal{A}^{s_3}_{sst,{\rm red}_1}/(k_1k_2k_3)$ as a function of $k_2/k_1$ and $k_3/k_1$. We took $c_t/c_s=0.01$ and set $\mathcal{A}/(k_1k_2k_3)$ to $1$ for the equilateral configuration.}\label{Fig: Cred1}
\end{figure}

Then, we investigate a shape of the scalar-scalar-tensor bispectrum. Note that the leading-order bispectrum is determined by the $c_3, c_6$, and $c_{\rm red}$ terms. 
First, Figure~\ref{Fig: C3} shows that the $c_3$ term for $c_t/c_s=1$ does not have any sharp peaks, while that for $c_t/c_s=100$ has a peak at the squeezed limit, $k_2/k_1\ll1$. 
The shape of the $c_3$ term for $c_t/c_s=0.01$ is similar to that for $c_t/c_s=100$ (the right panel of Figure~\ref{Fig: C3}) (but the opposite sign). Here, the sharp peak at the squeezed limit appears for $c_t/c_s\neq1$ in general.\footnote{We have
\begin{align}
\frac{\mathcal{A}^{s_3}_{sst,3}}{k_1k_2k_3}&\propto \frac{k_1^3-(c_t/c_s)^3k_3^3}{k_1k_2^3k_3^3}K(k_1-k_2-k_3)(k_1+k_2-k_3)(k_1-k_2+k_3)\notag\\
&\quad\ -\frac{(k_1^3+k_1^2k_2-k_2^3)+(c_t/c_s)^3k_3^3}{k_1^3k_2k_3^3}K(k_1-k_2-k_3)(k_1+k_2-k_3)(k_1-k_2+k_3).
\end{align}
The sharp peak at $k_2\ll k_1\simeq k_3$ comes from the first line of the above for $c_t/c_s\neq1$. Taking $c_t/c_s=1$, one can find that the first line is suppressed at $k_2\ll k_1\simeq k_3$. As a result, the sharp peak did not appear for $c_t/c_s=1$.} Then, we move to the shape of the $c_6$ term. Figure~\ref{Fig: C6}
 shows that the bispectrum for $c_t/c_s=1$ does not have a similar sharp peak. 
The shapes of the $c_6$ term for $c_t/c_s=100$ and $c_t/c_s=0.01$ are similar to that of the $c_3$ term for $c_t/c_s=100$ (the right panel of Figure~\ref{Fig: C3}).\footnote{Similarly to the $c_3$ term, one can understand the absence of the sharp peak for $c_t/c_s=1$ as follows. We can write $\mathcal{A}^{s_3}_{sst,6}/(k_1k_2k_3)$ as
\begin{align}
\frac{\mathcal{A}^{s_3}_{sst,6}}{k_1k_2k_3}&\propto \frac{k_1^3-(c_t/c_s)^3k_3^3}{k_1^3k_2^3k_3}K(k_1-k_2-k_3)(k_1+k_2-k_3)(k_1-k_2+k_3)\notag\\
&\quad\ +\frac{c_s^3}{c_t^3}\frac{1}{k_1^3k_3}K(k_1-k_2-k_3)(k_1+k_2-k_3)(k_1-k_2+k_3).
\end{align}
The peak at $k_2\ll k_1\simeq k_3$ comes from the first line of the above for $c_t/c_s\neq1$. Taking $c_t/c_s=1$, one can find that the first line is suppressed at $k_2\ll k_1\simeq k_3$. As a result, the sharp peak did not appear for $c_t/c_s=1$.} Last, we study the shapes of the $c_{\rm red}$ terms. Figure~\ref{Fig: Cred1} shows that the $c_{\rm red_1}$ term for $c_t/c_s=1$ has peaks at both $k_2/k_1\ll1$ and $k_3/k_1\ll1$, while that for $c_t/c_s=0.01$ has a peak only at $k_3/k_1\ll1$.\footnote{We can write $\mathcal{A}^{s_3}_{sst,{\rm red_1}}/(k_1k_2k_3)$ as
\begin{align}
\frac{\mathcal{A}^{s_3}_{sst,{\rm red_1}}}{k_1k_2k_3}&\propto\frac{K(k_1+k_2)}{k_1k_2k_3^3}(k_1-k_2-k_3)(k_1-k_2+k_3)(k_1+k_2-k_3)\notag\\
&\quad\ +\frac{c_t^3}{c_s^3}\frac{K(k_1^2+k_2^2)}{k_1^3k_3^3}(k_1-k_2-k_3)(k_1-k_2+k_3)(k_1+k_2-k_3).
\end{align}
The first line has a peak at $k_3/k_1\ll1$, while the second one has a peak at $k_2/k_1\ll1$. Then, $c_t/c_s\ll1$ suppresses the second line, which has resulted in the absence of the sharp peak at $k_3/k_1\ll1$ for $c_t/c_s\ll1$.
} 
The $c_{\rm red_1}$ term for $c_t/c_s=100$ has a peak at $k_2/k_1\ll1$ similarly to the $c_3$ term for $c_t/c_s=100$ (the right panel of Figure~\ref{Fig: C3}).
The $c_{\rm red_2}$ term does not have $c_t/c_s$ dependence, and this term has a peak at $k_2/k_1\ll1$ similarly to the $c_3$ term for $c_t/c_s=100$ (the right panel of Figure~\ref{Fig: C3}). As a result, we find that the scalar-scalar-tensor bispectrum peaks at $k_2/k_1\ll1$ and/or $k_3/k_1\ll1$ except for the $c_3$ and $c_6$ terms for $c_t/c_s=1$.

\begin{figure}\centering
\subfloat[$c_t/c_s=0.01$.]{\label{a}\includegraphics[width=.5\linewidth]{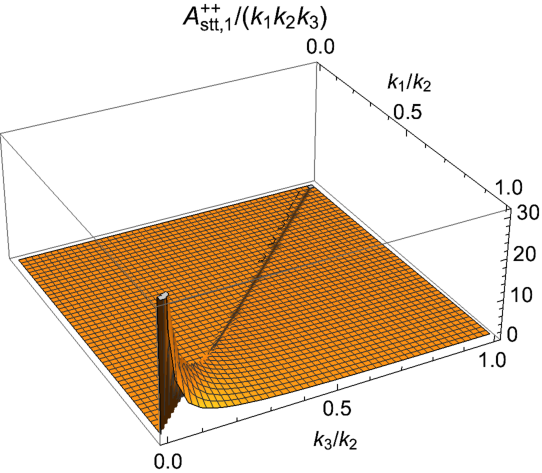}}\hfill
\subfloat[$c_t/c_s=100$.]{\label{b}\includegraphics[width=.5\linewidth]{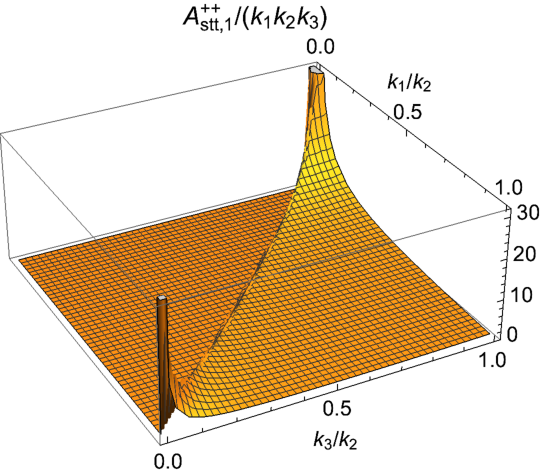}}\par 
\subfloat[$c_t/c_s=1$.]{\label{c}\includegraphics[width=.5\linewidth]{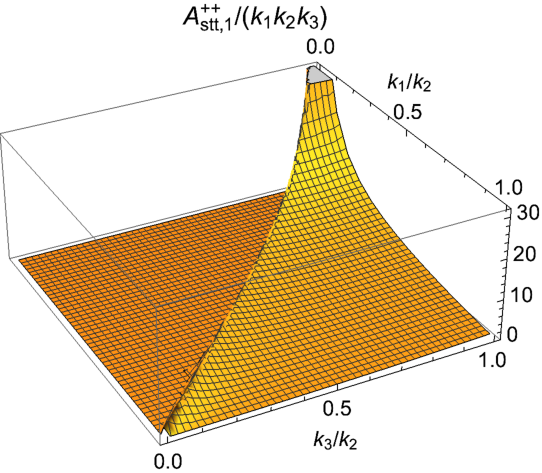}}
\caption{$\mathcal{A}^{++}_{stt,1}/(k_1k_2k_3)$, as a function of $k_2/k_1$ and $k_3/k_1$. We set $\mathcal{A}/(k_1k_2k_3)$ to $1$ for the equilateral configuration. The plots of (a), (b), and (c) correspond to $c_t/c_s=0.01, c_t/c_s=100$, and $c_t/c_s=1$, respectively.}
\label{Fig: B1}
\end{figure}

\begin{figure}[h]
\centering
\includegraphics[width=80mm]{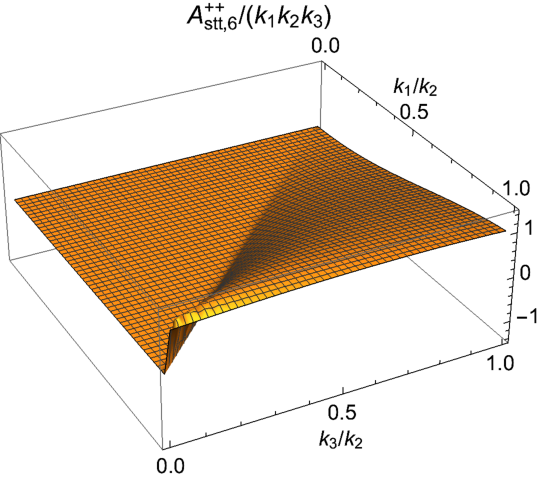}
\captionsetup{justification=raggedright}
\caption{$\mathcal{A}^{++}_{stt,6}/(k_1k_2k_3)$ as a function of $k_1/k_2$ and $k_3/k_2$. We took $c_t/c_s=0.01$ and set $\mathcal{A}/(k_1k_2k_3)$ to $1$ for the equilateral configuration.}\label{Fig: B6}
\end{figure}

We next study a shape of the scalar-tensor-tensor bispectrum in the same-helicity case. The $b_1$ term has a $c_t/c_s$-dependent shape. Figure~\ref{Fig: B1} shows that the bispectra for the three cases, $c_t/c_s=0.01, c_t/c_s=100$, and $c_t/c_s=1$, peak at squeezed limits. Depending on the value of $c_t/c_s$, the position and number of the peaks vary: the bispectra for $c_t/c_s=0.01, c_t/c_s=100$, and $c_t/c_s=1$ peak at $k_1/k_2\ll1$,  $k_1/k_2\ll1$ and $k_3/k_2\ll1$, and $k_3/k_2\ll1$, respectively.\footnote{Here, the $b_1$ term can be written as
\begin{align}
\frac{\mathcal{A}^{++}_{stt,1}}{k_1k_2k_3}&\propto \frac{[k_1^2-(k_2+k_3)^2]^2}{k_1k_3^3}\times\frac{k_1^3-(c_t/c_s)^3(k_2^3+k_3^3)}{k_2^3}.
\end{align}
The first $c_t/c_s$-independent part of the above peaks at both $k_1\ll k_2\simeq k_3$ and $k_3\ll k_1\simeq k_2$. For $c_t/c_s=1$ and $c_t/c_s\ll1$, the second $c_t/c_s$-dependent part of the above is suppressed at $k_3\ll k_1\simeq k_2$ and at $k_1\ll k_2\simeq k_3$, respectively. As a result, one of the peaks did not appear for those two cases. For $c_t/c_s\gg1$, the $c_t/c_s$-dependent part is not suppressed at $k_3\ll k_1\simeq k_2$ or $k_1\ll k_2\simeq k_3$, and hence the resultant bispectrum has two peaks originating from the $c_t/c_s$-independent part.\label{footnote: b1-nu-3/2}} 
The $b_3$ term exhibits a shape and $c_t/c_s$ dependence similar to those of the $b_1$ term (Figure~\ref{Fig: B1}). Similarly to the $b_1$ term for $c_t/c_s=100$ (Figure~\ref{Fig: B1}-(b)), the $b_4$ term peaks at both $k_1/k_2\ll1$ and $k_3/k_2\ll1$ for all three values of $c_t/c_s$. The $b_6$ term, for both $c_t/c_s=100$ and $c_t/c_s=1$, peaks at $k_1/k_2\ll1$ similarly to the $b_1$ term for $c_t/c_s=1$ (Figure~\ref{Fig: B1}-(c)), while that for $c_t/c_s=0.01$ does not peak as shown in Figure~\ref{Fig: B6}.\footnote{Here, the suppression of the $b_6$ term for $c_t/c_s\ll1$ can be understood by using the following expression:
\begin{align}
\frac{\mathcal{A}^{++,b_6}_{stt}}{k_1k_2k_3}&\propto \frac{c_t^3}{c_s^3}\frac{k_2^3+k_3^3}{k_1^3k_2^3k_3^3}K^2(k_1+k_2-k_3)(k_1-k_2+k_3)(k_1-k_2-k_3)^2\notag\\
&\quad\ +\frac{1}{k_2^3k_3^3}K^2(k_1+k_2-k_3)(k_1-k_2+k_3)(k_1-k_2-k_3)^2.
\end{align}
The sharp peak at $k_1\ll k_2\simeq k_3$ comes from the first line. Taking $c_t/c_s\ll1$, one can find that the first line is suppressed in proportion to $(c_t/c_s)^3$. As a result, the sharp peak appeared only for $c_t/c_s\geq1$.}
Then, we move to the bispectra originating from the field redefinitions. Both $b_{\rm red_1}$ and $b_{\rm red_2}$ terms peak at both $k_1/k_2\ll1$ and $k_3/k_2\ll1$ similarly to the $b_1$ term for $c_t/c_s=100$ (Figure~\ref{Fig: B1}-(b)). We thus find that the scalar-tensor-tensor bispectrum in the same-helicity case peaks at $k_1/k_2\ll1$ and/or $k_3/k_2\ll1$, except for the $b_6$ term for $c_t/c_s=0.01$. Here, we emphasize that the scalar-tensor-tensor cross-bispectrum for $\nu=-3/2$ can have the divergent peak(s) at the squeezed limits, which is in sharp contrast to that for $\nu=3/2$. This is a consequence of the violation of the non-Gaussianity consistency relation.

\begin{figure} [htb]
     \begin{tabular}{cc}
        \begin{minipage}{0.45\hsize}
            \centering
            \includegraphics[width=7.cm]{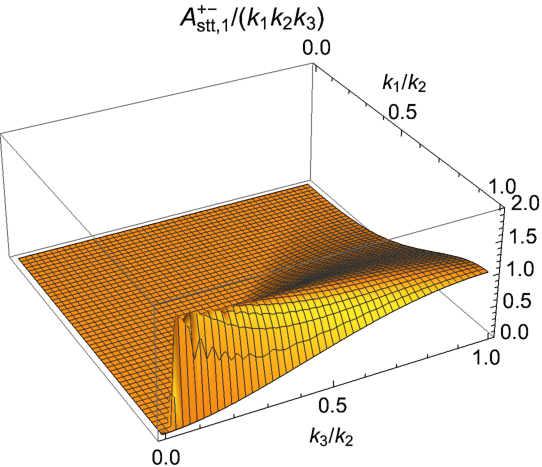}
        \end{minipage} &
        \begin{minipage}{0.45\hsize}
            \centering
            \includegraphics[width=7.cm]{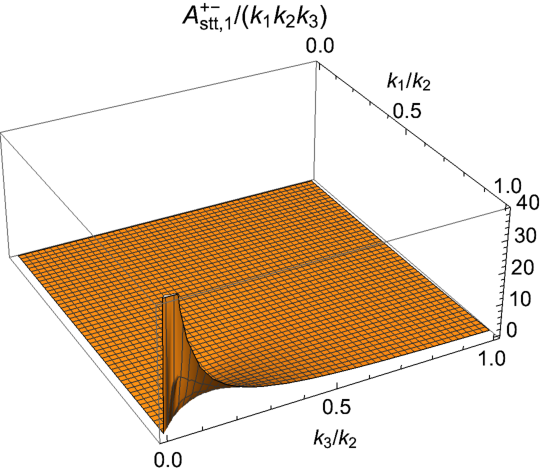}
        \end{minipage} 
    \end{tabular}
\caption{\emph{Left}:$\mathcal{A}^{+-}_{stt,1}/(k_1k_2k_3)$ as a function of $k_1/k_2$ and $k_3/k_2$. We took $c_t/c_s=1$ and set $\mathcal{A}/(k_1k_2k_3)$ to $1$ for the equilateral configuration.\emph{Right}: $\mathcal{A}^{+-}_{stt,1}/(k_1k_2k_3)$ as a function of $k_1/k_2$ and $k_3/k_2$. We took $c_t/c_s=100$ and set $\mathcal{A}/(k_1k_2k_3)$ to $1$ for the equilateral configuration.}\label{Fig: B1pm}
\end{figure}

\begin{figure}[htb]
\centering
\includegraphics[width=80mm]{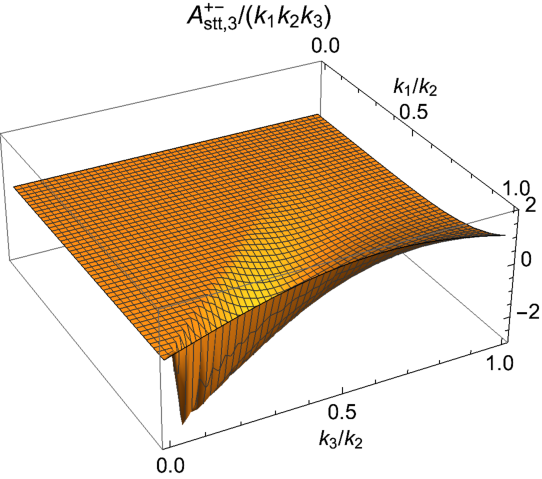}
\captionsetup{justification=raggedright}
\caption{$\mathcal{A}^{+-}_{stt,3}/(k_1k_2k_3)$ as a function of $k_1/k_2$ and $k_3/k_2$. We took $c_t/c_s=1$ and set $\mathcal{A}/(k_1k_2k_3)$ to $1$ for the equilateral configuration.}\label{Fig: B3pm}
\end{figure}

\begin{figure}[htb]
\centering
\includegraphics[width=80mm]{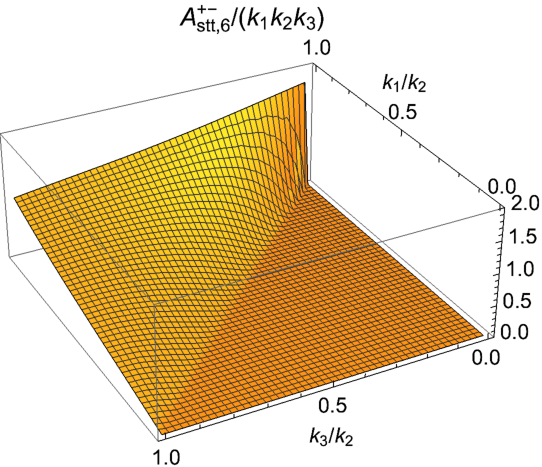}
\captionsetup{justification=raggedright}
\caption{$\mathcal{A}^{+-}_{stt,6}/(k_1k_2k_3)$ as a function of $k_1/k_2$ and $k_3/k_2$. We took $c_t/c_s=1$ and set $\mathcal{A}/(k_1k_2k_3)$ to $1$ for the equilateral configuration.}\label{Fig: B6pm}
\end{figure}

\begin{figure}[htb]
\centering
\includegraphics[width=80mm]{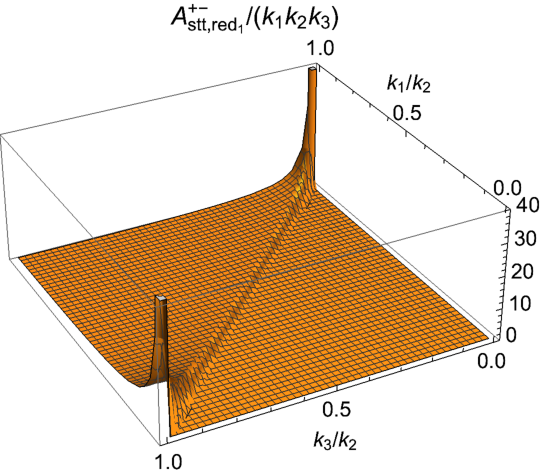}
\captionsetup{justification=raggedright}
\caption{$\mathcal{A}^{+-}_{stt,{\rm red}_1}/(k_1k_2k_3)$ as a function of $k_1/k_2$ and $k_3/k_2$. We took $c_t/c_s=1$ and set $\mathcal{A}/(k_1k_2k_3)$ to $1$ for the equilateral configuration.}\label{Fig: Bred1pm}
\end{figure}

We move on to the mixed-helicity case. Figure~\ref{Fig: B1pm} shows that the $b_1$ term for $c_t/c_s=100$ peaks at $k_3/k_2\ll1$, while that for $c_t/c_s=1$ does not. 
The shape for $c_t/c_s=0.01$ is similar to that for $c_t/c_s=100$. (The difference in the shape of the $b_3$ term in the mixed-helicity case between $c_t/c_s=1$ and $c_t/c_s\neq1$ occurs in a manner similar to that of the $b_1$ term in the same-helicity case, see footnote~\ref{footnote: b1-nu-3/2}.) Figure~\ref{Fig: B3pm} shows that the $b_3$ term for $c_t/c_s=1$ does not have a sharp peak. 
The $b_3$ term, for both $c_t/c_s=100$ and $c_t/c_s=0.01$, peaks at $k_3/k_2\ll1$ similarly to the $b_1$ term for these two cases (the right panel of Figure~\ref{Fig: B1pm}). The $b_4$ term for $c_t/c_s=1, 100, 0.01$ peaks at $k_3/k_2\ll1$ similarly to the $b_1$ term for $c_t/c_s=100$ (Figure~\ref{Fig: B3pm}). Figure~\ref{Fig: B6pm} shows that the $b_6$ term for $c_t/c_s=1$ does not have a sharp peak. 
The shapes for $c_t/c_s=100$ and $c_t/c_s=0.01$ are similar to that for $c_t/c_s=1$. Figure~\ref{Fig: Bred1pm} shows that the $b_{\rm red_1}$ term for $c_t/c_s=1$ peaks at both $k_1/k_2\ll1$ and $k_3/k_2\ll1$.
The shapes for $c_t/c_s=100$ and $c_t/c_s=0.01$ are similar to that for $c_t/c_s=1$. Similarly, the $b_{\rm red_2}$ term, for all three values of $c_t/c_s$, peaks at both $k_1/k_2\ll1$ and $k_3/k_2\ll1$.

\begin{figure}[htb]
\centering
\includegraphics[width=80mm]{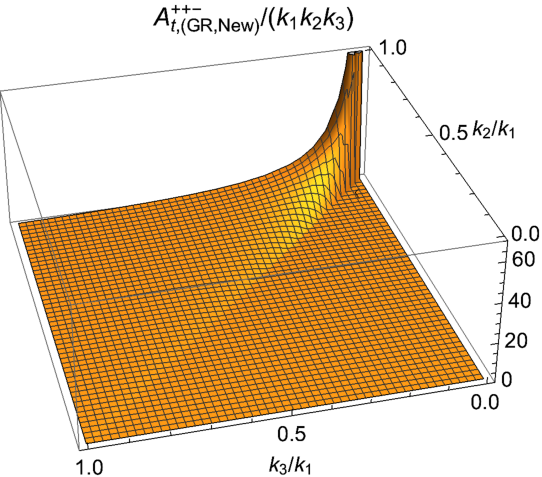}
\captionsetup{justification=raggedright}
\caption{$\mathcal{A}^{++-}_t$ as a function of $k_2/k_1$ and $k_3/k_1$. We set $\mathcal{A}/(k_1k_2k_3)$ to $1$ for the equilateral configuration and took $c_t^2(k_1+k_2-k_3)^2\eta_b^2=10^{-2}$.}\label{Fig: ttt-pm}
\end{figure}

Last, we study a shape of the tensor auto-bispectrum. The shape in the same-helicity case has been studied in Ref.~\cite{Akama:2019qeh} which has clarified that both GR and New terms peak at the squeezed limits, $k_2/k_1\ll1$ and $k_3/k_1\ll1$. Thus, we here show only the shape in the mixed-helicity case. Taking $c_t^2(k_1+k_2-k_3)^2\eta_b^2=10^{-2}\ll1$, one can find that the bispectra originating from both GR- and new-type cubic operators have the same shape. Figure~\ref{Fig: ttt-pm} shows that the tensor auto-bispectrum in the mixed-helicity case peaks only at $k_3/k_1\ll1$, which is in contrast to the same-helicity case where the bispectrum peaks at both $k_2/k_1\ll1$ and $k_3/k_1\ll1$.\footnote{In the same-helicity case, we have
\begin{align}
\frac{\mathcal{A}^{+++}_{t}}{k_1k_2k_3}&\propto\frac{1}{k_2^3k_3^3}(k_1-k_2-k_3)(k_1-k_2+k_3)(k_1+k_2-k_3)K^3,
\end{align}
which is enhanced at both $k_2/k_1\ll1$ and $k_3/k_1\ll1$. In the mixed-helicity case, we have
\begin{align}
\frac{\mathcal{A}^{++-}_{t}}{k_1k_2k_3}&\propto\frac{1}{k_2^3k_3^3}(k_1-k_2-k_3)(k_1-k_2+k_3)(k_1+k_2-k_3)^3K,
\end{align}
which is enhanced only at $k_3/k_1\ll1$ due to a higher power of $(k_1+k_2-k_3)$ compared to the same-helicity case.
}

We have thus found that the bispectra for $\nu=-3/2$ can peak at squeezed limits over a wider range of parameters than those for $\nu=3/2$.
A summary of the shapes for $\nu=-3/2$ is given in Table~\ref{table: auto-nu-3/2} for the auto-bispectra and Table~\ref{table: cross-nu-3/2} for the cross-bispectra. 

\begin{table}[htb]
  \centering
  \begin{tabular}{|c|c|c|c|} \hline
    \diagbox{\quad\quad\quad\quad\quad\ }{\quad\quad\quad\quad\quad\ } & \\ \hline
    $\Lambda_{(1,2,5,6,{\rm red_1},{\rm red_2})}$ & squeezed \\ \hline
    ${\rm GR}$\ ($s_1=s_2=s_3=+1$)& squeezed\\ \hline
    ${\rm New}$\ ($s_1=s_2=s_3=+1$)& squeezed\\ \hline
    ${\rm GR}$\ ($s_1=s_2=-s_3=+1$)& squeezed\\ \hline
    ${\rm New}$\ ($s_1=s_2=-s_3=+1$)& squeezed\\ \hline
  \end{tabular}
  \caption{Summary of the shapes of the auto-bispectra for $\nu=-3/2$. }
  \label{table: auto-nu-3/2}
\end{table}

\begin{table}[htb]
  \centering
  \begin{tabular}{|c|c|c|c|} \hline
    \diagbox{\quad\quad\quad\quad\quad }{\quad\quad\quad\quad\quad } &  $c_t/c_s=1$ &  $c_t/c_s=0.01$ &  $c_t/c_s=100$ \\ \hline
    $c_{(3,6)}$ & other & squeezed & squeezed \\ \hline
    $c_{({\rm red_1},{\rm red_2})}$ & squeezed & squeezed & squeezed \\ \hline
    $b_{(1,3,4,{\rm red_1},{\rm red_2})}$\ ($s_2=s_3=+1$) & squeezed & squeezed & squeezed \\ \hline
    $b_{6}$\ ($s_2=s_3=+1$) & squeezed & other & squeezed \\ \hline
    $b_{(1,3)}$\ ($s_2=-s_3=+1$) & other & squeezed & squeezed \\ \hline
    $b_{(4,{\rm red_1},{\rm red_2})}$\ ($s_2=-s_3=+1$) & squeezed & squeezed & squeezed \\ \hline
    $b_{6}$\ ($s_2=-s_3=+1$) & other & other & other \\ \hline
  \end{tabular}
  \caption{Summary of the shapes of the cross-bispectra for $\nu=-3/2$. The shapes that appear to be away from the squeezed or equilateral ones are denoted by ``other''.}
  \label{table: cross-nu-3/2}
\end{table}

\section{Theoretical and observational constraints}\label{Sec: const}
\subsection{Perturbativity}\label{Sec: pertur}
In this section, we study perturbativity at and after sound-horizon cross scales by evaluating ratios of cubic Lagrangians to quadratic ones. The ratios of the perturbed Lagrangians (actions) represent non-linearities of perturbations, and one can clarify a perturbative regime from those~\cite{Leblond:2008gg,Baumann:2011dt}. The ratio (relevant to the curvature perturbation) has also been used to discuss perturbativity from the viewpoint of loop corrections to tree-level correlation functions when evaluated at horizon crossing~\cite{Leblond:2008gg,Baumann:2011dt}. In this section, we use the following inequalities as criteria for the scale-invariant scalar and tensor perturbations to be weakly coupled during the contracting phase:
\begin{align}
\frac{\mathcal{L}_s^{(3)}}{\mathcal{L}_s^{(2)}},\ \frac{\mathcal{L}_{sst}^{(3)}}{\mathcal{L}_s^{(2)}},\ \frac{\mathcal{L}_{stt}^{(3)}}{\mathcal{L}_t^{(2)}},\ \frac{\mathcal{L}_t^{(3)}}{\mathcal{L}_t^{(2)}}<1.\label{eq: four-ratios}
\end{align}
As we will explicitly show below, the above four ratios for both $\nu=3/2$ and $\nu=-3/2$ are roughly of $\mathcal{O}(f^s_{\rm NL}|\zeta|), \mathcal{O}(f^{sst}_{\rm NL}|h_{ij}|), \mathcal{O}(f^{stt}_{\rm NL}|\zeta|)$, and $\mathcal{O}(f^{t}_{\rm NL}| h_{ij}|)$, respectively, where $f_{\rm NL}$ is a schematic form of each non-linearity parameter collecting all of the existing terms in each bispectrum as that evaluated at the equilateral limit. Note that $\mathcal{O}(f^s_{\rm NL}|\zeta|), \mathcal{O}(f^{sst}_{\rm NL}|h_{ij}|)$ are the nonlinear corrections to the linear curvature perturbation and $\mathcal{O}(f^{stt}_{\rm NL}|\zeta|), \mathcal{O}(f^{t}_{\rm NL}| h_{ij}|)$ are those to the linear tensor ones.

So far, we have discussed general predictions from the power-law contracting universe. From this section, we put several assumptions to simplify the subsequent discussions on perturbativity and a parameter search for viable models. First, the times of (sound-)horizon crossing for the curvature and tensor perturbations, $-c_sk\eta=1, -c_tk\eta=1$ are different for $c_s\neq c_t$. In this section, we assume $\mathcal{O}(c_s)=\mathcal{O}(c_t)$ and ignore this subtlety. Furthermore, we assume $(1-n)/n=\mathcal{O}(1)$ for simplicity, under which we have
\begin{align}
\zeta'&\sim c_s^2k^2\eta^2\frac{\zeta}{\eta}, h_{ij}'\sim c_t^2k^2\eta^2\frac{h_{ij}}{\eta}\ \ {\rm for}\ \ \nu=\frac{3}{2},\\
\zeta'&\sim \frac{\zeta}{\eta}, h_{ij}'\sim \frac{h_{ij}}{\eta}\ \ {\rm for}\ \ \nu=-\frac{3}{2},
\end{align}
at and after sound-horizon crossing. The above indicates that we can replace $\partial_t$ to $H$ or $\partial_\eta$ to $\eta^{-1}$ when we evaluate the ratios of the perturbed Lagrangians at sound-horizon crossing for the two cases of $\nu=\pm3/2$. For $\nu=3/2$, we need to take into account the factors $c_s^2k^2\eta^2, c_t^2k^2\eta^2$ after sound-horizon crossing, which comes from the fact that the superhorizon modes are conserved. Also, we replace $c_s\eta\partial_k\zeta$ and $c_t\eta\partial_k h_{ij}$
as quantities of $\mathcal{O}(-c_sk\eta|\zeta|)$ and $\mathcal{O}(-c_tk\eta|h_{ij}|)$, respectively. Schematically, the resultant ratios of the perturbed Lagrangians can be written as
\begin{align}
\frac{\mathcal{L}^{(3)}}{\mathcal{L}^{(2)}}\sim({\rm const.})\times\mathcal{O}((-k\eta)^\bullet)\times(\zeta\ {\rm or}\ h_{ij}),
\end{align}
where $\bullet\geq0$. For $\nu=3/2$, the perturbations become constant on the superhorizon scales, and it is enough to evaluate the above ratio at sound-horizon crossing since $\mathcal{O}((-k\eta)^\bullet)$ suppresses the ratio, and the amplitudes of the perturbations are conserved after the sound-horizon cross scales. 
Since the total-time-derivative interactions at sound-horizon crossing do not contribute to the three-point functions at the end of the contracting phase, we ignore those for $\nu=3/2$. On the other hand, for $\nu=-3/2$, the perturbations grow after sound-horizon crossing, and hence the above ratio takes its maximum value at the sound-horizon cross scales for $\bullet>0$ and on the superhorizon scales for $\bullet=0$. Given that $|\zeta(\eta_*)|\sim10^{-5}$ and $|h_{ij}(\eta_*)|\sim 10^{-6}$ for $r=0.01$ and the perturbations grow at the late time as $\zeta(\eta)=\zeta(\eta_c)(\eta_c/\eta)^3$ and $h_{ij}(\eta)=h_{ij}(\eta_c)(\eta_c/\eta)^3$ where $\eta_c$ is the conformal time at sound-horizon crossing, the amplitudes of the perturbations at $\eta_c$ are much smaller than $10^{-5}$. More explicitly, for a given $k$, the perturbations at the sound-horizon cross scales are suppressed in proportion to $(-k\eta_*)^3$ compared to those on the superhorizon scales. The ratios at horizon crossing would not exceed unity unless the cubic interactions are significantly enhanced. Since the coefficients of $\zeta$ or $h_{ij}$ in the four ratios are constant, if one enhances those coefficients, such a case would be excluded by the current CMB experiments. In light of this, we will evaluate the four ratios at the end of the contracting phase for $\nu=-3/2$. 
In particular, the total-time-derivative interactions at the end of the contracting phase contribute to the three-point functions, and hence we roughly take those into account simply by replacing $\partial_t[\cdots]$ to $H[\cdots]$.

 First, the four ratios in Eq.~(\ref{eq: four-ratios}) for $\nu=3/2$ at sound-horizon crossing read
\begin{align}
\frac{\mathcal{L}^{(3)}_s}{\mathcal{L}^{(2)}_s}&\sim \Lambda_1|\zeta|,\ \Lambda_2|\zeta|,\ \frac{\Lambda_3}{c_s^2}|\zeta|,\ \frac{\Lambda_4}{c_s^2}|\zeta|,\ \Lambda_5|\zeta|,\ \Lambda_6|\zeta|,\ \frac{\Lambda_7}{c_s^4}|\zeta|,\ \frac{\Lambda_8}{c_s^2}|\zeta|,\\
\frac{\mathcal{L}_{sst}^{(3)}}{\mathcal{L}_s^{(2)}}&\sim \frac{c_1}{\mathcal{F}_S}|h_{ij}|,\ \frac{Hc_2}{\mathcal{F}_S}|h_{ij}|,\ \frac{c_3c_s^2}{\mathcal{F}_S}|h_{ij}|,\ \frac{Hc_4}{\mathcal{F}_S}|h_{ij}|,\ \frac{H^2c_5}{c_s^2\mathcal{F}_S}|h_{ij}|,\ \frac{c_6c_s^2}{\mathcal{F}_S}|h_{ij}|,\\
\frac{\mathcal{L}_{stt}^{(3)}}{\mathcal{L}_t^{(2)}}&\sim \frac{b_1c_t^2}{\mathcal{F}_T}|\zeta|,\ \frac{b_2}{\mathcal{F}_T}|\zeta|,\ \frac{b_3c_t^2}{\mathcal{F}_T}|\zeta|,\ \frac{Hb_4c_t^2}{\mathcal{F}_T}|\zeta|,\ \frac{H^2b_5}{\mathcal{F}_T}|\zeta|,\ \frac{Hb_6}{\mathcal{F}_T}|\zeta|,\ \frac{H^2b_7}{\mathcal{F}_T}|\zeta|,\\
\frac{\mathcal{L}_{t}^{(3)}}{\mathcal{L}_t^{(2)}}&\sim |h_{ij}|,\ \frac{\mu H}{\mathcal{G}_T}|h_{ij}|,
\end{align}
where we assumed $\partial_i\sim1/(-c_s\eta)\sim1/(-c_t\eta)$. Note that one can verify from the above and Eqs.~(\ref{eq: 3pt-sss-const}),~(\ref{eq: 3pt-sst-const}),~(\ref{eq: 3pt-stt-const}),~(\ref{eq: ttt-bispectrum-nu3/2}), and Eq.~(\ref{eq: def-fnl}) that we have $\mathcal{L}_s^{(3)}/\mathcal{L}_s^{(2)}\sim f^s_{\rm NL}|\zeta|$, $\mathcal{L}_{sst}^{(3)}/\mathcal{L}_s^{(2)}\sim f^{sst}_{\rm NL}|h_{ij}|$, $\mathcal{L}_{stt}^{(3)}/\mathcal{L}_t^{(2)}\sim f^{stt}_{\rm NL}|\zeta|$, and $\mathcal{L}_t^{(3)}/\mathcal{L}_t^{(2)}\sim f^t_{\rm NL}|h_{ij}|$.

Next, the four ratios in Eq.~(\ref{eq: four-ratios}) for $\nu=-3/2$ at the end of the contracting phase read
\begin{align}
\frac{\mathcal{L}^{(3)}_s}{\mathcal{L}^{(2)}_s}&\sim \Lambda_1|\zeta|,\ \Lambda_2|\zeta|,\ \frac{\Lambda_3}{c_s^2}(c_sk\eta_*)^2|\zeta|,\ \frac{\Lambda_4}{c_s^2}(c_sk\eta_*)^2|\zeta|,\ \Lambda_5|\zeta|,\ \Lambda_6|\zeta|,\ \frac{\Lambda_7}{c_s^4}(c_sk\eta_*)^4|\zeta|,\ \frac{\Lambda_8}{c_s^2}(c_sk\eta_*)^2|\zeta|,\\
\frac{\mathcal{L}_{sst}^{(3)}}{\mathcal{L}_s^{(2)}}&\sim \frac{c_1}{\mathcal{F}_S}(c_sk\eta_*)^2|h_{ij}|,\ \frac{Hc_2}{\mathcal{F}_S}(c_sk\eta_*)^2|h_{ij}|,\ \frac{c_3c_s^2}{\mathcal{F}_S}|h_{ij}|,\ \frac{Hc_4}{\mathcal{F}_S}(c_sk\eta_*)^2|h_{ij}|,\notag\\
&\quad\ \frac{H^2c_5}{c_s^2\mathcal{F}_S}(c_sk\eta_*)^2(c_tk\eta_*)^2|h_{ij}|,\ \frac{c_6c_s^2}{\mathcal{F}_S}|h_{ij}|,\\
\frac{\mathcal{L}_{stt}^{(3)}}{\mathcal{L}_t^{(2)}}&\sim \frac{b_1c_t^2}{\mathcal{F}_T}|\zeta|,\ \frac{b_2}{\mathcal{F}_T}(c_tk\eta_*)^2|\zeta|,\ \frac{b_3c_t^2}{\mathcal{F}_T}|\zeta|,\ \frac{Hb_4c_t^2}{\mathcal{F}_T}|\zeta|,\ \frac{H^2b_5}{\mathcal{F}_T}(c_tk\eta_*)^2|\zeta|,\ \frac{Hb_6}{\mathcal{F}_T}|\zeta|,\ \frac{H^2b_7}{\mathcal{F}_T}(c_tk\eta_*)^2|\zeta|,\\
\frac{\mathcal{L}_{t}^{(3)}}{\mathcal{L}_t^{(2)}}&\sim (c_tk\eta_*)^2|h_{ij}|,\  \frac{\mu H}{\mathcal{G}_T}|h_{ij}|,
\end{align}
where we replaced the spatial derivative as $c_s\eta_*\partial_i\to\mathcal{O}(c_sk\eta_*)$.
Here, the terms proportional to a power of $(-c_sk\eta_*)\ll1$ are significantly suppressed, and hence we ignore those. Note that the suppressed terms come from the cubic interactions involving the spatial derivatives, that is, those are the ones that have not contributed to the leading-order bispectra in the previous section. Ignoring the similar suppressed terms, the magnitudes of the total-time-derivative cubic interactions are found to be
\begin{align}
\frac{\mathcal{L}^{(3)}}{\mathcal{L}^{(2)}}&\supset \Lambda_{\rm red_1}|\zeta|,\ \Lambda_{\rm red_2}|\zeta|,\ \frac{H^2c_{\rm red_1}c_s^2}{\mathcal{F}_S}|h_{ij}|,\ \frac{H^2c_{\rm red_2}c_s^2}{\mathcal{F}_S}|h_{ij}|,\ \frac{H^2b_{\rm red_1}c_t^2}{\mathcal{F}_T}|\zeta|,\ \frac{H^2b_{\rm red_2}c_t^2}{\mathcal{F}_T}|\zeta|.
\end{align}
Similarly to the case of $\nu=3/2$, one can verify that $\mathcal{L}_s^{(3)}/\mathcal{L}_s^{(2)}\sim f^s_{\rm NL}|\zeta|$, $\mathcal{L}_{sst}^{(3)}/\mathcal{L}_s^{(2)}\sim f^{sst}_{\rm NL}|h_{ij}|$, $\mathcal{L}_{stt}^{(3)}/\mathcal{L}_t^{(2)}\sim f^{stt}_{\rm NL}|\zeta|$, and $\mathcal{L}_t^{(3)}/\mathcal{L}_t^{(2)}\sim f^t_{\rm NL}|h_{ij}|$. 

Assuming $r=0.01$ for both $\nu=3/2$ and $\nu=-3/2$, the perturbativity conditions $\mathcal{L}^{(3)}/\mathcal{L}^{(2)}<1$ read $f^s_{\rm NL}<10^5, f^{sst}_{\rm NL}<10^6, f^{stt}_{\rm NL}<10^5$, and $f^t_{\rm NL}<10^6$. In the following subsection, we compare these constraints to those obtained from the CMB experiments.

\subsection{CMB experiments}\label{Sec: CMB}

Of the non-Gaussianities, the scalar non-Gaussianity has been the most severely constrained. The constraints on the non-linearity parameter of that non-Gaussianity are $f^{s,{\rm sq}}_{\rm NL}=0.8\pm5.0$ for the squeezed limit and $f^{s,{\rm eq}}_{\rm NL}=-4\pm43$ for the equilateral limit~\cite{Planck:2019kim}. These constraints are more stringent than that determined by the perturbativity condition. As we have discussed in Sec.~\ref{Sec: bispectra}, $f^{s,{\rm sq}}_{\rm NL}$ for $\nu=3/2$ is suppressed. Thus, when building models, it is necessary to explore models realizing a small $f^s_{\rm NL}$ at non-squeezed limits for $\nu=3/2$ and that at all of the limits considered for $\nu=-3/2$. The difficulty of constructing such successful models for $\nu=-3/2$ has been reported in Refs.~\cite{Quintin:2015rta,Li:2016xjb} within the Horndeski theory, but that difficulty is indeed model-dependent in beyond k-essence theories~\cite{Akama:2019qeh}. On the other hand, the authors of Ref.~\cite{Ageeva:2024knc} found a concrete example for $\nu\simeq3/2$ predicting enhanced non-Gaussianities. To what extent this enhancement happens in a wider class can be clarified by studying the scalar non-Gaussianity originating from the fluctuation whose power spectrum is slightly red-tilted, which is beyond the scope of our paper. Our purpose is to discuss the generic properties of non-Gaussianities, and hence we do not discuss explicit models in detail, but at least in the exact scale-invariant case, it is true that the present framework has sufficient functional degrees of freedom to construct such successful models for both $\nu=3/2$ and $\nu=-3/2$. Regarding this point, see Appendix~\ref{app: example} and Ref.~\cite{Akama:2019qeh}. 

In addition to the scalar non-Gaussianity, the current CMB experiments have also put constraints on the tensor non-Gaussianities through the following parameters~\cite{Planck:2015zrl,Shiraishi:2017yrq,Planck:2019kim,Shiraishi:2019yux,Philcox:2024wqx}:
\begin{align}
F^{{\rm sq}}_{{\rm NL},sst}&:=\lim_{\substack{k_3\to 0\\ k_1\to k_2}}\frac{\mathcal{B}^{+}_{sst}}{S^{\rm loc}},\ F^{{\rm sq}}_{{\rm NL},t}:=\lim_{\substack{k_3\to 0\\ k_1\to k_2}}\frac{\mathcal{B}^{+++}_{t}}{S^{\rm loc}},\ F^{{\rm eq}}_{{\rm NL},t}:=\lim_{k_i\to k}\frac{\mathcal{B}^{+++}_{t}}{S^{\rm eq}},
\end{align}
where
\begin{align}
S^{\rm loc}&:=\frac{6}{5}(2\pi^2\mathcal{P}_\zeta)^2\frac{\sum_i k_i^3}{k_1^3k_2^3k_3^3},\\
S^{\rm eq}&:=\frac{18}{5}(2\pi^2\mathcal{P}_\zeta)^2\biggl[-\biggl(\frac{1}{k_1^3k_2^3}+{2\ {\rm perm.}}\biggr)-\frac{2}{k_1^2k_2^2k_3^2}+\biggl(\frac{1}{k_1k_2^2k_3^2}+{5\ \rm perm.}\biggr)\biggr].
\end{align}
Observational constraints on the above parameters that we adopt are given by $F^{{\rm sq}}_{{\rm NL},sst}=-2\pm12, F^{{\rm sq}}_{{\rm NL},t}=90\pm80$, and $F^{{\rm eq}}_{{\rm NL},t}=100\pm300$ by Planck data~\cite{Philcox:2024wqx}.
In our notation, those constrain $r f^{sst}_{\rm NL}$ (at the squeezed limit) and $r^2 f^t_{\rm NL}$ (at the squeezed and equilateral limits) as $r f^{sst}_{\rm NL}\leq\mathcal{O}(10)$ and $r^2 f^t_{\rm NL}\leq\mathcal{O}(10^2)$.
For $r=0.01$, the perturbativity conditions yield $r f^{sst}_{\rm NL}<\mathcal{O}(10^4)$ and $r^2 f^t_{\rm NL}<\mathcal{O}(10^2)$. The former gives the condition more lenient than the observational constraint, while the latter gives a similar constraint as the observational ones. The observational constraints on the tensor auto-bispectrum would thus be satisfied in the perturbative regime. On the other hand, the scalar-scalar-tensor cross-bispectrum is still severely constrained by the CMB observations even in the perturbative regime. Here, as we have discussed in Sec.~\ref{Sec: bispectra}, the scalar-scalar-tensor cross-bispectrum at the squeezed limit for $\nu=3/2$ cannot be enhanced (see Eq.~(\ref{eq: fnl-consistency})). Therefore, only a certain class for $\nu=-3/2$ such that $f^{sst}_{\rm NL}|_{k_3\to0}\geq\mathcal{O}(10^3)$ is constrained by the current CMB experiments.

\subsection{Viable parameter space}\label{Sec: model-space}
So far, we have evaluated the theoretical and observational constraints. In this subsection, we evaluate the magnitudes of the coefficients of the cubic interaction terms by order estimation and derive constraints on parameters from both constraints.

Under the assumptions imposed above (i.e., $\mathcal{O}(c_s^2)=\mathcal{O}(c_t^2)$ and $(1-n)/n=\mathcal{O}(1)$), we first estimate the magnitudes of the coefficients of the scalar-scalar-scalar cubic interactions. For the non-linearity parameter at the squeezed limit, an amplitude of up to $\mathcal{O}(1)$ is allowed, while for the other non-linearity parameters, e.g., the equilateral one, amplitudes of $\mathcal{O}(10)$ are still allowed in light of the constraints from the CMB experiments. However, in this section, we restrict our analysis to the case in which any momentum triangle limits yield $f^s_{\rm NL}$ of $\mathcal{O}(1)$. In this case, it is conservative to explore a parameter space in which each of $\Lambda_i$ is at most $\mathcal{O}(1)$. To do so, let us introduce the following dimensionless parameters:
\begin{align}
\sigma_1&:=\frac{H\mathcal{G}_T}{\Theta}, \sigma_2:=\frac{\Xi}{\Theta}, \sigma_3:=\frac{\mu H}{\mathcal{G}_T},\notag\\
\sigma_4&:=\frac{\Gamma}{\mathcal{G}_T}, \sigma_5:=\frac{\Sigma-X\Sigma_X}{\Sigma}. \label{eq: five-para}
\end{align}
Note that the above five parameters are constant for both $\nu=3/2$ and $\nu=-3/2$. Here, $\sigma_1$ and $\sigma_2$ are written by the $G_{(3,4,5)}$ terms, $\sigma_3$ and $\sigma_4$ are written by the $G_{(4,5)}$ terms, and $\sigma_5$ is written by the $G_{(2,3,4,5)}$ terms. In general, each of them takes a different functional form. For instance, in a cubic Galileon theory, we have $\sigma_1=\mpl^2 H/(-\dot\phi X G_{3X}+\mpl^2 H)$ and $\sigma_3=6(2\dot\phi X G_{3X}+\dot\phi X^2G_{3XX}-\mpl^2 H)/(-\dot\phi X G_{3X}+\mpl^2 H)$, both of which can take different values due to the functional degree of freedom of $G_3$. In a more general theory, the functional forms are more complex. We thus proceed with our analysis by assuming that the magnitudes of all five parameters can be controlled independently. Then, we rewrite the coefficients of the scalar-scalar-scalar cubic interactions as
\begin{align}
\Lambda_1&=\frac{\mathcal{G}_T}{\mathcal{G}_S}\left(3\sigma_1+\sigma_1\sigma_2+2\sigma_3+3\sigma_1\sigma_4-2\sigma_1\sigma_5-\frac{1}{3}\sigma_1^2\sigma_2\right)\notag\\
&\quad\ +\sigma_1\left(\frac{1}{c_s^2}-1\right)-\frac{1}{3}\sigma_1\sigma_2-2\sigma_1\sigma_4-2\sigma_3+\frac{2}{3}\sigma_1\sigma_5,\\
\Lambda_2&=3-\frac{\sigma_1}{c_s^2}\frac{1+2\alpha+3n}{n},\\
\Lambda_3&=\frac{\mathcal{G}_T}{\mathcal{G}_S}\left(c_t^2-\sigma_1\frac{3+2\alpha+n}{n}\right)+\sigma_1\frac{3+2\alpha+n}{n},\\
\Lambda_4&=\frac{\mathcal{G}_T}{\mathcal{G}_S}\sigma_1\left(\frac{1}{3}\sigma_1\sigma_2+6\sigma_3+3\sigma_1\sigma_4-\sigma_1\right),\\
\Lambda_5&=-\frac{\mathcal{G}_S}{2\mathcal{G}_T}\left[1+(\sigma_1\sigma_4+2\sigma_3)\frac{1+2\alpha+3n}{n}\right],\\
\Lambda_6&=\frac{\mathcal{G}_S}{4\mathcal{G}_T}\left[3-(\sigma_1\sigma_4+2\sigma_3)\frac{1+2\alpha+3n}{n}\right],\\
\Lambda_7&=\frac{\mathcal{G}_T}{6\mathcal{G}_S}\sigma_1^2\left[1+(\sigma_1\sigma_4+6\sigma_3)\frac{5+2\alpha-n}{n}\right]-\frac{c_s^2}{2}\sigma_1(\sigma_1\sigma_4+4\sigma_3),\\
\Lambda_8&=\sigma_1\left[-1+(\sigma_1\sigma_4+4\sigma_3)\frac{n-1}{n}\right]+\frac{\mathcal{G}_S}{\mathcal{G}_T}c_s^2(2\sigma_3+\sigma_1\sigma_4),\\
\Lambda_{\rm red_1}&=(\sigma_1\sigma_4+2\sigma_3)\frac{\mathcal{G}_S}{\mathcal{G}_T},\\
\Lambda_{\rm red_2}&=\frac{\sigma_1}{c_s^2}.
\end{align}
Now we have assumed $c_s\sim c_t$, which means that $\mathcal{F}_S/\mathcal{F}_T=\mathcal{O}(r/10)$. Some of the above coefficients include the terms proportional to $1/c_s^2$ or $\mathcal{F}_T/\mathcal{F}_S=\mathcal{O}(10/r)$. In light of the current constraint on $r$, $r\leq\mathcal{O}(10^{-2})$, we have $\mathcal{F}_T/\mathcal{F}_S\gg1$. To realize a small scalar non-Gaussianity, it is ideal to reduce the number of enhancement factors, and hence we assume $c_s^2=\mathcal{O}(1)$, which yields $c_t^2=\mathcal{O}(1)$ coming from $c_s\sim c_t$. 
At least when the coefficient of the $\mathcal{G}_T/\mathcal{G}_S$ term in each of $\Lambda_1, \Lambda_3, \Lambda_4$, and $\Lambda_7$ is $\mathcal{O}(\mathcal{G}_S/\mathcal{G}_T)\ll1$, any momentum triangle limits yield $f_{\rm NL}$ of $\mathcal{O}(1)$. For $\nu=3/2$ where the $\Lambda_{\rm red_1}$ and $\Lambda_{\rm red_2}$ terms are absent, we can achieve this by choosing four of the parameters in Eq.~(\ref{eq: five-para}) appropriately, while for $\nu=-3/2$ where the $\Lambda_3, \Lambda_4, \Lambda_7$, and $\Lambda_8$ terms are absent, we can achieve by choosing one of those parameters appropriately. 
To make all of the coefficients $\mathcal{O}(1)$, it is reasonable to require that $\sigma_i$ are $\mathcal{O}(1)$. In the following, we estimate the magnitude of each term in the other non-linearity parameters under the conditions, $\sigma_i=\mathcal{O}(1)$.

The terms appearing in the non-linearity parameters of the other bispectra read
\begin{align}
\frac{Hc_2}{\mathcal{F}_S}&=-\frac{\sigma_1}{4}\frac{\mathcal{F}_T}{\mathcal{F}_S}\biggl[\sigma_4+\frac{1}{c_t^2}\biggl(2-\sigma_1\sigma_4\frac{1+2\alpha+3n}{n}+8\frac{1-n}{n}\sigma_3\biggr)\biggr]+\mathcal{O}(1),\\
\frac{H^2c_5}{c_s^2\mathcal{F}_S}&=\frac{\sigma_1}{4c_s^2}\frac{\mathcal{F}_T}{\mathcal{F}_S}\biggl(\sigma_1\frac{1-\sigma_4 c_t^2}{c_t^2}+\frac{4\sigma_1\sigma_3}{c_t^2}\frac{n-1}{n}-6\sigma_3\biggr)+\mathcal{O}(1),
\end{align}
and some of the other terms, $c_1/\mathcal{F}_S$, $c_3c_s^2/\mathcal{F}_S$, $Hc_4/\mathcal{F}_S$, $H^2c_s^2c_{\rm red_1}/\mathcal{F}_S$, $b_1c_t^2/\mathcal{F}_T$, $Hb_4c_t^2/\mathcal{F}_T$, $H^2b_5/\mathcal{F}_T$, $H^2b_7/\mathcal{F}_T$, $H^2b_{\rm red_1}c_t^2/\mathcal{F}_T$, $H^2b_{\rm red_2}c_t^2/\mathcal{F}_T$, are $\mathcal{O}(1)$, and the remaining terms, $c_6c_s^2/\mathcal{F}_S$, $H^2c_{\rm red_2}c_s^2/\mathcal{F}_S$, $b_2/\mathcal{F}_T$, $b_3c_t^2/\mathcal{F}_T$, $Hb_6/\mathcal{F}_T$, are $\mathcal{O}(r/{10})$. Here, the coefficients of the $\mathcal{G}_T/\mathcal{G}_S$ terms in the scalar auto-bispectrum are different from those of the $\mathcal{F}_T/\mathcal{F}_S$ terms in the scalar-scalar-tensor bispectrum, which implies that there is a possibility that $f^{sst}_{\rm NL}$ can increase up to $\mathcal{O}(10/r)$, while keeping $f^s_{\rm NL}$ $\mathcal{O}(1)$ even in the present limited parameter space. More explicitly, for $\nu=3/2$ where the $c_{\rm red}$ and $b_{\rm red}$ terms are absent, we find $f^{sst}_{\rm NL}\leq\mathcal{O}(10/r)\leq\mathcal{O}(10^3)$ and $f^{stt}_{\rm NL}\leq\mathcal{O}(1)$ where we used $r\leq\mathcal{O}(10^{-2})$. Note that we have $f^{sst}_{\rm NL}=\mathcal{O}(10^{-1})$ and $f^{stt}_{\rm NL}=\mathcal{O}(10^{-2})$ for the squeezed limit model independently (see Section~\ref{SubSubSec: consistency-const}).  For $\nu=-3/2$ where the $c_{(1,2,4,5)}$ and $b_{(2,5,7)}$ terms are absent, we find $f^{sst}_{\rm NL}\leq\mathcal{O}(1)$ and $f^{stt}_{\rm NL}\leq\mathcal{O}(1)$. These upper values for the cross-bispectra are below both theoretical and observational constraints. For $\sigma_3=\mathcal{O}(1)$, we have $f^t_{\rm NL}=\mathcal{O}(1)$ which satisfies both theoretical and observational constraints as well.

We have thus found that the present bounce model allows for a parameter region satisfying both perturbativity conditions and observational constraints. Here, we have put conservative but severe constraints on the model parameters $\sigma_i$ in such a way that each of $\Lambda_i$ is at most of $\mathcal{O}(1)$. If one allows for the case in which some of $\Lambda_i$ are (much) greater than unity, while the resultant non-linearity parameters (except for those at the squeezed limit for $\nu=3/2$) are of $\mathcal{O}(1)$ due to cancellations, then the constraints on $\sigma_i$ would become weaker than those assumed in this section. This could expand the possibility of keeping $f^s_{\rm NL}$ at most $\mathcal{O}(1)$, while enhancing the non-linearity parameters of the scalar-tensor cross-bispectra and tensor auto-bispectrum. For instance, a possibility of enhancing the scalar-tensor-tensor bispectrum due to a small $c_t$ has been discussed in the context of inflation in Ref.~\cite{Noumi:2014zqa}. It is therefore interesting to build examples having enhanced scalar-tensor cross-bispectra and tensor auto-bispectrum in the context of bouncing cosmologies, which is beyond the scope of the present paper.

Before closing this section, we mention the observability of the tensor non-Gaussianities. Whether the tensor non-Gaussian amplitudes below the current constraints are detectable or not is clarified after evaluating signal-to-noise ratios of, e.g., CMB bispectra. So far, a smallness of the signal-to-noise ratio of the CMB B-mode bispectrum has been found in the contexts of both inflation~\cite{Tahara:2017wud} ($\mathcal{O}(10^7)$ fine-tuning is required for the signal-to-noise ratio to reach unity in the absence of any experimental noises) and bounce~\cite{Kothari:2019yyw,Akama:2024vtu}. (See also Refs.~\cite{Shiraishi:2011dh,Shiraishi:2012rm,Shiraishi:2013kxa} for an observable tensor auto-bispectrum from inflation.) In particular, we have found that, in a LiteBIRD-like realistic setup, the tensor auto-bispectrum in the perturbative regime predicts a signal-to-noise ratio smaller than unity~\cite{Akama:2024vtu}. 
Therefore, it would be interesting to repeat a similar analysis (forecast) for the scalar-tensor cross-bispectra. We leave this for our follow-up research~\cite{Akama:prep}.

\section{Conclusion and outlook}\label{Sec: summary}
In this paper, we have computed the auto- and cross-bispectra of scale-invariant curvature and tensor perturbations from the general bounce cosmology where the amplitudes of the scale-invariant perturbations either become constant (for $\nu=3/2$) or grow ($\nu=-3/2$) on the superhorizon scales. In particular, the bispectra for $\nu=3/2$ can mimic the momentum dependence of those from inflation where the superhorizon modes are constant, and we have found the suppression of squeezed-type non-Gaussianities as the Maldacena's consistency relation. The non-Gaussianity consistency relation for the tensor auto-bispectrum has been studied in the context of bouncing cosmology in Ref.~\cite{Nandi:2019xag}, while that for the other bispectra has been discussed in the same context for the first time in the present paper. Furthermore, the cross-bispectra for $\nu=-3/2$ have been computed in that context for the first time in the present paper as well. In computing the scalar auto-bispectrum and the cross-bispectra for $\nu=-3/2$, we have computed the total-time-derivative interaction terms and verified that the contributions from the total-time-derivative terms are identical to those from the field redefinitions as shown in Appendix~\ref{Sec: app-boundary}.

We have also studied the shapes of the bispectra, the summary of which is in Tables~\ref{table: auto-nu3/2},~\ref{table: cross-nu3/2},~\ref{table: auto-nu-3/2}, and~\ref{table: cross-nu-3/2}. We have found that the shapes of the bispectra for $\nu=3/2$ are generally mixed among the squeezed, equilateral, and other types. In particular, the bispectra include the squeezed-type terms only in certain cases: the scalar-scalar-tensor bispectrum for $c_t/c_s\leq1$, the scalar-tensor-tensor bispectrum for $c_t/c_s\ll1$ in the same-helicity case, and the tensor auto-bispectrum from the GR-type operator. On the other hand, the shapes of the bispectra for $\nu=-3/2$ are generally mixed between the squeezed and other types. In particular, all of the bispectra always include the terms peaking at the squeezed limits. Therefore, a peak at the equilateral limit is a characteristic feature of the branch for $\nu=3/2$, while sharp (divergent) peaks at the squeezed limits are characteristic features of the other branch. Furthermore, the shapes for $\nu=3/2$ are identical to those in the generalized G-inflation, and hence a similar conclusion can be applied to the differences between the generalized G-inflation and the general bounce cosmology for $\nu=-3/2$.

Then, we have studied the perturbativity conditions and the CMB constraints as the theoretical and observational constraints. As a result, we have found that the CMB constraints have put more stringent constraints on the scalar auto-bispectrum and the scalar-scalar-tensor bispectrum than the perturbativity conditions. We have also performed order estimation in a simplified setup such as $c_t\sim c_s$ etc., and clarified a parameter space where both theoretical and observational constraints are satisfied. We have presented a concrete example in the viable model space for $\nu=3/2$. A similar example for $\nu=-3/2$ has been presented in Ref.~\cite{Akama:2019qeh}.

We also mention possible extensions of the present work. In the present paper, we have ignored the subsequent bouncing and expanding phases. It is important to clarify in which parameter space the observational signatures from the contracting phase can change and to explore the models that yield any deviations from the observational predictions studied in the present paper, e.g., enhancements of the tensor non-Gaussianities towards the expected future detections of the CMB B-mode polarization. In particular, the present model for $\nu=-3/2$ suffers from growth of the spacetime anisotropies (if those were present initially), and see also Refs.~\cite{Grain:2020wro,Zhu:2024ffj} which studied anisotropic components sourced by quantum fluctuations. Therefore, a concrete realization of suppressing these anisotropies, especially from a later stage of the contracting phase to the end of the bouncing phase, is a key ingredient to complement the general bounce cosmology. Furthermore, we have worked with the Horndeski action, but one may invoke beyond-Horndeski operators to avoid the no-go theorem for nonsingular cosmologies in the case of $\nu=-3/2$. It is therefore important to study the impacts of beyond-Horndeski operators on primordial correlation functions during or after a contracting phase. Depending on each model construction, the additional ingredients such as a concrete realization of a successful bounce and the addition of beyond-Horndeski operators could affect the primordial correlation functions studied in the general model. Thus, if future experiments report any deviations from the observational predictions studied in the present paper, those deviations could be used to test these additional ingredients, which would be of help for model construction.

\section*{Acknowledgements}
I thank Paola C. M. Delgado and Giorgio Orlando for the collaboration at an initial stage of this work. I thank Shin'ichi Hirano, Tsutomu Kobayashi, and Shuichiro Yokoyama for helpful comments on the manuscript. I thank Chunshan Lin, Giorgio Orlando, and Pavel Petrov for fruitful discussions. 
This work was supported by the grant No.~UMO-2021/42/E/ST9/00260 from the National Science Centre, Poland.

\appendix
\section{Quadratic and cubic actions for scalar and tensor perturbations}\label{Sec: app-perturbed-actions}
In this section, we show the explicit forms of the perturbed actions. First, by substituting the metric perturbed by the scalar and tensor perturbations into the Horndeski action, one obtains~\cite{Kobayashi:2011nu}
\begin{align}
S^{(2)}&=\int\D t\D^3xa^3\biggl[-3\mathcal{G}_T\dot\zeta^2+\frac{\mathcal{F}_T}{a^2}(\partial_i\zeta)^2+\Sigma\delta n^2-\frac{2}{a^2}\Theta\delta n\partial^2\chi+\frac{2}{a^2}\mathcal{G}_T\dot\zeta\partial^2\chi+6\Theta\delta n\dot\zeta-\frac{2}{a^2}\mathcal{G}_T\delta n\partial^2\zeta\notag\\
&\quad\ +\frac{\mathcal{G}_T}{8}\dot h_{ij}^2-\frac{\mathcal{F}_T}{8a^2}(\partial_k h_{ij})^2\biggr], \label{eq: quadratic-action}
\end{align}
where 
\begin{align}
\mathcal{G}_T&:=2[G_4-2XG_{4X}-X(H\dot\phi G_{5X}-G_{5\phi})] \, ,\\
\mathcal{F}_T&:=2[G_4-X(\ddot\phi G_{5X}+G_{5\phi})],\\
\Theta&:=-\dot{\phi}XG_{3X}+2HG_4-8HXG_{4X}\notag-8HX^2G_{4XX}+\dot{\phi}G_{4\phi}+2X\dot{\phi}G_{4\phi{X}}\notag\\&
\quad-H^2\dot{\phi}(5XG_{5X}+2X^2G_{5XX})+2HX(3G_{5\phi}+2XG_{5\phi{X}}), \label{Theta-ap}\\
\Sigma&:=XG_{2X}+2X^2G_{2XX}+12H\dot{\phi}XG_{3X}+6H\dot{\phi}X^2G_{3XX}-2XG_{3\phi}-2X^2G_{3\phi{X}}-6H^2G_4\notag\\ &
\quad +6\bigl[H^2(7XG_{4X}+16X^2G_{4XX}+4X^3G_{4XXX})-H\dot{\phi}(G_{4\phi}+5XG_{4\phi{X}}+2X^2G_{4\phi{X}X})\bigr]\notag\\ &
\quad +2H^3\dot{\phi}\left(15XG_{5X}+13X^2G_{5XX}+2X^3G_{5XXX}\right)-6H^2X(6G_{5\phi}+9XG_{5\phi{X}}+2X^2G_{5\phi{X}X}), \label{Sigma-ap}
\end{align}
The compacted forms of $\Theta$ and $\Sigma$ are
\begin{align}
\Theta&=-\frac{1}{6}\frac{\partial\cal{E}}{\partial H}, \label{eq: theta-compc}\\
\Sigma&=X\frac{\partial\mathcal{E}}{\partial X}+\frac{H}{2}\frac{\partial\mathcal{E}}{\partial H}. \label{eq: sigma-compac}
\end{align}
These compacted forms with the scaling $\mathcal{E}\propto(-t)^{2\alpha}$ lead to $\Theta\propto(-t)^{2\alpha+1}$ and $\Sigma\propto(-t)^{2\alpha}$.
From variations of the quadratic action in Eq.~(\ref{eq: quadratic-action}) with respect to the auxiliary fields $\delta n$ and $\chi$, the constraint equations are obtained as
\begin{align}
\delta n&=\frac{\mathcal{G}_T}{\Theta}\dot\zeta,\\
\chi&=\frac{1}{a\mathcal{G}_T}\biggl(a^3\mathcal{G}_S\psi-\frac{a\mathcal{G}_T^2}{\Theta}\zeta\biggr),
\end{align}
where $\psi:=\partial^{-2}\dot\zeta$. Then, after substituting the solutions of $\delta n$ and $\chi$ back into Eq.~(\ref{eq: quadratic-action}), the quadratic action in Eq.~(\ref{eq: quadratic-action}) can be simplified as
\begin{align}
S^{(2)}=\int\D t\D^3x a^3\biggl[\mathcal{L}^{(2)}_s+\mathcal{L}^{(2)}_t\biggr],
\end{align}
where
\begin{align}
\mathcal{L}^{(2)}_s&=\mathcal{G}_S\dot\zeta^2-\frac{\mathcal{F}_S}{a^2}(\partial_i\zeta)^2,\\
\mathcal{L}^{(2)}_t&=\frac{1}{8}\biggl[\mathcal{G}_T\dot h_{ij}^2-\frac{\mathcal{F}_T}{a^2}(\partial_k h_{ij})^2\biggr],
\end{align}
with
\begin{align}
\mathcal{G}_S&:=\mathcal{G}_T\biggl(3+\frac{\mathcal{G}_T\Sigma}{\Theta^2}\biggr),\\
\mathcal{F}_S&:=\frac{1}{a}\frac{\D}{\D t}\biggl(\frac{a\mathcal{G}_T^2}{\Theta}\biggr)-\mathcal{F}_T.
\end{align}
Here, our assumption on the time scaling of each term leads to $\mathcal{G}_T, \mathcal{F}_T\propto(-t)^{2(\alpha+1)}$ since the same functions appear in both $\mathcal{G}_T, \mathcal{F}_T$ and $\mathcal{E}_{(4,5)}/H^2, \mathcal{P}_{(4,5)}/H^2\propto(-t)^{2(\alpha+1)}$.
Next, by expanding the Horndeski action up to the cubic order in the perturbations, substituting the solutions of the constraint equations, and integrating by parts, one obtains all cubic Lagrangians~\cite{Gao:2012ib},
\begin{align}
S^{(3)}=\int\D t\D^3x a^3\biggl[\mathcal{L}^{(3)}_s+\mathcal{L}^{(3)}_{sst}+\mathcal{L}^{(3)}_{stt}+\mathcal{L}^{(3)}_t\biggr],
\end{align}
where
\begin{align}
\mathcal{L}^{(3)}_s&=a^3\mathcal{G}_S\biggl[\frac{\Lambda_1}{H}\dot\zeta^3+\Lambda_2\zeta\dot\zeta^2+\Lambda_3\zeta\frac{(\partial_i\zeta)^2}{a^2}+\frac{\Lambda_4}{H^2}\dot\zeta^2\frac{\partial^2\zeta}{a^2}+\Lambda_5\dot\zeta\partial_i\zeta\partial_i\psi+\Lambda_6\partial^2\zeta(\partial_i\psi)^2\notag\\
&\quad\ +\frac{\Lambda_7}{H^2}\frac{1}{a^4}[\partial^2\zeta(\partial_i\zeta)^2-\zeta\partial_i\partial_j(\partial_i\zeta\partial_j\zeta)]+\frac{\Lambda_8}{H}\frac{1}{a^2}[\partial^2\zeta\partial_i\zeta\partial_i\psi-\zeta\partial_i\partial_j(\partial_i\zeta\partial_j\psi)]\biggr]+\bar E_s,\\
\mathcal{L}^{(3)}_{sst}&=a^3\bigg[\frac{c_1}{a^2}h_{ij}\partial_i\zeta\partial_j\zeta+\frac{c_2}{a^2}\dot h_{ij}\partial_i\zeta\partial_j\zeta+c_3\dot h_{ij}\partial_i\zeta\partial_j\psi+\frac{c_4}{a^2}\partial^2h_{ij}\partial_i\zeta\partial_j\psi\notag\\
&\quad +\frac{c_5}{a^4}\partial^2h_{ij}\partial_i\zeta\partial_j\zeta+c_6\partial^2h_{ij}\partial_i\psi\partial_j\psi\biggr]+\bar E_{sst},\\
\mathcal{L}^{(3)}_{stt}&=a^3\biggl[b_1\zeta\dot h_{ij}^2+\frac{b_2}{a^2}\zeta(\partial_k h_{ij})^2+b_3(\partial_k\psi)\dot h_{ij}\partial_k h_{ij}+b_4\dot\zeta\dot h_{ij}^2+\frac{b_5}{a^2}(\partial^2\zeta)\dot h_{ij}^2\notag\\
&\quad +b_6(\partial_i\partial_j\psi) \dot h_{ik}\dot h_{jk}+\frac{b_7}{a^2}(\partial_i\partial_j\zeta)\dot h_{ik}\dot h_{jk}\biggr]+\bar E_{stt},\\
\mathcal{L}^{(3)}_t&=a^3\biggl[\frac{\mathcal{F}_T}{4a^2}\biggl(h_{ik}h_{jl}-\frac{1}{2}h_{ij}h_{kl}\biggr)\partial_k\partial_l h_{ij}+\frac{\mu}{12}\dot h_{ij}\dot h_{jk}\dot h_{ki}\biggr],
\end{align}
with
\begin{align}
\Lambda_1&=H\biggl[\frac{\mathcal{G}_T}{\Theta}\biggl(\frac{\mathcal{G}_S}{\mathcal{F}_S}+3\frac{\mathcal{G}_T}{\mathcal{G}_S}-1\biggr)+\frac{\Xi\mathcal{G}_T}{3\Theta^2}\biggl(3\frac{\mathcal{G}_T}{\mathcal{G}_S}-1\biggr)+2\mu\biggl(\frac{1}{\mathcal{G}_S}-\frac{1}{\mathcal{G}_T}\biggr)+\frac{\Gamma}{\Theta}\biggl(3\frac{\mathcal{G}_T}{\mathcal{G}_S}-2\biggr)\notag\\
&\quad\ +\frac{2}{3}\frac{\mathcal{G}_T^3}{\Theta^3\mathcal{G}_S}(\Sigma-X\Sigma_X)-\frac{H}{3}\frac{\mathcal{G}_T^3\Xi}{\Theta^3\mathcal{G}_S}\biggr],\\
\Lambda_2&=3-\frac{H\mathcal{G}_T\mathcal{G}_S}{\mathcal{F}_S\Theta}(3-g_T+f_S+f_\Theta),\\
\Lambda_3&=\frac{\mathcal{F}_T}{\mathcal{G}_S}+\frac{H\mathcal{G}_T}{\Theta}(1+g_T+g_S-f_\Theta)-\frac{H\mathcal{G}_T^2}{\mathcal{G}_S\Theta}(1+2g_T-f_\Theta),\\
\Lambda_4&=H^2\biggl[\frac{\Xi}{3}\frac{\mathcal{G}_T^3}{\mathcal{G}_S\Theta^3}+6\mu\frac{\mathcal{G}_T}{\mathcal{G}_S\Theta}+(3\Gamma-\mathcal{G}_T)\frac{\mathcal{G}_T^2}{\mathcal{G}_S\Theta^2}\biggr],\\
\Lambda_5&=-\frac{1}{2}\frac{\mathcal{G}_S}{\mathcal{G}_T}-\frac{H}{2}\frac{\Gamma\mathcal{G}_S}{\mathcal{G}_T\Theta}(3+g_T-f_\Gamma+f_\Theta)-\mu H\frac{\mathcal{G}_S}{\mathcal{G}_T^2}(3+2g_T-f_\mu),\\
\Lambda_6&=\frac{3}{4}\frac{\mathcal{G}_S}{\mathcal{G}_T}-\frac{\mathcal{G}_S}{4\mathcal{G}_T}\frac{H\Gamma}{\Theta}(3+g_T-f_\Gamma+f_\Theta)-\mu H\frac{\mathcal{G}_S}{\mathcal{G}_T^2}\biggl(\frac{3}{2}+g_T-\frac{1}{2}f_\mu\biggr),\\
\Lambda_7&=\frac{H^2}{6}\biggl[\frac{\mathcal{G}_T^3}{\mathcal{G}_S\Theta^2}-\frac{H\Gamma\mathcal{G}_T^3}{\mathcal{G}_S\Theta^3}\biggl(1-3g_T+3f_\Theta-f_\Gamma+3\frac{\Theta\mathcal{F}_S}{H\mathcal{G}_T^2}\biggr)\notag\\
&\quad\ -6\mu H\frac{\mathcal{G}_T^2}{\mathcal{G}_S\Theta^2}\biggl(1-2g_T-f_\mu+2f_\Theta+2\frac{\Theta\mathcal{F}_S}{H\mathcal{G}_T^2}\biggr)\biggr],\\
\Lambda_8&=H\biggl[-\frac{\mathcal{G}_T}{\Theta}+\frac{2\mu H}{\Theta}\biggl(2+f_\Theta-f_\mu+\frac{\Theta\mathcal{F}_S}{H\mathcal{G}_T^2}\biggr)+H\frac{\Gamma\mathcal{G}_T}{\Theta^2}\biggl(1-\frac{1}{2}g_T-\frac{1}{2}f_\Gamma+f_\Theta+\frac{\Theta\mathcal{F}_S}{H\mathcal{G}_T^2}\biggr)\biggr],\\
c_1&=\mathcal{F}_S,\\
c_2&=\frac{\Gamma}{4\Theta}(\mathcal{F}_S-\mathcal{F}_T)+\frac{\mathcal{G}_T^2}{\Theta}\biggl[-\frac{1}{2}+\frac{H\Gamma}{4\Theta}(3+g_T-f_\Gamma+f_\Theta)\biggr]+\frac{\mu\mathcal{F}_S}{\mathcal{G}_T}+\frac{\mu H\mathcal{G}_T}{\Theta}(2-f_\mu+f_\Theta),\\
c_3&=\mathcal{G}_S\biggl[\frac{3}{2}-\frac{H\Gamma}{2\Theta}(3+g_T-f_\Gamma+f_\Theta)-\frac{\mu H}{\mathcal{G}_T}(3+2g_T-f_\mu)\biggr],\\
c_4&=\mathcal{G}_S\biggl[-\frac{\mathcal{G}_T^2-\Gamma\mathcal{F}_T}{2\Theta\mathcal{G}_T}-\frac{\mu H}{\Theta}(2-f_\mu+f_\Theta)+\frac{\mu}{\mathcal{G}_T^2}(\mathcal{F}_T-\mathcal{F}_S)\biggr],\\
c_5&=\frac{\mathcal{G}_T^2}{2\Theta}\biggl[\frac{\mathcal{G}_T^2-\Gamma\mathcal{F}_T}{2\Theta\mathcal{G}_T}+\frac{\mu H}{\Theta}(2-f_\mu+f_\Theta)-\frac{\mu}{\mathcal{G}_T^2}(3\mathcal{F}_T-\mathcal{F}_S)\biggr],\\
c_6&=\frac{\mathcal{G}_S^2}{4\mathcal{G}_T}\biggl[1+\frac{2\mu H}{\mathcal{G}_T}(3+2g_T-f_\mu)\biggr],\\
b_1&=\frac{3\mathcal{G}_T}{8}\biggl[1-\frac{H\mathcal{G}_T}{3\Theta c_t^2}(3+f_\Theta+2s_T)\biggr],\\
b_2&=\frac{\mathcal{F}_S}{8},\\
b_3&=-\frac{\mathcal{G}_S}{4},\\
b_4&=\frac{\mathcal{G}_T}{8\Theta\mathcal{F}_T}(\mathcal{G}_T^2-\Gamma\mathcal{F}_T)+\frac{\mu}{4}\biggl[\frac{\mathcal{G}_S}{\mathcal{G}_T}-1-\frac{H\mathcal{G}_T^2}{\Theta\mathcal{F}_T}(6+g_S+f_T+f_\Theta-f_\mu)\biggr],\\
b_5&=\frac{\mu \mathcal{G}_T}{4\Theta}\biggl(\frac{c_s^2}{c_t^2}-1\biggr),\\
b_6&=-\frac{\mu}{2}\frac{\mathcal{G}_S}{\mathcal{G}_T},\\
b_7&=\frac{\mu}{2}\frac{\mathcal{G}_T}{\Theta}.
\end{align}
and
\begin{align}
\Gamma&:=2G_4-8XG_{4X}-8X^2G_{4XX}-2H\dot\phi(5XG_{5X}+2X^2G_{5XX})+2X(3G_{5\phi}+2XG_{5\phi X}), \label{Gamma-ap}\\
\Xi&:=12\dot\phi{X}G_{3X}+6\dot\phi{X^2}G_{3XX}-12HG_4+6\biggl[2H(7XG_{4X}+16X^2G_{4XX}+4X^3G_{4XXX})-\dot\phi(G_{4\phi}\notag\\
&\quad+5XG_{4\phi{X}}+2X^2G_{4\phi{XX}})\biggr]+90H^2\dot\phi{X}G_{5X}+78H^2\dot\phi{X^2}G_{5XX}+12H^2\dot\phi{X^3}G_{5XXX}\notag\\
&\quad-12HX(6G_{5\phi}+9XG_{5\phi X}+2X^2G_{5\phi XX}), \label{Xi-ap}\\
\mu&:=\dot\phi X G_{5X}, \label{eq: def-mu}
\end{align}
and
\begin{align}
g_T&:=\frac{\dot{\mathcal{G}}_T}{H\mathcal{G}_T},\ f_T:=\frac{\dot{\mathcal{F}}_T}{H\mathcal{F}_T},\ g_S:=\frac{\dot{\mathcal{G}}_S}{H\mathcal{G}_S},\ f_S:=\frac{\dot{\mathcal{F}}_S}{H\mathcal{F}_S},\ s_T:=\frac{\dot{c}_t}{Hc_t},\notag\\
f_\Theta&:=\frac{\dot{\Theta}}{H\Theta},\ f_\Gamma:=\frac{\dot{\Gamma}}{H\Gamma},\ f_\mu:=\frac{\dot\mu}{H\mu}.
\end{align}
For the cross-interaction terms obtained in Ref.~\cite{Gao:2012ib}, we expressed several quantities with a time derivative by using $g_T, f_\Gamma, f_\Theta, f_\mu$, and $s_T$.
The compacted forms of $\Xi, \Gamma$, and $\mu$ are
\begin{align}
\Xi&=\frac{\partial\Sigma}{\partial H}, \label{eq: xi-compac}\\
\Gamma&=\frac{\partial\Theta}{\partial H}, \label{eq: gamma-compac}\\
\mu&=-\frac{1}{2}\frac{\partial\mathcal{G}_T}{\partial H}. \label{eq: mu-compac}
\end{align}
From these compacted forms, one has $\Xi\propto(-t)^{2\alpha+1}, \Gamma\propto(-t)^{2(\alpha+1)}$, and $\mu\propto(-t)^{2\alpha+3}$.
Here, the terms multiplied by the equations of motion for the perturbations in the cubic Lagrangians are of the form,
\begin{align}
\bar E_s&:=-2F(\zeta)E^s,\\
\bar E_{sst}&:=\bar f_i \partial^{-2}\partial_i E^s+\bar f_{ij}E^t_{ij},\\
\bar E_{stt}&:=\frac{\mu}{4\mathcal{G}_S}\frac{\mathcal{G}_T^2}{\Theta\mathcal{F}_T}\dot h_{ij}^2E^s+\frac{\mathcal{G}_T^2}{2\Theta\mathcal{F}_T}\biggl(\frac{\zeta}{2}+\frac{\mu}{\mathcal{G}_T}\dot\zeta\biggr)\dot h_{ij}E^t_{ij},
\end{align}
where
\begin{align}
F(\zeta)&:=-\frac{\mathcal{G}_T\mathcal{G}_S}{\Theta\mathcal{F}_S}\zeta\dot\zeta-\frac{1}{2}\biggl(\frac{\Gamma\mathcal{G}_S}{\Theta\mathcal{G}_T}+2\mu\frac{\mathcal{G}_S}{\mathcal{G}_T^2}\biggr)(\partial_i\zeta\partial_j\psi-\partial^{-2}\partial_i\partial_j(\partial_i\zeta\partial_j\psi))\notag\\
&\quad\ +\frac{1}{4a^2}\biggl(\frac{\Gamma\mathcal{G}_T}{\Theta^2}+\frac{4\mu}{\Theta}\biggr)[(\partial_i\zeta)^2-\partial^{-2}\partial_i\partial_j(\partial_i\zeta\partial_j\zeta)],\\
\bar f_i&:=\frac{1}{\Theta}\biggl(\frac{\Gamma}{2}+\frac{\mu\Theta}{\mathcal{G}_T}\biggr)(\partial_j\zeta)\dot h_{ij}+\frac{\mu}{a^2\Theta}(\partial_j\zeta)\partial^2 h_{ij}-\frac{\mu \mathcal{G}_S}{\mathcal{G}_T^2}(\partial_j\psi)\partial^2 h_{ij},\\
\bar f_{ij}&:=\frac{\mathcal{G}_S}{\Theta\mathcal{G}_T}\biggl(\frac{\Gamma}{2}+\frac{\mu\Theta}{\mathcal{G}_T}\biggr)(\partial_i\zeta)\partial_j\psi-\frac{\mathcal{G}_T}{a^2\Theta^2}\biggl(\frac{\Gamma}{4}+\frac{\mu\Theta}{\mathcal{G}_T}\biggr)(\partial_i\zeta)\partial_j\zeta.
\end{align}
One can eliminate the $\bar E_s$ term in the scalar-scalar-scalar cubic action by a field redefinition,
\begin{align}
\zeta&\to \zeta-F(\zeta). \label{eq:red-Es}
\end{align}
Similarly, one can eliminate the $\bar E_{sst}$ and $\bar E_{stt}$ terms in the scalar-tensor cubic actions by field redefinitions as
\begin{align}
h_{ij}&\to h_{ij}+4\bar f_{ij}, \label{eq:red-Esst-h}\\
\zeta&\to \zeta-\frac{1}{2}\partial^{-2}\partial_i\bar f_i, \label{eq:red-Esst-zeta}\\
\end{align}
and 
\begin{align}
h_{ij}\ &\to\ h_{ij}+\frac{\mathcal{G}_T^2}{\Theta\mathcal{F}_T}\biggl(\zeta+\frac{2\mu}{\mathcal{G}_T}\dot\zeta\biggr)\dot h_{ij}, \label{eq:red-Estt-h}\\
\zeta\ &\to\ \zeta+\frac{\mu}{8\mathcal{G}_S}\frac{\mathcal{G}_T^2}{\Theta\mathcal{F}_T}\dot h_{ij}^2, \label{eq:red-Estt-zeta}\\
\end{align}
respectively.

In the above computations of the cubic actions, all boundary terms were ignored. Among the boundary terms, if a field with a time derivative is included, the boundary terms with respect to a time derivative can contribute to the tree-level bispectra~\cite{Arroja:2011yj, Adshead:2011bw, Burrage:2011hd, Rigopoulos:2011eq, Celoria:2021cxq,
Braglia:2024zsl, Kawaguchi:2024lsw}. The boundary term $S_B$ involving a time derivative is of the form,
\begin{align}
S_B=S_B^s+S_B^{sst}+S_B^{stt},
\end{align}
where
\begin{align}
S_B^s&=\int\D t\D ^3x\frac{\D}{\D t}\biggl[-a^3\frac{\mathcal{G}_T\mathcal{G}_S^2}{\Theta\mathcal{F}_S}\zeta\dot\zeta^2+a^3\frac{\mathcal{G}_S^2}{2\mathcal{G}_T^2}\biggl(2\mu+\frac{\Gamma\mathcal{G}_T}{\Theta}\biggr)(\zeta\dot\zeta^2-\zeta(\partial_i\partial_j\psi)^2)\notag\\
&\quad\quad -\frac{a\mathcal{G}_S}{2\Theta}\biggl(4\mu+\frac{\Gamma\mathcal{G}_T}{\Theta}\biggr)(\zeta\dot\zeta\partial^2\zeta-\zeta\partial_i\partial_j\psi\partial_i\partial_j\zeta)\biggr], \label{eq: SB-s}\\
S_B^{sst}&=\int\D t\D^3 x\frac{\D}{\D t}\biggl[\frac{a}{4}\frac{\mathcal{G}_T^2}{\Theta}\biggl(\frac{\Gamma}{\Theta}+\frac{4\mu}{\mathcal{G}_T}\biggr)(\partial_i\zeta\partial_j\zeta)\dot h_{ij}-\frac{a^3}{2}\mathcal{G}_S\biggl(\frac{\Gamma}{\Theta}+\frac{2\mu}{\mathcal{G}_T}\biggr)(\partial_i\zeta\partial_j\psi)\dot h_{ij}\notag\\
&\quad\quad\ -a\frac{\mu\mathcal{G}_S}{\Theta}(\partial_i\zeta\partial_j\psi)\partial^2 h_{ij}+\frac{a^3}{2}\frac{\mu\mathcal{G}_S^2}{\mathcal{G}_T^2}(\partial_i\psi\partial_j\psi)\partial^2 h_{ij}\biggr], \label{eq: SB-sst}\\
S_B^{stt}&=\int\D t\D ^3x\frac{\D}{\D t}\biggl[-\frac{a^3}{4}\frac{\mu\mathcal{G}_T^2}{\Theta\mathcal{F}_T}\dot\zeta\dot h_{ij}^2-\frac{a^3}{8}\frac{\mathcal{G}_T^3}{\Theta\mathcal{F}_T}\zeta\dot h_{ij}^2\biggr]. \label{eq: SB-stt}
\end{align}
Here, regarding the explicit form of $S_B$ derived from the full Horndeski action, $S_B^s$ has been obtained in Ref.~\cite{Akama:2019qeh}, while the others have been computed for the first time in the present paper. In the following section, we will explicitly show that the contributions from the above boundary terms to the leading-order bispectra are identical to those computed by means of the field redefinitions. 

The time dependence of coefficients of the interaction Hamiltonians expressed by the cubic Lagrangians is as follows,
\begin{align}
\frac{a\mathcal{G}_S}{H}\Lambda_1&\propto(-\eta)^{2(1-\nu)},\ a^2\mathcal{G}_S\Lambda_2\propto(-\eta)^{1-2\nu},\ a^2\mathcal{G}_S\Lambda_3\propto(-\eta)^{1-2\nu},\ \frac{\mathcal{G}_S}{H^2}\Lambda_4\propto(-\eta)^{3-2\nu},\notag\\
a^2\mathcal{G}_S\Lambda_5&\propto(-\eta)^{1-2\nu},\ a^2\mathcal{G}_S\Lambda_6\propto(-\eta)^{1-2\nu},\ \frac{\mathcal{G}_S}{H^2}\Lambda_7\propto(-\eta)^{3-2\nu},\ \frac{a\mathcal{G}_S}{H}\Lambda_8\propto(-\eta)^{2(1-\nu)},\notag\\
a^2c_1&\propto(-\eta)^{1-2\nu},\ ac_2\propto(-\eta)^{2(1-\nu)},\ a^2c_3\propto(-\eta)^{1-2\nu},\ ac_4\propto(-\eta)^{2(1-\nu)},\notag\\ c_5&\propto(-\eta)^{3-2\nu},\ a^2c_6\propto(-\eta)^{1-2\nu},\notag\\
a^2b_1&\propto(-\eta)^{1-2\nu},\ a^2b_2\propto(-\eta)^{1-2\nu},\ a^2b_3\propto(-\eta)^{1-2\nu},\ ab_4\propto(-\eta)^{2(1-\nu)},\notag\\
b_5&\propto(-\eta)^{3-2\nu},\ ab_6\propto(-\eta)^{2(1-\nu)},\ b_7\propto(-\eta)^{3-2\nu},\notag\\
a^2\mathcal{F}_T&\propto(-\eta)^{1-2\nu},\ a\mu\propto(-\eta)^{2(1-\nu)}.
\end{align}
All of the above time dependence depend on only one parameter $\nu$. The integrations for $\nu=\pm3/2$ can be performed without any complexities  originating from an incomplete Gamma function.

\section{Three-point functions from the field redefinitions and boundary terms}\label{Sec: app-boundary}
The terms relevant to the three-point functions include a time derivative, and hence the contributions from the boundary terms are significant only when the perturbations grow on the superhorizon scales, i.e., for the case of $\nu=-3/2$. (See also the forms of the field redefinitions that include a time derivative.) In what follows, we thus compute the three-point functions originating from both the field redefinitions and the time boundary terms for $\nu=-3/2$. To compute the bispectra originating from the field redefinitions, we use the following forms of the perturbations on the superhorizon scales:
\begin{align}
\dot\zeta({\bf k})&\simeq -\frac{3(1-n)}{n}H\zeta({\bf k}),\\
\dot h_{ij}({\bf k})&\simeq -\frac{3(1-n)}{n}H h_{ij}({\bf k}).
\end{align}
For example, the first condition is derived from
\begin{align}
\frac{\dot\zeta({\bf k})}{H\zeta({\bf k})}=\frac{\zeta'({\bf k})}{aH\zeta({\bf k})}\simeq-\frac{3(1-n)}{n}.
\end{align}
In the same way, one can obtain the second one.

\subsection{Scalar-scalar-scalar interaction}\label{SubSec: app-boundary-sss}
In Fourier space, one can write the field redefinition in Eq.~(\ref{eq:red-Es}) as
\begin{align}
\zeta({\bf k})&\to \zeta({\bf k})-\frac{3(1-n)}{n}\int\frac{\D^3 k'}{(2\pi)^3}\biggl\{\frac{\Lambda_{{\rm red}_1}}{2}\biggl[\frac{{\bf k}'\cdot ({\bf k}-{\bf k}')}{k'^2}-\frac{({\bf k}\cdot{\bf k}')({\bf k}\cdot({\bf k}-{\bf k}'))}{k^2k'^2}\biggr]+\Lambda_{{\rm red}_2}\biggr\}\notag\\
&\quad\ \times\zeta({\bf k}')\zeta({\bf k}-{\bf k}'),
\end{align}
where
\begin{align}
\Lambda_{{\rm red}_1}&:=\frac{H\Gamma\mathcal{G}_S}{\Theta\mathcal{G}_T}+2\frac{\mu H\mathcal{G}_S}{\mathcal{G}_T^2}={\rm const},\\
\Lambda_{{\rm red}_2}&:=\frac{H\mathcal{G}_T\mathcal{G}_S}{\Theta\mathcal{F}_S}={\rm const},
\end{align}
and we dropped the other terms suppressed on the superhorizon scales.
The contribution from the above to the three-point function has been computed as~\cite{Akama:2019qeh}
\begin{align}
\langle\zeta({\bf k}_1)\zeta({\bf k}_2)\zeta({\bf k}_3)\rangle_{\rm redefinition}&=(2\pi)^3\delta({\bf k}_1+{\bf k}_2+{\bf k}_3)\frac{(2\pi)^4\mathcal{P}_\zeta^2}{k_1^3k_2^3k_3^3}\notag\\
&\quad\ \times\frac{3}{8}\frac{1-n}{n}\biggl[(\Lambda_{\rm red_1}-4\Lambda_{\rm red_2})\sum_i k_i^3+\frac{\Lambda_{\rm red_1}}{4}\sum_{i\neq j}k_i^2k_j\notag\\
&\quad\quad\ -\frac{\Lambda_{\rm red_1}}{4}\frac{1}{k_1^2k_2^2k_3^2}\biggl(\sum_{i\neq j}k_i^7k_j^2+\sum_{i\neq j}k_i^6k_j^3-2\sum_{i\neq j}k_i^5k_j^4\biggr)\biggr].
\end{align}
Here, one can straightforwardly show that the contribution from the boundary term to the leading-order bispectrum is identical to the above:
\begin{align}
\langle\zeta({\bf k}_1)\zeta({\bf k}_2)\zeta({\bf k}_3)\rangle_{\rm redefinition}=\langle\zeta({\bf k}_1)\zeta({\bf k}_2)\zeta({\bf k}_3)\rangle_{\rm Boundary}.
\end{align}

\subsection{Scalar-scalar-tensor interaction}\label{SubSec: app-boundary-sst}
In Fourier space, the redefined perturbations in Eq.~(\ref{eq:red-Esst-h}) and Eq.~(\ref{eq:red-Esst-zeta}) on the superhorizon scales take the following forms,
\begin{align}
\xi^{(s)}({\bf k})\ &\to\ \xi^{(s)}({\bf k})+C_1\int\frac{\D^3k'}{(2\pi)^3}\frac{e^{(s)*}_{ij}({\bf k})k'_ik'_j}{{k'}^2}\zeta({\bf k}-{\bf k'})\zeta({\bf k}'),\\
\zeta({\bf k})\ &\to\ \zeta({\bf k})+C_2\frac{1}{k^2}\int\frac{\D^3k'}{(2\pi)^3}\zeta({\bf k}')k_ik'_jh_{ij}({\bf k}-{\bf k}')\notag\\
&\quad\ +C_3\frac{1}{k^2}\int\frac{\D^3k'}{(2\pi)^3}\frac{1}{{k'_j}^2}|{\bf k}-{\bf k}'|^2\zeta({\bf k}')k_ik'_jh_{ij}({\bf k}-{\bf k}'),
\end{align}
where
\begin{align}
C_1&:=4\cdot\frac{3(1-n)}{n}\frac{H\mathcal{G}_S}{\Theta\mathcal{G}_T}\biggl(\frac{\Gamma}{2}+\frac{\mu\Theta}{\mathcal{G}_T}\biggr)={\rm const.},\\
C_2&:=\frac{1}{2}\cdot\frac{3(1-n)}{n}\frac{H}{\Theta}\biggl(\frac{\Gamma}{2}+\frac{\mu\Theta}{\mathcal{G}_T}\biggr)={\rm const.},\\
C_3&:=-\frac{1}{2}\cdot\frac{3(1-n)}{n}\frac{\mu H\mathcal{G}_S}{\mathcal{G}_T^2}={\rm const.},
\end{align}
and we dropped the other terms suppressed on the superhorizon scales.

First, the $C_1$ term yields
\begin{align}
\langle\zeta({\bf k}_1)\zeta({\bf k}_2)\xi^{(s_3)}({\bf k}_3)\rangle_{C_1}&=C_1 e^{(s_3)*}_{ij}({\bf k}_3)\int\frac{\D^3 k'}{(2\pi)^3}\frac{k_i'k_j'}{k'^2}\langle\zeta({\bf k}_1)\zeta({\bf k}_2)\zeta({\bf k}')\zeta({\bf k}-{\bf k}')\rangle
\notag\\
&=(2\pi)^3\delta({\bf k}_1+{\bf k}_2+{\bf k}_3)\frac{(2\pi)^4\mathcal{P}_\zeta^2}{k_1^3k_2^3k_3^3}\cdot\frac{C_1}{4}k_3^3\frac{k_{1i}k_{1j}}{k_1^2}e^{(s_3)*}_{ij}({\bf k}_3)+({\bf k}_1\leftrightarrow {\bf k}_2).
\end{align}
Next, the $C_2$ term yields
\begin{align}
\langle\zeta({\bf k}_1)\zeta({\bf k}_2)\xi^{(s_3)}({\bf k}_3)\rangle_{C_2}&=C_2\frac{k_{1i}}{k_1^2}\int\frac{\D^3 k'}{(2\pi)^3}k_j'\langle\zeta({\bf k}')h_{ij}({\bf k}_1-{\bf k}')\zeta({\bf k}_2)\xi^{(s_3)}({\bf k}_3)\rangle+({\bf k}_1\leftrightarrow{\bf k}_2)\notag\\
&=(2\pi)^3\delta({\bf k}_1+{\bf k}_2+{\bf k}_3)\frac{(2\pi)^4\mathcal{P}_\zeta^2}{k_1^3k_2^3k_3^3}\cdot 2C_2\frac{c_s\mathcal{F}_S}{c_t\mathcal{F}_T}k_1^3\frac{k_{1i}k_{1j}}{k_1^2}e^{(s_3)*}_{ij}({\bf k}_3)\notag\\
&\quad\ +({\bf k}_1\leftrightarrow {\bf k}_2),
\end{align}
where we used
\begin{align}
e^{(s)*}_{kl}({\bf k}')\Pi_{ij,kl}(-{\bf k}')=e^{(s)*}_{ij}({\bf k}'),
\end{align}
with
\begin{align}
\langle h_{ij}({\bf k})h_{kl}({\bf k}')\rangle&=:(2\pi)^3\delta({\bf k}+{\bf k}')\frac{\pi^2}{k^3}\mathcal{P}_h\Pi_{ij,kl}(-{\bf k}'),\\
\Pi_{ij,kl}({\bf k})&:=\sum_s e^{(s)}_{ij}({\bf k})e^{(s)*}_{kl}({\bf k}).
\end{align}
Last, similarly to the $C_2$ term, the contribution from the $C_3$ term is obtained as
\begin{align}
\langle\zeta({\bf k}_1)\zeta({\bf k}_2)\xi^{(s_3)}({\bf k}_3)\rangle_{C_3}&=C_3\frac{k_{1i}}{k_1^2}\int\frac{\D^3 k'}{(2\pi)^3}\frac{k_j'}{k'^2}|{\bf k}_1-{\bf k}'|^2\langle\zeta({\bf k}')h_{ij}({\bf k}_1-{\bf k}')\zeta({\bf k}_2)\xi^{(s_3)}({\bf k}_3)\rangle+({\bf k}_1\leftrightarrow{\bf k}_2)\notag\\
&=(2\pi)^3\delta({\bf k}_1+{\bf k}_2+{\bf k}_3)\frac{(2\pi)^4\mathcal{P}_\zeta^2}{k_1^3k_2^3k_3^3}\cdot 2C_3\frac{c_s\mathcal{F}_S}{c_t\mathcal{F}_T}\frac{k_3^2}{k_2^2}k_1^3\frac{k_{1i}k_{1j}}{k_1^2}e^{(s_3)*}_{ij}({\bf k}_3)\notag\\
&\quad\ +({\bf k}_1\leftrightarrow {\bf k}_2),
\end{align}
Here, the boundary terms in Eq.~(\ref{eq: SB-sst}) give the following contributions to the three-point function:
\begin{align}
&\langle\zeta({\bf k}_1)\zeta({\bf k}_2)\xi^{(s_3)}({\bf k}_3)\rangle_{\rm Boundary}\notag\\
\ &=(2\pi)^3\delta({\bf k}_1+{\bf k}_2+{\bf k}_3)\frac{(2\pi)^4\mathcal{P}_\zeta^2}{k_1^3k_2^3k_3^3}\biggl[\frac{3}{2}\frac{1-n}{n}\frac{H\mathcal{G}_S}{c_t\mathcal{F}_T}\biggl(\frac{\Gamma}{\Theta}+\frac{2\mu}{\mathcal{G}_T}\biggr)\biggr|_{\eta=\eta_b}(c_s^3k_2^3+c_t^3k_3^3)\frac{k_{1i}k_{1i}}{k_2^2}e^{(s_3)*}_{ij}({\bf k}_3)\notag\\
&\quad\ -\frac{3}{2}\frac{1-n}{n}\frac{\mu H\mathcal{G}_S^2}{\mathcal{G}_T^3}\biggr|_{\eta=\eta_b}\frac{c_s^3}{c_t^3}(k_1^3+k_2^3)\frac{k_3^2}{k_1^2}\frac{k_{1i}k_{1j}}{k_2^2}e^{(s_3)*}_{ij}({\bf k}_3)+({\bf k}_1\leftrightarrow{\bf k}_2)\biggr].
\end{align}
By summing over the permutations, one can show the following relation,
\begin{align}
\sum_{i=1}^3\langle\zeta({\bf k}_1)\zeta({\bf k}_2)\xi^{(s_3)}({\bf k}_3)\rangle_{C_i}&=\langle\zeta({\bf k}_1)\zeta({\bf k}_2)\xi^{(s_3)}({\bf k}_3)\rangle_{{\rm Boundary}}\notag\\
&=(2\pi)^3\delta({\bf k}_1+{\bf k}_2+{\bf k}_3)\frac{(2\pi)^4\mathcal{P}_\zeta\mathcal{P}_h}{k_1^3k_2^3k_3^3}\frac{\pi^2}{2}\biggl(\frac{n}{1-n}\biggr)^2\mathcal{P}_\zeta\notag\\
&\quad\ \biggl[c_{\rm red_1}\mathcal{J}^{({\rm red},1)}|_{\eta=\eta_b}\mathcal{V}^{({\rm red},1)}_{s_3}+c_{\rm red_2}\mathcal{J}^{({\rm red},2)}|_{\eta=\eta_b}\mathcal{V}^{({\rm red},2)}_{s_3}\biggr]\notag\\
&\quad\ +({\bf k}_1\leftrightarrow{\bf k}_2),
\end{align}
where
\begin{align}
\mathcal{J}^{({\rm red},1)}&:=c_{\rm red_1}(c_s^3k_2^3+c_t^3k_3^3),\\
\mathcal{J}^{({\rm red},2)}&:=c_{\rm red_2}c_s^3(k_1^3+k_2^3),\\
c_{\rm red_1}&:=\frac{3}{2}\frac{1-n}{n}\frac{\mathcal{G}_S}{H^2}\biggl(\frac{\Gamma H}{\Theta}+\frac{2\mu H}{\mathcal{G}_T}\biggr),\\
c_{\rm red_2}&:=-\frac{3}{2}\frac{1-n}{n}\frac{\mu\mathcal{G}_S^2}{H\mathcal{G}_T^2},\\
\mathcal{V}^{({\rm red},1)}_{s_3}&:=-\frac{1}{k_2^2}\mathcal{V}^{(1)}_{s_3},\\
\mathcal{V}^{({\rm red},2)}_{s_3}&:=\frac{k_3^2}{k_1^2}\mathcal{V}^{({\rm red},1)}_{s_3}.
\end{align}

\subsection{Scalar-tensor-tensor interaction}\label{SubSec: app-boundary-stt}
The field redefinitions in Eq.~(\ref{eq:red-Estt-h}) and Eq.~(\ref{eq:red-Estt-zeta}) on the superhorizon scales read
\begin{align}
\xi^{(s)}({\bf k})&\to \xi^{(s)}({\bf k})+B_1e^{(s)*}_{ij}({\bf k})\int\frac{\D^3k'}{(2\pi)^3}\zeta({\bf k}')h_{ij}({\bf k}-{\bf k}'),\\
\zeta({\bf k})&\to\zeta({\bf k})+B_2\int\frac{\D^3k'}{(2\pi)^3}h_{ij}({\bf k}')h_{ij}({\bf k}-{\bf k}'),
\end{align}
where
\begin{align}
B_1&:=\frac{\mathcal{G}_T^2}{\Theta\mathcal{F}_T}\biggl[-\frac{3(1-n)}{n}H\biggr]\biggl[1-\frac{2\mu H}{\mathcal{G}_T}\frac{3(1-n)}{n}\biggr],\\
B_2&:=\frac{\mu}{8\mathcal{G}_S}\frac{\mathcal{G}_T^2}{\Theta\mathcal{F}_T}\frac{9(1-n)^2}{n^2}H^2.
\end{align}
First, the $B_1$ term yields
\begin{align}
\langle\zeta({\bf k}_1)\xi^{(s_2)}({\bf k}_2)\xi^{(s_3)}({\bf k}_3)\rangle_{B_1}&=B_1e^{(s_2)*}_{ij}({\bf k}_2)e^{(s_3)*}_{kl}({\bf k}_3)\int\frac{\D^3 k'}{(2\pi)^3}\langle\zeta({\bf k}_1)\zeta({\bf k}')h_{ij}({\bf k}_2-{\bf k}')h_{kl}({\bf k}_3)\rangle\notag\\
&\quad\ +(s_2,{\bf k}_2\leftrightarrow s_3,{\bf k}_3)\notag\\
&=(2\pi)^3\delta({\bf k}_1+{\bf k}_2+{\bf k}_3)B_1\frac{2\pi^2}{k_1^3}\mathcal{P}_\zeta\cdot \frac{\pi^2}{k_3^3}\mathcal{P}_h e^{(s_2)*}_{ij}({\bf k}_2)e^{(s_3)*}_{kl}({\bf k}_3)\notag\\
&\quad\ \times\Pi_{ij,kl}({\bf k}_1+{\bf k}_2)+(s_2,{\bf k}_2\leftrightarrow s_3,{\bf k}_3).
\end{align}
By using the normalization condition of the polarization tensor, we obtain
\begin{align}
e^{(s_2)}_{ij}({\bf k}_2)e^{(s_3)*}_{kl}({\bf k}_3)\Pi_{ij,kl}({\bf k}_1+{\bf k}_2)=e^{(s_2)*}_{ij}({\bf k}_2)e^{(s_3)*}_{ij}({\bf k}_3),
\end{align}
and hence the resultant contribution from the $B_1$ term reads
\begin{align}
\langle\zeta({\bf k}_1)\xi^{(s_2)}({\bf k}_2)\xi^{(s_3)}({\bf k}_3)\rangle_{B_1}&=(2\pi)^3\delta({\bf k}_1+{\bf k}_2+{\bf k}_3)\frac{(2\pi)^4\mathcal{P}_\zeta\mathcal{P}_h}{k_1^3k_2^3k_3^3}\frac{H^2}{2c_t\mathcal{F}_T}\cdot B_1\frac{c_t\mathcal{F}_T}{4H^2}\biggr|_{\eta=\eta_b}\notag\\
&\quad\ \times k_2^3e^{(s_2)*}_{ij}({\bf k}_2)e^{(s_3)*}_{ij}({\bf k}_3)+(s_2,{\bf k}_2\leftrightarrow s_3,{\bf k}_3).
\end{align}
Similarly, the $B_2$ term yields
\begin{align}
\langle\zeta({\bf k}_1)\xi^{(s_2)}({\bf k}_2)\xi^{(s_3)}({\bf k}_3)\rangle_{B_2}&=B_2\int\frac{\D^3 k'}{(2\pi)^3}\langle h_{ij}({\bf k}') h_{ij}({\bf k}_1-{\bf k}')\xi^{(s_2)}({\bf k}_2)\xi^{(s_3)}({\bf k}_3)\rangle\notag\\
&=(2\pi)^3\delta({\bf k}_1+{\bf k}_2+{\bf k}_3)\frac{(2\pi)^4\mathcal{P}_\zeta\mathcal{P}_h}{k_1^3k_2^3k_3^3}\frac{H^2}{2c_t\mathcal{F}_T}\cdot 2B_2\frac{c_s\mathcal{F}_S}{H^2}\biggr|_{\eta=\eta_b}\notag\\
&\quad \times k_1^3e^{(s_2)*}_{ij}({\bf k}_2)e^{(s_3)*}_{ij}({\bf k}_3)+(s_2,{\bf k}_2\leftrightarrow s_3,{\bf k}_3).
\end{align}
Here, from the boundary terms, one can compute the three-point function as
\begin{align}
&\langle\zeta({\bf k}_1)\xi^{(s_2)}({\bf k}_2)\xi^{(s_3)}({\bf k}_3)\rangle_{{\rm Boundary}}\notag\\
&=(2\pi)^3\delta({\bf k}_1+{\bf k}_2+{\bf k}_3)\frac{(2\pi)^4\mathcal{P}_\zeta\mathcal{P}_h}{k_1^3k_2^3k_3^3}\frac{H^2}{2c_t\mathcal{F}_T}\cdot\biggl\{\frac{9}{4}\frac{(1-n)^2}{n^2}\frac{\mu \mathcal{F}_T}{c_t^4\Theta}\biggr|_{\eta=\eta_b}[c_s^3k_1^3+c_t^3(k_2^3+k_3^3)]\notag\\
&\quad\ -\frac{3}{8}\frac{1-n}{n}\frac{\mathcal{F}_T^2}{c_t^3H\Theta }\biggr|_{\eta=\eta_b}(k_2^3+k_3^3)\biggr\}e^{(s_2)*}_{ij}({\bf k}_2)e^{(s_3)*}_{ij}({\bf k}_3)+(s_2,{\bf k}_2\leftrightarrow s_3,{\bf k}_3).
\end{align}
As a result, we can verify that the field redefinitions properly pick up the contributions from the boundary terms:
\begin{align}
\sum_{i=1}^2\langle\zeta({\bf k}_1)\xi^{(s_2)}({\bf k}_2)\xi^{(s_3)}({\bf k}_3)\rangle_{B_i}&=\langle\zeta({\bf k}_1)\xi^{(s_2)}({\bf k}_2)\xi^{(s_3)}({\bf k}_3)\rangle_{{\rm Boundary}}\notag\\
&=(2\pi)^3\delta({\bf k}_1+{\bf k}_2+{\bf k}_3)\frac{(2\pi)^4\mathcal{P}_\zeta\mathcal{P}_h}{k_1^3k_2^3k_3^3}\frac{\pi^2}{4}\biggl(\frac{n}{1-n}\biggr)^2\mathcal{P}_h\notag\\
&\quad\ \biggl[\mathcal{I}^{({\rm red},1)}|_{\eta=\eta_b}\mathcal{V}^{({\rm red},1)}_{s_2s_3}+\mathcal{I}^{({\rm red},2)}|_{\eta=\eta_b}\mathcal{V}^{({\rm red},2)}_{s_2s_3}\biggr]\notag\\
&\quad\ +(s_2,{\bf k}_2\leftrightarrow s_3, {\bf k}_3),
\end{align}
where
\begin{align}
\mathcal{I}^{({\rm red},1)}&:=b_{\rm red_1}[c_s^3k_1^3+c_t^3(k_2^3+k_3^3)],\\
\mathcal{I}^{({\rm red},2)}&:=b_{\rm red_2}c_t^3(k_2^3+k_3^3),\\
\mathcal{V}^{({\rm red},1)}_{s_3}&=\mathcal{V}^{({\rm red},2)}_{s_3}=\mathcal{V}^{(1)}_{s_2s_3}.
\end{align}
with
\begin{align}
b_{\rm red_1}&:=\frac{9}{4}\frac{(1-n)^2}{n^2}\frac{\mu \mathcal{F}_T}{c_t^4\Theta},\\
b_{\rm red_2}&:=-\frac{3}{8}\frac{1-n}{n}\frac{\mathcal{F}_T^2}{c_t^6H\Theta }.
\end{align}

\section{Polarization tensor and its products}\label{Sec: app-polarization-tensor}
In this section, we show the explicit form of the polarization tensor and its products.
All momenta can be placed on the same plane, which is from the momentum conservation ${\bf k}_1+{\bf k}_2+{\bf k}_3={\bf 0}$. Then, by following Ref.~\cite{Soda:2011am}, let us parametrize the momenta as
\begin{align}
{\bf k}_1&=k_1(1,0,0),\\
{\bf k}_2&=k_2(\cos\theta,\sin\theta,0),\\
{\bf k}_3&=k_3(\cos\phi,\sin\phi,0),
\end{align}
where $\theta$ and $\phi$ are related to the momenta as
\begin{align}
\cos\theta&=\frac{k_3^2-k_1^2-k_2^2}{2k_1k_2},\ \sin\theta=\frac{1}{2k_1k_2}\sqrt{K(-k_1+k_2+k_3)(k_1-k_2+k_3)(k_1+k_2-k_3)},\notag\\
\cos\phi&=\frac{k_2^2-k_1^2-k_3^2}{2k_1k_3},\ \sin\phi=-\frac{1}{2k_1k_3}\sqrt{K(-k_1+k_2+k_3)(k_1-k_2+k_3)(k_1+k_2-k_3)}.
\end{align}
Then, one can write down the explicit expressions of the polarizations tensors as
\begin{align}
e^{(s_1)}({\bf k}_1)&=\frac{1}{2}
\begin{pmatrix}
0 & 0 & 0 \\
0 & 1 & is_1 \\
0 & is_1 & -1
\end{pmatrix},\\
e^{(s_2)}({\bf k}_2)&=\frac{1}{2}
\begin{pmatrix}
\sin^2\theta & -\sin\theta\cos\theta & -is_2\sin\theta \\
-\sin\theta\cos\theta & \cos^2\theta & is_2\cos\theta \\
-is_2\sin\theta & is_2\cos\theta & -1
\end{pmatrix}
,\\
e^{(s_3)}({\bf k}_3)&=\frac{1}{2}
\begin{pmatrix}
\sin^2\phi & -\sin\phi\cos\phi & -is_3\sin\phi \\
-\sin\phi\cos\phi & \cos^2\phi & is_3\cos\phi \\
-is_3\sin\phi & is_3\cos\phi & -1
\end{pmatrix}.
\end{align}
To obtain the products of the polarization tensor appearing in the tensor cross- and auto-bispectra, we explicitly show the following quantities:
\begin{align}
\mathcal{V}^{(1)}_{s_3}&=k_{1i}k_{2j}e^{(s_3)*}_{ij}({\bf k}_3)\notag\\
&=\frac{K}{8k_3^2}(k_1-k_2-k_3)(k_1-k_2+k_3)(k_1+k_2-k_3),\\
\mathcal{V}^{(1)}_{s_2s_3}&=e^{(s_2)*}_{ij}({\bf k}_2)e^{(s_3)*}_{ij}({\bf k}_3)\notag\\
&=\frac{1}{16k_2^2k_3^2}[k_1^2-(s_2k_2+s_3k_3)^2]^2,\\
\mathcal{V}^{(6)}_{s_2s_3}&=\hat k_{1m}\hat k_{1n}e^{(s_2)*}_{mm'}({\bf k}_2)e^{(s_3)*}_{nm'}({\bf k}_3)\notag\\
&=\frac{K}{32k_1^2k_2^2k_3^2}(k_1-k_2-k_3)(k_1-k_2+k_3)(k_1+k_2-k_3)[k_1^2-(s_2k_2+s_3k_3)^2]
\end{align}
for the cross-bispectra and
\begin{align}
F_{\rm GR}&=e^{(s_1)*}_{ik}({\bf k}_1)e^{(s_2)*}_{jl}({\bf k}_2)\biggl[k_{3k}k_{3l}e^{(s_3)*}_{ij}({\bf k}_3)-\frac{1}{2}k_{3i}k_{3k}e^{(s_3)*}_{jl}({\bf k}_3)\biggr]+(5\ {\rm perm.})\notag\\
&=\frac{1}{2}(s_1k_1+s_2k_2+s_3k_3)^2F(s_1k_1,s_2k_2,s_3k_3),\\
F_{\rm New}&=e^{(s_1)*}_{ij}({\bf k}_1)e^{(s_2)*}_{jk}({\bf k}_2)e^{(s_3)*}_{ki}({\bf k}_3)\notag\\
&=F(s_1k_1,s_2k_2,s_3k_3),
\end{align}
for the auto-bispectrum where
\begin{align}
F(x,y,z):=\frac{1}{64x^2y^2z^2}(x+y+z)^3(x-y-z)(x-y+z)(x+y-z).
\end{align}
The other products can be derived from the above.

\section{Non-Gaussianity consistency relations}\label{Sec: app-consistency}
In this section, let us review the derivation of the consistency relation by following Ref.~\cite{Kundu:2013gha} (see also Ref.~\cite{Sreenath:2014nka}). When we compute the three-point functions at squeezed limits characterized by two short-wavelength modes and one long-wavelength mode in the setup such that the amplitude of the latter is constant, we may first evaluate two-point functions of the short-wavelength modes in the spacetime perturbed by the long-wavelength perturbation, that is, we may compute
\begin{align}
\langle\langle\psi({\bf k}_1)\psi({\bf k}_2)\rangle_{{\bf k}_3}\psi({\bf k}_3)\rangle|_{k_1\simeq k_2\gg k_3},
\end{align}
where $\psi({\bf k})$ is a scalar or tensor modes, and $\langle\psi({\bf k}_1)\psi({\bf k}_2)\rangle_{{\bf k}_3}$ is the two-point function of the short-wavelength modes evaluated in the FLRW spacetime perturbed by the long-wavelength $k_3$ mode. In the presence of the curvature and tensor perturbations whose long-wavelength modes perturb the FLRW background, let us absorb those into the spatial coordinate as
\begin{align}
\D s^2\to \D s'^2&=-\D t^2+a^2 e^{2\bar \zeta}(e^{\bar h})_{ij}\D x^i\D x^j\notag\\
&=:-\D t^2+a^2 \delta_{ij}(\D x')^i(\D x')^j,
\end{align}
where $\bar\zeta$ and $\bar h$ denote the perturbations on large scales, and we defined
\begin{align}
x'_i:=S_{ij}x_j,
\end{align}
with
\begin{align}
S_{ij}:=e^{\bar\zeta}(e^{\bar h/2})_{ij}.
\end{align}
Here, $\bar\zeta$ and $\bar h_{ij}$ are expressed as
\begin{align}
\bar\zeta({\bf x})&=\int_{k\ll k_S}\frac{\D^3k}{(2\pi)^3}\zeta({\bf k})e^{i{\bf k}\cdot{\bf x}},\\
\bar h_{ij}({\bf x})&=\int_{k\ll k_S}\frac{\D^3k}{(2\pi)^3}h_{ij}({\bf k})e^{i{\bf k}\cdot{\bf x}},
\end{align}
where $k_S$ denotes the wavenumbers of the two short-wavelength modes. The above rescaling of the spatial coordinate leads to the following relation:
\begin{align}
\zeta({\bf k})&\to{\rm det}({\bf S}^{-1})\zeta({\bf S}^{-1}{\bf k}),\\
\xi^{(s)}({\bf k})&\to{\rm det}({\bf S}^{-1})\sum_{s'}\xi^{(s')}({\bf S}^{-1}{\bf k})e^{(s)*}_{ij}({\bf k})e^{(s')}_{ij}({{\bf S}^{-1}}{\bf k}),
\end{align}
where
\begin{align}
(S^{-1})_{ij}=e^{-\bar\zeta}(e^{-\bar h/2})_{ij}.
\end{align}
One can derive from the above that
\begin{align}
|{\bf S}^{-1}{\bf k}|&=[(S^{-1})_{ij}(S^{-1})_{ik}k^jk^k]^{1/2}\notag\\
&\simeq k\biggl(1-\bar\zeta-\frac{1}{2}\frac{k^ik^j}{k^2}\bar h_{ij}\biggr),
\end{align}
where we have ignored the subleading terms of higher order in the perturbations. Under the squeezed limits, one can rewrite the three-point functions as
\begin{align}
\langle\zeta({\bf k}_1)\zeta({\bf k}_2)\zeta({\bf k}_3)\rangle&\simeq \langle\langle\zeta({\bf k}_1)\zeta({\bf k}_2)\rangle_{{\bf k}_3}\zeta({\bf k}_3)\rangle,\\
\langle\zeta({\bf k}_1)\zeta({\bf k}_{2})\xi^{(s_{3})}({\bf k}_{3})\rangle&\simeq \langle\langle\zeta({\bf k}_1)\zeta({\bf k}_{2})\rangle_{{\bf k}_{3}}\xi^{(s_{3})}({\bf k}_{3})\rangle,\\
\langle\zeta({\bf k}_1)\xi^{(s_{2})}({\bf k}_{2})\xi^{(s_{3})}({\bf k}_{3})\rangle&\simeq \langle\zeta({\bf k}_1)\langle\xi^{(s_{2})}({\bf k}_{2})\xi^{(s_{3})}({\bf k}_{3})\rangle_{{\bf k}_1}\rangle,\\
\langle\xi^{(s_1)}({\bf k}_1)\xi^{(s_{2})}({\bf k}_{2})\xi^{(s_{3})}({\bf k}_{3})\rangle&\simeq \langle\langle\xi^{(s_1)}({\bf k}_1)\xi^{(s_2)}({\bf k}_{2})\rangle_{{\bf k}_{3}}\xi^{(s_{3})}({\bf k}_{3})\rangle.
\end{align}
First, the power spectrum of the curvature perturbation can be obtained as
\begin{align}
\langle\zeta({\bf k}_1)\zeta({\bf k}_2)\rangle_{{\bf k}_3}&\simeq (2\pi)^3\delta({\bf k}_1+{\bf k}_{2})\frac{2\pi^2}{k_1^3}\mathcal{P}_\zeta(k_1)\cdot{\rm det}({\bf S}^{-1})\biggl(1-\zeta({\bf k}_{3})-\frac{1}{2}\frac{k_{1l}k_{1m}}{k_1^2} h_{lm}({\bf k}_{3})\biggr)^{n_s-4}\notag\\
&\simeq(2\pi)^3\delta({\bf k}_1+{\bf k}_{2})\frac{2\pi^2}{k_1^3}\mathcal{P}_\zeta(k_1)\biggl[1+(1-n_s)\zeta({\bf k}_{3})+\biggl(2-\frac{n_s}{2}\biggr)\frac{k_{1l}k_{1m}}{k_1^2}h_{lm}({\bf k}_{3})\biggr],
\end{align}
where we parametrized the scalar power spectrum as $\mathcal{P}_\zeta(k)\propto k^{1-n_s}$ and used the relations about ${\bf S}$,
${\rm det}({\bf S})=e^{3\bar\zeta}=1/{\rm det}({\bf S}^{-1})$ and $\delta({\bf S}^{-1}{\bf k}_1+{\bf S}^{-1}{\bf k}_2)={\rm det}({\bf S})\delta({\bf k}_1+{\bf k}_2)$. Similarly to the above, the power spectrum of the tensor perturbations on the perturbed spacetime is evaluated as
\begin{align}
\langle\xi^{(s_1)}({\bf k}_1)\xi^{(s_2)}({\bf k}_{2})\rangle_{{\bf k}_3}
&\simeq (2\pi)^3\delta({\bf k}_1+{\bf k}_2)\delta_{s_1s_2}\frac{\pi^2}{k_1^3}\mathcal{P}_h(k_1)\biggl[1-n_t\zeta({\bf k}_{3})+\frac{1}{2}(3-n_t)\frac{k_{1l}k_{1m}}{k_1^2} h_{lm}({\bf k}_{3})\biggr],
\end{align}
where we parametrized the tensor power spectrum as $\mathcal{P}_h(k)\propto k^{n_t}$ and ignored the terms of higher-order in the perturbations. As a result, one can obtain the three-point functions at the squeezed limits given in Eqs.~(\ref{eq: sss-consistency-relation})--(\ref{eq: ttt-consistency-relation}).

\section{Shapes for $\nu=3/2$}\label{Appendix: shape}
In this section, we summarize the shapes for $\nu=3/2$ including the mixed-helicity case. Since we have explained the shape of the scalar auto-bispectrum, we study the other bispectra. We first study the scalar-scalar-tensor bispectrum.

\begin{figure} [htb]
     \begin{tabular}{cc}
        \begin{minipage}{0.45\hsize}
            \centering
            \includegraphics[width=7.cm]{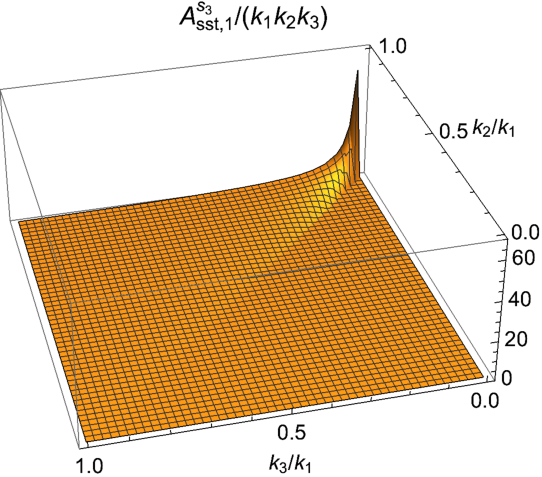}
        \end{minipage} &
        \begin{minipage}{0.45\hsize}
            \centering
            \includegraphics[width=7.cm]{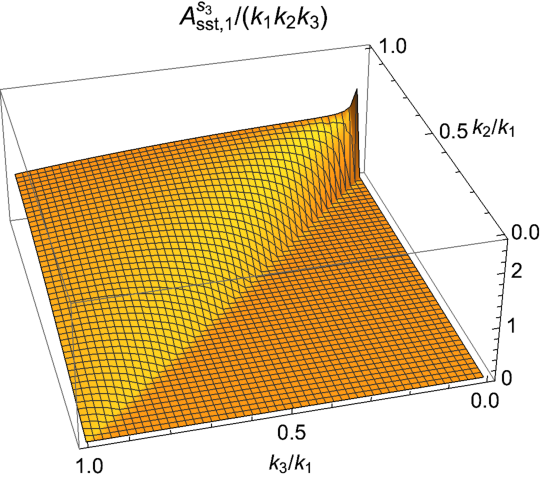}
        \end{minipage} 
    \end{tabular}
\caption{\emph{Left}:$\mathcal{A}^{s_3}_{sst,1}/(k_1k_2k_3)$ as a function of $k_2/k_1$ and $k_3/k_1$. We assumed $c_t/c_s=1$ and normalized the plot by setting $\mathcal{A}/(k_1k_2k_3)$ to $1$ for the equilateral configuration.\emph{Right}: $\mathcal{A}^{s_3}_{sst,1}/(k_1k_2k_3)$ as a function of $k_2/k_1$ and $k_3/k_1$. We assumed $c_t/c_s=100$ and normalized the plot by setting $\mathcal{A}/(k_1k_2k_3)$ to $1$ for the equilateral configuration.}\label{Fig: c1}
\end{figure}

\begin{figure} [htb]
     \begin{tabular}{cc}
        \begin{minipage}{0.45\hsize}
            \centering
            \includegraphics[width=7.cm]{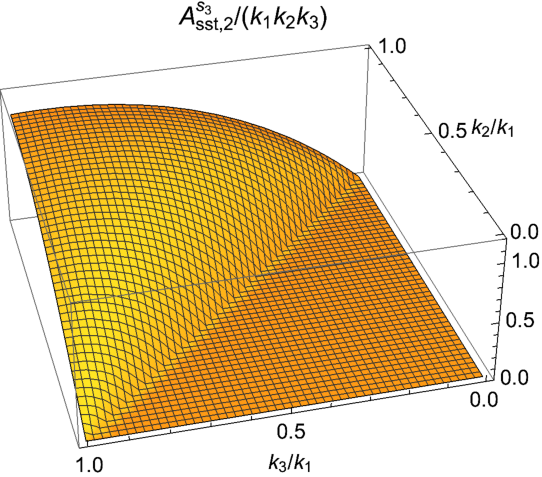}
        \end{minipage} &
        \begin{minipage}{0.45\hsize}
            \centering
            \includegraphics[width=7.cm]{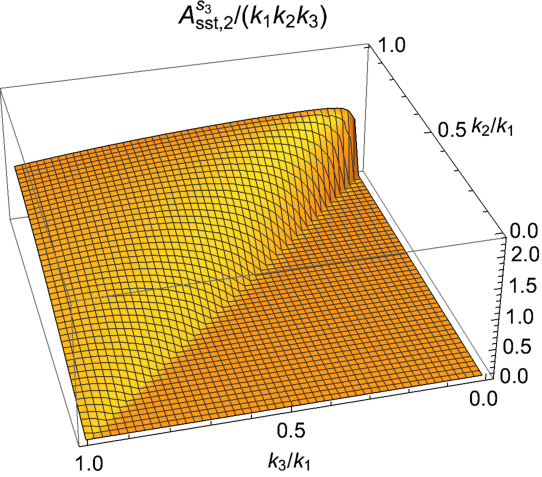}
        \end{minipage} 
    \end{tabular}
\caption{\emph{Left}:$\mathcal{A}^{s_3}_{sst,2}/(k_1k_2k_3)$ as a function of $k_2/k_1$ and $k_3/k_1$. We assumed $c_t/c_s=1$ and normalized the plot by setting $\mathcal{A}/(k_1k_2k_3)$ to $1$ for the equilateral configuration.\emph{Right}: $\mathcal{A}^{s_3}_{sst,2}/(k_1k_2k_3)$ as a function of $k_2/k_1$ and $k_3/k_1$. We assumed $c_t/c_s=100$ and normalized the plot by setting $\mathcal{A}/(k_1k_2k_3)$ to $1$ for the equilateral configuration.}\label{Fig: c2}
\end{figure}

As shown in Figure~\ref{Fig: c1}, the $c_1$ term for $c_t/c_s=0.01$ has a finite but sharp peak at $k_3/k_1\ll1$, while that peak becomes smaller for $c_t/c_s=100$.\footnote{We can rewrite $\mathcal{A}^{s_3}_{sst,1}/(k_1k_2k_3)$ as
\begin{align}
\frac{\mathcal{A}^{s_3}_{sst,1}}{k_1k_2k_3}&\propto \frac{K(k_1+k_2-k_3)}{k_1k_2k_3^3[k_1+k_2+(c_t/c_s)k_3]^2}[(k_1-k_2)^2-k_3^2]\notag\\
&\quad\times[(k_1+k_2)(k_1^2+k_1k_2+k_2^2)+2(c_t/c_s)(k_1^2+k_1k_2+k_2^2)k_3+2(c_t/c_s)^2(k_1+k_2)k_3^2+(c_t/c_s)^3k_3^3],
\end{align}
where we have ignored a proportional coefficient. 
For $k_1\simeq k_2\gg k_3$, the first line of the above is enhanced in proportion to $k_3^{-3}[k_1+k_2+(c_t/c_s)k_3]^{-2}$, but suppressed in proportion to $[(k_1-k_2)^2-k_3^2]$. For $c_t/c_s\leq1$, the second line is proportional to $k_1^3$, and the resultant bispectrum has a peak at the squeezed limit. On the other hand, for $c_t/c_s\gg1$, the $k_3$ term in the second line ($\propto k_3, k_3^2, k_3^3$) can become non-negligible at the squeezed limit, which made the enhancement behavior ($\sim k_3^{-3}$) in the first line weaker.}
Next, as shown in Figure~\ref{Fig: c2}, the $c_2$ term for $c_t/c_s=1$ has a shape close to the equilateral one, while that for $c_t/c_s=100$ does not. In particular, the latter takes a larger value around $k_3\ll k_2\simeq k_1$ than the former.\footnote{Now $\mathcal{A}^{s_3}_{sst,2}$ behaves as
\begin{align}
\frac{\mathcal{A}^{s_3}_{sst,2}}{k_1k_2k_3}&\propto \frac{(k_1+k_2-k_3)}{k_1k_2k_3}[(k_1-k_2)^2-k_3^2]\notag\\
&\quad\ \times\frac{K}{[k_1+k_2+(c_t/c_s)k_3]^3}[2(k_1^2+3k_1k_2+k_2^2)+3(c_t/c_s)(k_1+k_2)k_3+(c_t/c_s)^2k_3^2].
\end{align}
The first line has a peak at the equilateral limit. For $k_1\simeq k_2\gg k_3$, the second line has a sharp peak for $c_t/c_s\gg1$, while not for $c_t/c_s\leq1$. Therefore, $c_t/c_s\gg1$ has deformed the equilateral shape in such a way that the $c_2$ term for $c_t/c_s\gg1$ has a larger value at $k_2\simeq k_1\gg k_3$ than that for $c_t/c_s\leq1$.}  
The $c_3, c_4, c_5$, and $c_6$ terms have similar shapes and $c_t/c_s$ dependence to the $c_2$ term, and hence we do not show those. We thus find that the $c_1$ term for $c_t/c_s=0.01,1$ has a peak at the squeezed limit, the $c_i$ ($i\neq1$) terms for $c_t/c_s=0.01,1$ peak at the equilateral limit, and the others appear to be away from the squeezed or the equilateral ones. 

\begin{figure}[htb]
\centering
\includegraphics[width=70mm]{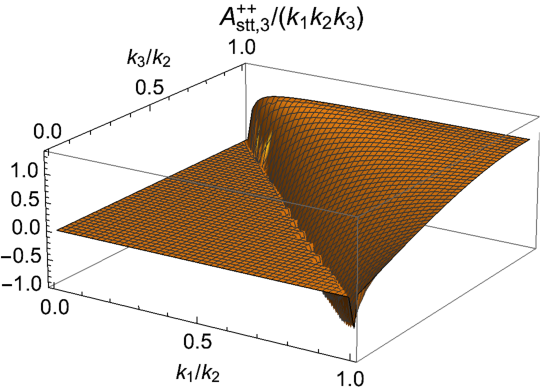}
\captionsetup{justification=raggedright}
\caption{$\mathcal{A}^{++}_{stt,3}/(k_1k_2k_3)$ as a function of $k_1/k_2$ and $k_3/k_2$. We normalized the plot by setting $\mathcal{A}/(k_1k_2k_3)$ to $1$ for the equilateral configuration.}\label{Fig: b3csch001}
\end{figure}

We then move to the scalar-tensor-tensor bispectrum. We have studied the $b_{(1,2,4,5,6,7)}$ terms in the same-helicity case in the main text, and we start with the $b_3$ term in the same-helicity case.
The $b_3$ term for $c_t/c_s=1$ and $c_t/c_s=100$ has a shape similar to that of the $b_{1+2}$ term for $c_t/c_s=100$ (Figure~\ref{Fig: b12}). Figure~\ref{Fig: b3csch001} shows the shape of the $b_3$ term for $c_t/c_s\ll1$.\footnote{The $b_3$ term for $c_t/c_s\ll1$ is slightly enhanced at $k_1\ll k_2\simeq k_3$ than that for $c_t/c_s=1, 100$. For $k_3\simeq k_2$, we have
\begin{align}
\frac{\mathcal{A}^{++}_{stt,3}}{k_1k_2k_3}&\propto \frac{k_1[k_1+3(c_t/c_s)k_2]}{[k_1+2(c_t/c_s)k_2]^2}.
\end{align}
For $k_1\ll k_2$, the above can slightly be enhanced only for $c_t/c_s\ll1$ which makes the denominator smaller.
}

\begin{figure} [htb]
     \begin{tabular}{cc}
        \begin{minipage}{0.45\hsize}
            \centering
            \includegraphics[width=7.cm]{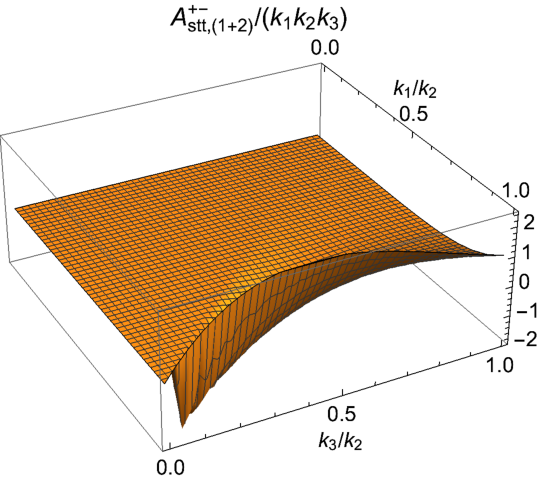}
        \end{minipage} &
        \begin{minipage}{0.45\hsize}
            \centering
            \includegraphics[width=7.cm]{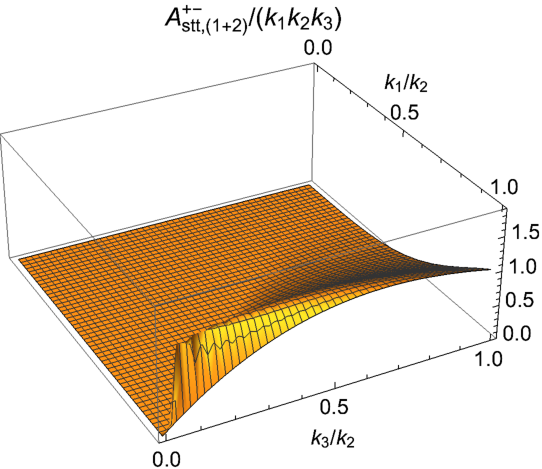}
        \end{minipage} 
    \end{tabular}
\caption{\emph{Left}:$(\mathcal{A}^{+-}_{stt,1}+\mathcal{A}^{+-}_{stt,2})/(k_1k_2k_3)$ as a function of $k_1/k_2$ and $k_3/k_2$. We took $c_t/c_s=1$ and normalized the plot by setting $\mathcal{A}/(k_1k_2k_3)$ to $1$ for the equilateral configuration.\emph{Right}: $(\mathcal{A}^{+-}_{stt,1}+\mathcal{A}^{+-}_{stt,2})/(k_1k_2k_3)$ as a function of $k_1/k_2$ and $k_3/k_2$. We took $c_t/c_s=0.01$ and normalized the plot by setting $\mathcal{A}/(k_1k_2k_3)$ to $1$ for the equilateral configuration.}\label{Fig: b12pm}
\end{figure}

\begin{figure}[htb]
\centering
\includegraphics[width=70mm]{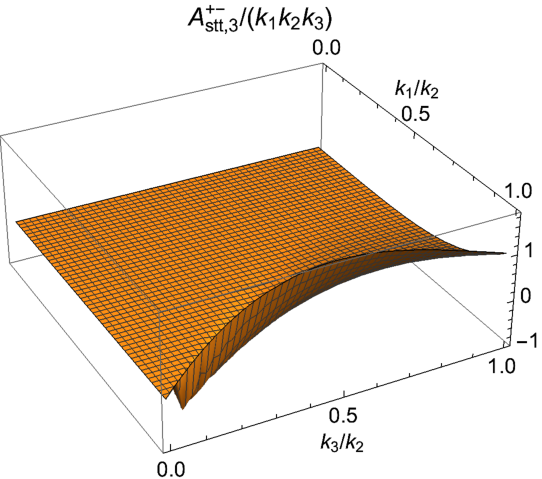}
\captionsetup{justification=raggedright}
\caption{$\mathcal{A}^{+-}_{stt,3}/(k_1k_2k_3)$ as a function of $k_1/k_2$ and $k_3/k_2$. We took $c_t/c_s=1$ and normalized the plot by setting $\mathcal{A}/(k_1k_2k_3)$ to $1$ for the equilateral configuration.}\label{Fig: b3csch1pm}
\end{figure}

\begin{figure} [htb]
     \begin{tabular}{cc}
        \begin{minipage}{0.45\hsize}
            \centering
            \includegraphics[width=7.cm]{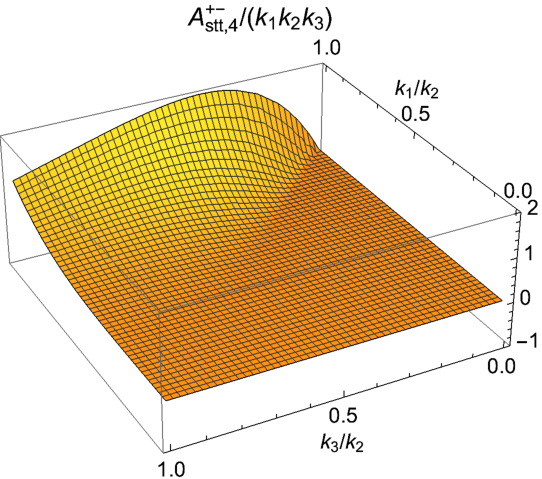}
        \end{minipage} &
        \begin{minipage}{0.45\hsize}
            \centering
            \includegraphics[width=7.cm]{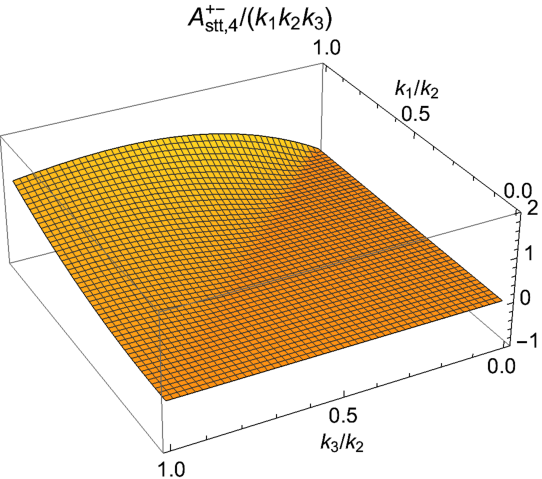}
        \end{minipage} 
    \end{tabular}
\caption{\emph{Left}:$\mathcal{A}^{+-}_{stt,4}/(k_1k_2k_3)$ as a function of $k_1/k_2$ and $k_3/k_2$. We took $c_t/c_s=1$ and normalized the plot by setting $\mathcal{A}/(k_1k_2k_3)$ to $1$ for the equilateral configuration.\emph{Right}: $\mathcal{A}^{+-}_{stt,4}/(k_1k_2k_3)$ as a function of $k_1/k_2$ and $k_3/k_2$. We took $c_t/c_s=0.01$ and normalized the plot by setting $\mathcal{A}/(k_1k_2k_3)$ to $1$ for the equilateral configuration.}\label{Fig: b4pm}
\end{figure}

\begin{figure}[htb]
\centering
\includegraphics[width=70mm]{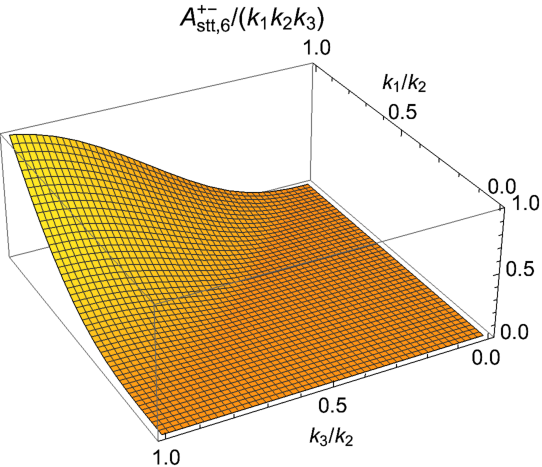}
\captionsetup{justification=raggedright}
\caption{$\mathcal{A}^{+-}_{stt,6}/(k_1k_2k_3)$ as a function of $k_1/k_2$ and $k_3/k_2$. We have taken $c_t/c_s=1$ and normalized the plot by setting $\mathcal{A}/(k_1k_2k_3)$ to $1$ for the equilateral configuration.}\label{Fig: b6csch1pm}
\end{figure}

Below, we consider the mixed-helicity case, $s_2=-s_3=+1$. Figure~\ref{Fig: b12pm} shows that the $b_{1+2}$ terms for $c_t/c_s=1$ and $c_t/c_s=0.01$ do not have any sharp peaks at the squeeezed limit $k_1\to0$.
Then, Figure~\ref{Fig: b3csch1pm} shows that the $b_3$ term for $c_t/c_s=1$ does not have a sharp peak at the squeezed limit $k_1\to0$.
Figure~\ref{Fig: b4pm} shows that the $b_4$ terms for both $c_t/c_s=1$ and $c_t/c_s=0.01$ do not have any sharp peaks. 
The $b_5$ term has a similar shape to the $b_4$ one.
Figure~\ref{Fig: b6csch1pm} shows that the $b_6$ term for $c_t/c_s=1$ has a peak at the equilateral limit. 
In addition, the $b_6$ term for $c_t/c_s=100$ and $c_t/c_s=0.01$ and the $b_7$ term for $c_t/c_s=1, 100, 0.01$ have similar shapes to the $b_6$ term for $c_t/c_s=1$ (Figure~\ref{Fig: b6csch1pm}).

\begin{figure}[htb]
\centering
\includegraphics[width=70mm]{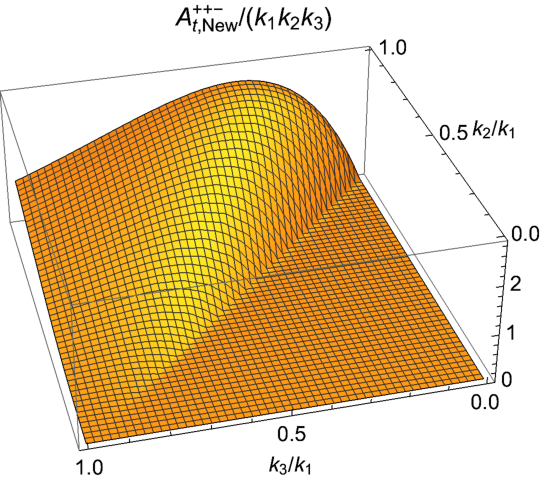}
\captionsetup{justification=raggedright}
\caption{$\mathcal{A}^{++-}_{t,{\rm New}}/(k_1k_2k_3)$ as a function of $k_2/k_1$ and $k_3/k_1$. We normalized the plot by setting $\mathcal{A}/(k_1k_2k_3)$ to $1$ for the equilateral configuration.}\label{Fig: tNewpm}
\end{figure}

Last, we study the shapes of the tensor auto-bispectrum. Both same-helicity ($s_1=s_2=s_3=+1$) and mixed-helicity ($s_1=s_2=-s_3=+1$) cases have been studied in Refs.~\cite{Gao:2011vs,Gao:2012ib,Akita:2015mho}. Thus, we here recap their results. In the same-helicity case, the GR and New terms peak at the equilateral and squeezed limits, respectively. In the mixed-helicity case, the GR term has a peak at the squeezed limit, while that of the New term appears to be away from the squeezed or equilateral ones as shown in Figure~\ref{Fig: tNewpm}.

\section{Concrete example for $\nu=3/2$}\label{app: example}
The purpose of this section is to show that the general bounce model for $\nu=3/2$ includes sufficient functional degrees of freedom to achieve the small tensor-to-scalar ratio and the small scalar non-Gaussianity at the same time. As an example, we choose the arbitrary functions $G_i$ in the Lagrangian as
\begin{align}
G_2&=f_2(Y)e^{-4\lambda\phi},\ G_3=f_3(Y)e^{-2\lambda\phi},\\
G_4&=M_4^2e^{-2\lambda\phi}, G_5=0,
\end{align}
where $Y:=X e^{2\lambda\phi}$, and $\lambda$ and $M_4$ are constant. Then, let us explore the solution with the following ansatz:
\begin{align}
\phi=\frac{1}{\lambda}\ln(-M t),
\end{align}
where $M$ is a constant. On the contracting background, the function $Y$ is constant: $Y=\bar Y=M^2/(2\lambda^2)$. With the ansatz, one can solve the Friedmann and evolution equations under the following conditions:
\begin{align}
f_2(\bar Y)&=-4\lambda\bar Y g_3(\bar Y)-12\lambda^2 \bar Y(n^2-2n-2)M_4^2,\\
\bar Y f_{2Y}(\bar Y)&=-2\lambda\bar Y[2f_3(\bar Y)+(3n-1)\bar Y f_{3Y}(\bar Y)]-12\lambda^2 \bar YM_4^2.
\end{align}
First, to obtain a small tensor-to-scalar ratio, we require $c_t/c_s=1$ and $\mathcal{F}_S/\mathcal{F}_T\ll1$. Since $c_t=1$, the first condition reads $c_s=1$ which can be achieved by imposing
\begin{align}
\bar Y^2 f_{2YY}(\bar Y)&=-2(n+3)\lambda\bar Y^2 g_{3Y}(\bar Y)-\frac{2\bar Y^3f_{3Y}^2(\bar Y)}{M_4^2}+2(1-3n)\lambda\bar Y^3 f_{3YY}(\bar Y).
\end{align}
The second condition, $\mathcal{F}_S/\mathcal{F}_T\ll1$, is realized under the condition,
\begin{align}
\bar Y f_{3Y}(\bar Y)=\frac{2\delta_1}{1+\delta_1}(n-1)\lambda M_4^2,
\end{align}
where $\delta_1(=\mathcal{F}_S/\mathcal{F}_T)\ll1$. As a result, the tensor-to-scalar ratio reads
\begin{align}
r=16\delta_1.
\end{align}
Here, the parameter $\delta_1$ is not related to the spectral index of the tensor power spectrum, and hence the standard consistency relation $r=-8c_sn_t$ is clearly violated. Also, the coefficients of the quadratic actions are of the form,
\begin{align}
\mathcal{G}_S&=\mathcal{F}_S=\frac{2M_4^2}{M^2t^2}\delta_1,\\
\mathcal{G}_T&=\mathcal{F}_T=\frac{2M_4^2}{M^2t^2}.
\end{align}
As long as $\delta_1>0$, both scalar and tensor perturbations satisfy the stability conditions, $\mathcal{G}_S, \mathcal{F}_S, \mathcal{G}_T, \mathcal{F}_T>0$.

Then, we investigate the scalar non-Gaussianity and derive conditions to make the non-linearity parameters at most $\mathcal{O}(1)$.
First, we have
\begin{align}
\Lambda_2&=-3\delta_1, \Lambda_3=\delta_1, \Lambda_5=-\frac{1}{2}\delta_1(4+3\delta_1),\ \Lambda_6=-\frac{3}{4}\delta_1^2,\notag\\
\Lambda_7&=-\frac{2}{3}\biggl(\frac{n}{1-n}\biggr)^2(1+\delta_1)^2,\ \Lambda_8=-\frac{2n}{1-n}\delta_1(1+\delta_1).
\end{align}
The $\Lambda_1$ and $\Lambda_4$ terms can be enhanced under $\mathcal{F}_S/\mathcal{F}_T\ll1$. One can avoid such enhancements provided that
\begin{align}
\bar Y^2 f_{3YY}(\bar Y)&=-\frac{(1-n)^3}{n^2}\lambda M_4^2\delta_2,\\
(3n-1)\bar Y^3f_{3YYY}(\bar Y)&=-\frac{\bar Y^2f_{2YYY}(\bar Y)}{2\lambda},
\end{align}
under which we have $\Lambda_1, \Lambda_4=\mathcal{O}(1)$ for $1/n=\mathcal{O}(1)$.
Thus, by choosing $\delta_1, \delta_2\ll1$, any momentum triangle limits yield $f_{\rm NL}$ at most of $\mathcal{O}(1)$. See Ref.~\cite{Akama:2019qeh} for a concrete example to realize such $f_{\rm NL}$ for $\nu=-3/2$. 

The coefficients for the cross interactions are obtained as
\begin{align}
\frac{c_1}{\mathcal{F}_S}&=1, \frac{Hc_2}{\mathcal{F}_S}=\frac{n}{n-1}(1+\delta_1),\ \frac{c_3c_s^2}{\mathcal{F}_S}=-\frac{3\delta_1}{2},\ c_4=c_5=0,\ \frac{c_6c_s^2}{\mathcal{F}_S}=\frac{\delta_1}{4},\\
\frac{b_1c_t^2}{\mathcal{F}_T}&=-\frac{3\delta_1}{8},\ \frac{b_2}{\mathcal{F}_T}=\frac{\delta_1}{8},\ \frac{b_3c_t^2}{\mathcal{F}_T}=-\frac{\delta_1}{4},\  b_4=b_5=b_6=b_7=0.
\end{align}
The above agrees with the order estimation in Sec.~\ref{Sec: model-space}. The order estimation has not excluded the possibility of enhancing the $c_2$ and $c_5$ terms, but in the above example the $c_2$ term was not enhanced and $c_5=0$. Therefore, more functional degrees of freedom are necessary to achieve the enhancements.

\bibliography{main}
\bibliographystyle{JHEP}
\end{document}